\PassOptionsToPackage{table}{xcolor}
\documentclass{article} 
\usepackage{iclr2027_conference,times}

\usepackage{amsmath,amsfonts,bm}

\def\eqref#1{equation~\ref{#1}}

\def\1{\bm{1}}

\DeclareMathAlphabet{\mathsfit}{\encodingdefault}{\sfdefault}{m}{sl}
\SetMathAlphabet{\mathsfit}{bold}{\encodingdefault}{\sfdefault}{bx}{n}

\usepackage[utf8]{inputenc}
\usepackage[T1]{fontenc}
\usepackage[hypertexnames=false]{hyperref}
\usepackage{url}
\usepackage{graphicx}
\usepackage{booktabs}
\usepackage{multirow}
\usepackage{array}
\usepackage{longtable}
\usepackage{pdflscape}
\usepackage{amsmath}
\usepackage{amsfonts}
\usepackage{amssymb}
\usepackage{nicefrac}
\usepackage{microtype}
\usepackage{xcolor}
\usepackage{float}
\usepackage{enumitem}
\usepackage{algorithm}
\usepackage{algpseudocode}
\usepackage{wrapfig}
\usepackage{subcaption}
\usepackage{caption}
\usepackage{makecell}
\usepackage{tabularx}
\usepackage{amsthm}
\usepackage[table]{xcolor}

\usepackage{soul}
\definecolor{mireuse}{RGB}{252,240,198}
\definecolor{miepara}{RGB}{225,235,246}
\definecolor{miestory}{RGB}{226,240,227}
\definecolor{mietone}{RGB}{249,226,233}

\theoremstyle{definition}

\theoremstyle{plain}
\newtheorem{theorem}{Theorem}

\newtheorem{lemma}{Lemma}

\setlist[itemize]{nosep, left=0pt, labelsep=0.5em, itemsep=0pt, topsep=0pt, parsep=0pt, partopsep=0pt}
\setlist[enumerate]{nosep, left=0pt, labelsep=0.5em, itemsep=0pt, topsep=0pt, parsep=0pt, partopsep=0pt}
\newcolumntype{L}[1]{>{\raggedright\arraybackslash}p{#1}}

\title{ThreadBench: Benchmarking Realism in Agentic Simulation of Real-world Discussions}

\author{
\textbf{Yaoning Yu$^{1}$, Ye Yu$^{1}$, John Sciarappo$^{2}$, Haojing Luo$^{3}$}\\
\textbf{Kai-min Chang$^{2}$, Yi-Chia Wang$^{4}$, Haohan Wang$^{1}$}\\
{ $^{1}$University of Illinois Urbana-Champaign}
{ $^{2}$U.S. Bank}\\
{ $^{3}$Starc.Institute}
{ $^{4}$Carnegie Mellon University}
}

\iclrfinalcopy

\begin{document}

\maketitle

\lhead{}

\begingroup
\renewcommand{\thefootnote}{}
\footnotetext{Preprint.}
\endgroup

\begin{abstract}
LLM agents are increasingly used to simulate real-world interactions, but it remains unclear whether simulated behaviors preserve the content patterns and interaction dynamics of real human behaviors. 
Existing evaluations often focus on individual comments or
task performance, rather than whether simulated behaviors match human behavior at thread level. 
In this paper, we use Reddit discussions
as a concrete testbed for real-world social simulation. Reddit provides public, topic-grounded, multi-party interactions with diverse opinions, experiences, and interaction patterns, making it a useful setting for evaluating whether LLM agents reproduce the distributional and structural properties of real online discussions.
We introduce ThreadBench, a benchmark for Reddit discussion simulation built from 9{,}817 real Reddit threads spanning 8 diverse domains and 116 subdomains. ThreadBench uses statistical analyses to compare generated and real discussions across 12 metrics grouped in 4 families: uniformity, expression, tone, and interaction. 
We further develop Thread-\textbf{P}rompt-based \textbf{D}iscussion
\textbf{G}eneration (ThreadPDG), a Planner--Writer framework that learns how a
real discussion grows. The Planner produces thread- and comment-level plans
from a real seed discussion, and the Writer generates comments from these plans.
For each seed, we generate multiple candidate threads and combine their best branches
using thread-level statistics. Across all domains and multiple LLMs, ThreadPDG achieves ThreadBench pass rates of 58--100\%, compared with 0--17\% for the baselines. Human evaluation with 436 annotators further shows that ThreadPDG is not distinguished from real discussions in 78.7\% of pairs, outperforming existing simulation methods.



\end{abstract}
\section{Introduction}

LLM-based agents are increasingly used as simulators of human behavior, social interaction, and online communities~\citep{park2023generative, zhou2025sotopia, yang2411oasis}. These systems provide a controllable way to study how people discuss products,
react to events, and exchange opinions. For such simulations to be useful as a controllable and observable window for
studying how people communicate, express opinions, and interact with others,
simulated discussions should reproduce real-world discussion behavior at the
thread level. In real discussions, users bring different points and use different wording;
some share personal experiences, while others disagree or respond with
different tones. Replies also grows unevenly, forming different
depths and comments of very different lengths. However, existing studies often focus on individual comments, controlled role-playing settings, or task performance in simulated environments\citep{zhou2025sotopia, yang2411oasis}. As a result, it remains difficult to measure whether a simulator can reproduce real-world thread-level discussion behavior, leading to a practical need for such measurement.

Reddit provides a practical setting for studying this problem
~\citep{amaya2021new, proferes2021studying}. As a large online platform,
Reddit reflects a broad range of real-world topics, opinions, experiences,
and interactions among users, while preserving both comment content and reply
structure. This makes Reddit a useful testbed for studying thread-level
communication and interaction, and for measuring how closely simulated
discussions reproduce real online discussion behavior.

Therefore, we introduce ThreadBench, a benchmark for Reddit discussion simulation. ThreadBench combines 9{,}817 real Reddit threads
and an evaluation framework that compares generated and real discussions through both comment-derived signals and reply-tree structure. Using ThreadBench, we evaluate whether sets of simulated threads match real Reddit discussions across comment content and thread structure. We find that current simulators remain mismatched to real Reddit threads even when individual comments are fluent. Figure~\ref{fig:excerpt-pair} shows one example of this gap.


To reduce this gap, we introduce ThreadPDG, a prompt-based planner-writer discussion
generation method framework. The Planner first studies a real seed thread
and specifies how its branches and comments should develop, while the Writer generates comments from these plans. We produce multiple candidate threads and select their best branches(comments and sub-comments). Evaluated with ThreadBench, ThreadPDG matches real discussion behavior substantially more closely
than existing baselines: it achieves pass rates of 58-100\% with multiple LLMs, while OASIS and SynthPAI remain at 0--17\%. A separate human evaluation from 436 annotators further shows that ThreadPDG is not distinguished from the real discussion in 78.7\% of evaluated pairs.

\begin{figure}[t]
\centering
\scriptsize
\setlength{\tabcolsep}{0pt}
\providecommand{\micomment}[3]{\par\hangindent=#1\hangafter=0\noindent
  \hspace*{#1}#2~#3\par\vspace{1.6pt}}
\providecommand{\mrk}[1]{\textcolor{black!45}{#1}}
\providecommand{\mkreuse}[1]{{\sethlcolor{mireuse}\hl{#1}}}
\providecommand{\mkpara}[1]{{\sethlcolor{miepara}\hl{#1}}}
\providecommand{\mkstory}[1]{{\sethlcolor{miestory}\hl{#1}}}
\providecommand{\mktone}[1]{{\sethlcolor{mietone}\hl{#1}}}
\begin{tabular}{@{}p{0.465\textwidth}@{\hspace{0.7em}}|@{\hspace{0.7em}}p{0.465\textwidth}@{}}
\noindent\hfil\textbf{Real discussion}\hfil &
\noindent\hfil\textbf{Simulated discussion}\hfil \\
\addlinespace[2pt]
\begin{minipage}[t]{0.465\textwidth}\raggedright
\micomment{0.0em}{\textbf{User~1:}}{Streams all on at the same time, or just different devices now and then? I would call support and ask. \mrk{\mbox{\scriptsize(20 words)}}}
\micomment{0.0em}{\textbf{User~3:}}{So I found a loophole that works with Fire Sticks. \mkstory{I had been signed in through my parents' account an hour away, until one day it forced me out and said I was not the main profile on the account.} If you open a show from the home screen instead of the app, it lets you straight in. Six months now, no new subscription. \mrk{\mbox{\scriptsize(64 words)}}}
\micomment{0.0em}{\textbf{User~5:}}{Mine just sends a code to my phone every time. \mrk{\mbox{\scriptsize(10 words)}}}
\micomment{1.1em}{$\hookrightarrow$\,\textbf{User~7:}}{Why not sign in on your phone and cast to the TV. That way the sign in is always the same. \mrk{\mbox{\scriptsize(21 words)}}}
\micomment{2.2em}{$\hookrightarrow$\,\textbf{User~3:}}{Casting drops to 720p on my setup, so that is not a fix. \mrk{\mbox{\scriptsize(13 words)}}}
\micomment{0.0em}{\textbf{User~8:}}{\mktone{Just get your own subscription!} No wonder my subscription price keeps going up. \mrk{\mbox{\scriptsize(13 words)}}}
\end{minipage} &
\begin{minipage}[t]{0.465\textwidth}\raggedright
\micomment{0.0em}{\textbf{User~1:}}{Then why did it work yesterday? Keep seeing these weird sign-in loops too. Something's off. \mrk{\mbox{\scriptsize(15 words)}}}
\micomment{0.0em}{\textbf{User~2:}}{\mktone{Yeah, same here.} \mkreuse{If it was working yesterday and now it's doing this,} feels like the app is messing up again. \mrk{\mbox{\scriptsize(21 words)}}}
\micomment{0.0em}{\textbf{User~3:}}{\mktone{Ugh, same.} \mkreuse{If it worked yesterday and now it's suddenly a mess, that's not user error, that's the app being weird again.} Did it on me too. \mrk{\mbox{\scriptsize(27 words)}}}
\micomment{0.0em}{\textbf{User~6:}}{\mktone{Yeah,} \mkpara{that sounds like a platform or session issue, not you.} \mkreuse{If it worked yesterday and now it's suddenly doing this, that's the app being weird again.} \mrk{\mbox{\scriptsize(27 words)}}}
\micomment{0.0em}{\textbf{User~7:}}{\mkpara{This one feels like a classic platform gremlin situation, not a random user-error thing.} I'd suspect expired session tokens or cached login weirdness first. \mrk{\mbox{\scriptsize(24 words)}}}
\micomment{0.0em}{\textbf{User~8:}}{\mktone{Ugh, same here.} \mkreuse{If it worked yesterday and now it's suddenly a mess,} that sounds like the app being weird, not something you did. \mrk{\mbox{\scriptsize(24 words)}}}
\end{minipage} \\
\addlinespace[1pt]
\midrule
\scriptsize\mrk{self-BLEU-4}~0.05\quad
  \mrk{length CV}~0.79\quad
  \mrk{avg.\ depth}~1.50 &
\scriptsize\mrk{self-BLEU-4}~0.18\quad
  \mrk{length CV}~0.18\quad
  \mrk{avg.\ depth}~1.00 \\
\bottomrule
\end{tabular}
\caption{Highlights mark \mkreuse{reused wording}, \mkpara{paraphrases},
\mkstory{personal narratives}, and \mktone{tone}. Indentation indicates replying to the top comment, and word counts indicate comment length. Below each discussion, self-BLEU-4 measures phrase reuse
across comments (higher is more repetitive), length CV measures how much comment
lengths vary (higher is more varied), and avg.\ depth is the mean reply depth
(higher is more nested). The simulated discussion shows
same wording/interpretation and a tendency to agree, whereas the
real discussion contains more diverse responses, personal experiences, disagreement,
and deeper interactions.}
\label{fig:excerpt-pair}
\vspace{-12pt}
\end{figure}

Our contributions are:

\begin{itemize}
    \item We introduce \textbf{ThreadBench}, a benchmark for evaluating discussion realism that combines 9{,}817 real Reddit threads from 8 domains and 116 subdomains with 12 thread-level evaluation metrics.
    
    \item We develop \textbf{ThreadPDG}, a \textbf{P}rompt-based Planner--Writer \textbf{D}iscussion \textbf{G}eneration method that plans thread development before comment generation and further refines complete discussions through multi-draft branch composition.
    
    \item Across domains and multiple LLMs, ThreadPDG substantially outperforms
    existing baselines on ThreadBench, with human evaluation also showing that its
    discussions are harder to distinguish from real ones.
\end{itemize}
\section{Related Work}

\textbf{LLM-based Social Simulation.} Recent work uses LLMs as agents to simulate people, social interaction, and group behavior. Generative Agents studies everyday routines and social behavior in a small town-like environment \citep{park2023generative}. SOTOPIA and SOTOPIA-$\pi$ move to role-based multi-turn interaction and social intelligence evaluation, where agents pursue goals and are judged by how they behave in social settings \citep{zhou2310sotopia, wang2024sotopia}. These systems show that LLM agents can keep interactions going over time, but they do not mainly study whether generated online discussions match the distribution of real ones.

Other systems move closer to larger online-platform settings. OASIS studies social-media simulation with dynamic networks, user actions, posts, and recommendation systems \citep{yang2411oasis}. AgentSociety and SOTOPIA-S4 scale up the number of agents and support larger social settings and more flexible control \citep{piao2025agentsociety, zhou2025sotopia}. Y Social and MOSAIC focus more directly on online platforms, user behavior, content spread, engagement, and moderation \citep{rossetti2024social, liu2025mosaic}. Unlike these systems, our work focuses on whether simulated discussion threads
reflect how people actually communicate and interact, and reducing the gap
between simulated and real discussions.

\textbf{Simulation and Social Media Benchmarks} Recent benchmarks for evaluating whether LLMs can simulate human and social behavior remain limited.
SimBench tests whether LLMs can match human response distributions in mostly static, single-turn settings \citep{hu2025simbench}.
SoMe evaluates LLM-based social media agents across tasks such as user behavior prediction, recommendation, and comment simulation \citep{xue2026some}.

ThreadBench addresses this missing evaluation setting that evaluates Reddit discussion simulation across eight domains and asks whether a set of simulated threads matches real Reddit threads across semantic uniformity, expression, tone, interaction, and form.

\section{ThreadBench: Dataset and Evaluation Metrics}
\label{sec:benchmark}

\begin{figure}[H]
    \centering
    \includegraphics[width=1\linewidth]{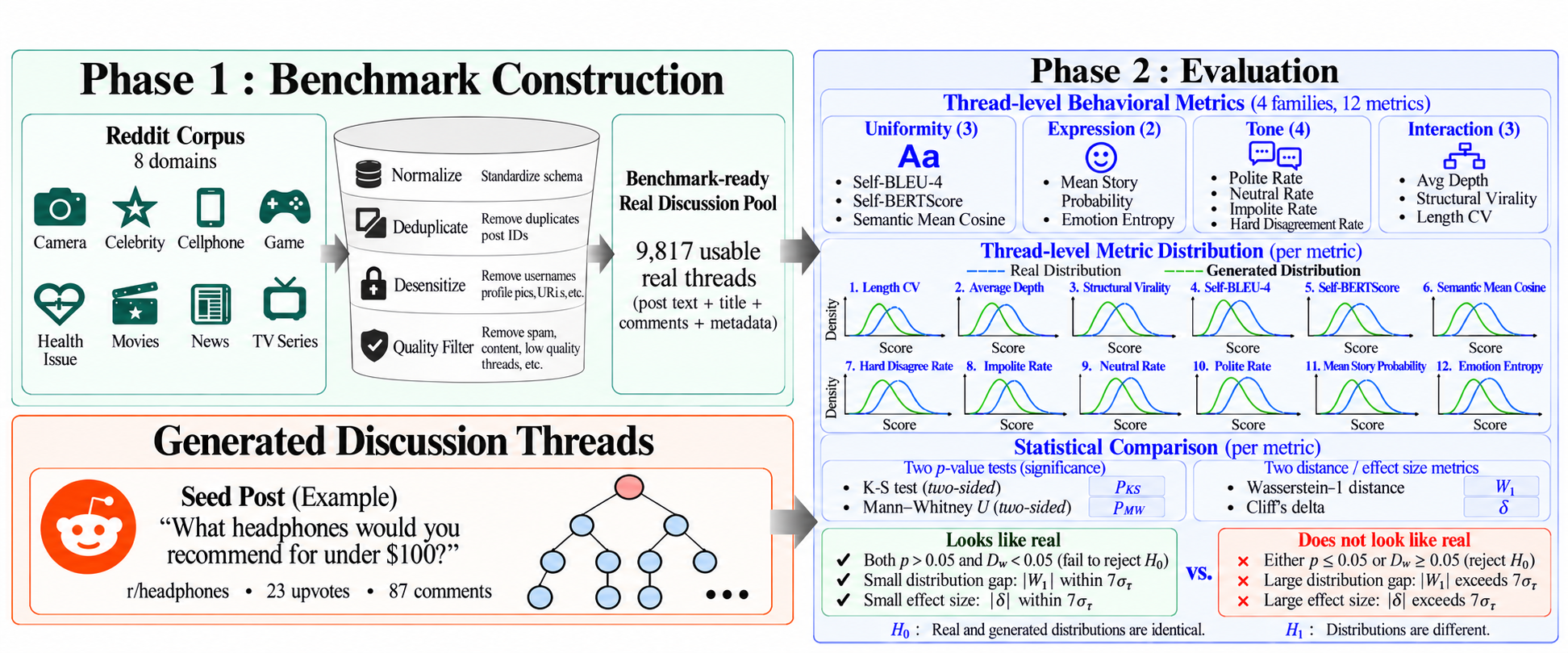}
    \vspace{-14pt}
    \caption{\textbf{Overview of the ThreadBench pipeline.} \textbf{Phase 1} constructs the benchmark by collecting Reddit discussions from 8 domains and processing them through normalization, deduplication, desensitization, and quality filtering, resulting in 9,817 real threads. \textbf{Phase 2} evaluates real and simulated discussions using 12 metrics, grouped into four families, and compares their score distributions using statistical tests and distance measures to assess how closely simulated discussions match real discussion behavior.}
    \label{fig:SimBench}
    \vspace{-14pt}

\end{figure}

ThreadBench consists of two components: (1) a benchmark corpus of 9{,}817 real
Reddit discussions spanning 8 domains and 116 subdomains, and (2) an evaluation protocol built
around 12 thread-level metrics for comparing simulated and real discussion
threads. Figure~\ref{fig:SimBench} provides an overview of ThreadBench, from
benchmark construction to its use in evaluating simulated discussions.

\subsection{Benchmark Construction}
\label{sec:benchmark_construction}

ThreadBench is constructed from real Reddit discussions spanning 8 domains and 116 subdomains, providing broad topical variation.
We select these domains to cover a diverse set of common discussion settings:
product-oriented topics such as cameras and cellphones; entertainment and
popular culture, including celebrities, games, movies, and TV series; current
events through news; and health-related discussions around symptoms,
conditions, and personal health experiences. Together, these domains cover a broad range of topics that people commonly
discuss in everyday life, including products, entertainment, current events,
and health-related concerns. Each discussion preserves the root post, its comments, the parent--reply
structure, and metadata.

\begin{wraptable}{r}{0.28\textwidth}
\centering
\scriptsize
\setlength{\tabcolsep}{3pt}
\renewcommand{\arraystretch}{0.92}
\caption{The benchmark
contains 9,817 threads and 116 subdomains in 8 domains.}
\label{tab:benchmark_inventory}
\vspace{-8pt}
\begin{tabular}{@{}lrr@{}}
\toprule
\textbf{Domain} & \textbf{Threads} & \textbf{Subdomains} \\
\midrule
Camera        & 1,922 & 44 \\
Celebrity     &   700 &  4 \\
Cellphone     & 1,395 & 44 \\
Game          & 1,000 &  5 \\
Health Issue  & 2,100 &  5 \\
Movies        &   700 &  5 \\
News          & 1,000 &  5 \\
TV Series     & 1,000 &  5 \\
\midrule
\textbf{Total} & \textbf{9,817} & \textbf{116} \\
\bottomrule
\end{tabular}

\vspace{-20pt}
\end{wraptable}

The discussion collection stage contains four preparation steps.
Normalization maps posts and comments into a unified thread schema
and reconstructs reply trees from parent identifiers.
Deduplication removes repeated posts and comments based on source
IDs. Desensitization replaces identifiable Reddit user information
with anonymized representations.
Quality filtering removes incomplete or unusable discussions
according to predefined content and structural requirements. The resulting benchmark contains 9,817 real discussion threads.
Table~\ref{tab:benchmark_inventory} provides an overview of the benchmark
composition, while Appendix~\ref{app:benchmark_inventory} details the complete Threadbench statistic.

\subsection{Benchmark Metrics}
\label{sec:scores}

ThreadBench evaluates each discussion thread using 12 metrics organized into
four families: Uniformity, Expression, Tone, and Interaction.



\textbf{Uniformity.}
A realistic discussion should contain variation not only in wording but also
in the ideas expressed by different users. We therefore measure similarity at
three levels: exact lexical overlap, contextual expression, and semantic
content. These levels are complementary because comments can avoid repeating
the same words while still expressing nearly identical ideas.

\begin{itemize}
    \item \texttt{self\_bleu\_4} measures pairwise 4-gram overlap between
    comments. We include it to detect direct phrase reuse and lexical
    repetition, which semantic measures may overlook
    \citep{zhu2018texygen}.

    \item \texttt{self\_bertscore\_mean\_f1} measures contextual similarity
    between comments. It complements lexical overlap by detecting similar
    expressions and paraphrases even when the exact wording differs
    \citep{zhang2019bertscore, zhang2024improving}.

    \item \texttt{semantic\_mean\_cosine} measures similarity in semantic
    content using \texttt{all-mpnet-base-v2}. We include it to capture whether
    comments converge on the same meanings or introduce different points,
    which may not be reflected by either lexical overlap or local contextual
    similarity \citep{reimers2019sentence, song2020mpnet}.
\end{itemize}

\textbf{Expression.}
Discussion realism also depends on how users express their opinions.
We focus on two sources of variation: whether users bring personal
experiences into the discussion and how emotional expression varies across
comments. These capture different aspects of expression: a comment can be
highly personal without being emotional, or emotional without telling a
personal story.

\begin{itemize}
    \item \texttt{mean\_story\_probability} measures the prevalence of personal
    stories and first-person experiences using StorySeeker. We include it
    because real discussions often contain users grounding their opinions or
    advice in personal experience, whereas generated discussions may remain
    more generic or impersonal
    \citep{antoniak2024people}.

    \item \texttt{emotion\_entropy} measures the diversity of emotions across
    comments using \texttt{SamLowe/roberta-base-go\_emotions}. We use entropy
    rather than the prevalence of a single emotion because realistic threads
    can contain a mixture of emotional responses rather than one dominant
    emotional style
    \citep{sam_lowe_2024}.
\end{itemize}

\textbf{Tone.}
Users differ not only in what they say but also in how they address one
another. We capture two related but distinct properties: interpersonal manner
and disagreement. They should be measured separately because disagreement does
not necessarily imply impoliteness; users can disagree politely, while an
impolite comment does not necessarily express an opposing position.

\begin{itemize}
    \item \texttt{polite\_rate}, \texttt{neutral\_rate}, and
    \texttt{impolite\_rate} jointly describe the distribution of interpersonal
    tone across comments using \texttt{Intel/polite-guard}. We use all three
    rates because simulated discussions may shift the balance among polite,
    neutral, and impolite responses rather than simply becoming more or less polite overall.

    \item \texttt{hard\_disagree\_rate} measures explicit disagreement between
    parent--reply pairs. We include disagreement separately from politeness
    because it captures opposing stance rather than social manner
    \citep{luo2023improving}.
    
\end{itemize}

\textbf{Interaction.}
Finally, realistic discussions differ in how interaction develops across the
thread. We measure three complementary properties: how deeply replies continue,
how broadly interaction is distributed across the reply tree, and how much
users differ in the length of their contributions. These properties are not
interchangeable: two threads can have the same maximum or average depth while
having very different overall branching structures, and similar reply
structures can still contain very different patterns of contribution length.

\begin{itemize}
    \item \texttt{avg\_depth} measures how deeply replies develop in the
    discussion tree. We include it as a direct measure of whether conversations
    remain mostly at the top level or continue through multi-step exchanges.

    \item \texttt{structural\_virality} measures the average shortest-path
    distance between comments in the reply tree. It complements depth by
    capturing how broadly interaction is distributed across the entire tree
    rather than only how far individual reply chains extend
    \citep{goel2016structural}.

    \item \texttt{length\_cv} measures variation in comment length. We include
    it because real discussions mix short reactions with longer explanations,
    stories, or arguments, whereas simulated users may contribute comments of
    unusually similar length.
\end{itemize}


\subsection{Scoring Setup}
\label{sec:scoring_setup}
Let
\(
\mathcal{F}=\{f_1,f_2,\ldots,f_{12}\}
\)
denote the set of thread-level evaluation metrics used in ThreadBench, with
$f\in\mathcal{F}$ denoting an individual metric. Each discussion thread $t$
receives score $f(t)$ for each metric. The complete metric formulas are provided in
Appendix~\ref{app:metric_calculation}. Within a domain, write the real threads as $r_1,\ldots,r_{n_R}$ and the generated threads as $g_1,\ldots,g_{n_G}$. For each metric $f$, the comparison uses the observation sequences
\[
R_f=(f(r_1),\ldots,f(r_{n_R})),\qquad
G_f=(f(g_1),\ldots,f(g_{n_G})).
\]

For each metric \(f\in\mathcal{F}\), ThreadBench compares the empirical metric samples \(R_f\) and \(G_f\) using two complementary nonparametric two-sample tests: the Mann--Whitney U (MWU) test and the Kolmogorov--Smirnov (KS) test.

\noindent\textbf{Statistical Measures.}
The Mann--Whitney U (MWU) test is a rank-based two-sample test that evaluates
whether observations from one sample tend to be larger or smaller than those
from the other. The Kolmogorov--Smirnov (KS) test measures the maximum
absolute difference between the two empirical cumulative distribution
functions (CDFs), making it sensitive to broader differences in distributional
shape.
Let \(p_f^{\mathrm{MWU}}\) and \(p_f^{\mathrm{KS}}\) denote the corresponding
two-sided \(p\)-values. At significance level \(\alpha=0.05\), a metric
passes when
\[
p_f^{\mathrm{MWU}}\geq\alpha,
\qquad
p_f^{\mathrm{KS}}\geq\alpha,
\]
which indicates that neither test rejects its null hypothesis. We also report the Wasserstein distance and absolute Cliff's delta to quantify the magnitude of the discrepancy between \(R_f\) and \(G_f\). We describe these measures in Appendix~\ref{app:distance_measures}.

\noindent\textbf{Evaluation Interpretation.}
The two types of measures provide complementary evidence. The MWU and KS
tests assess whether the samples show a statistically detectable difference
at \(\alpha=0.05\), whereas the distance measures quantify the magnitude of
the discrepancy. We therefore favor generated distributions that pass more
statistical tests while exhibiting smaller distributional distances.

\begin{figure*}[t]
    \centering
    \vspace{-8pt}
    \includegraphics[
        width=0.98\textwidth,
        trim={0 0 0 0},
        clip
    ]{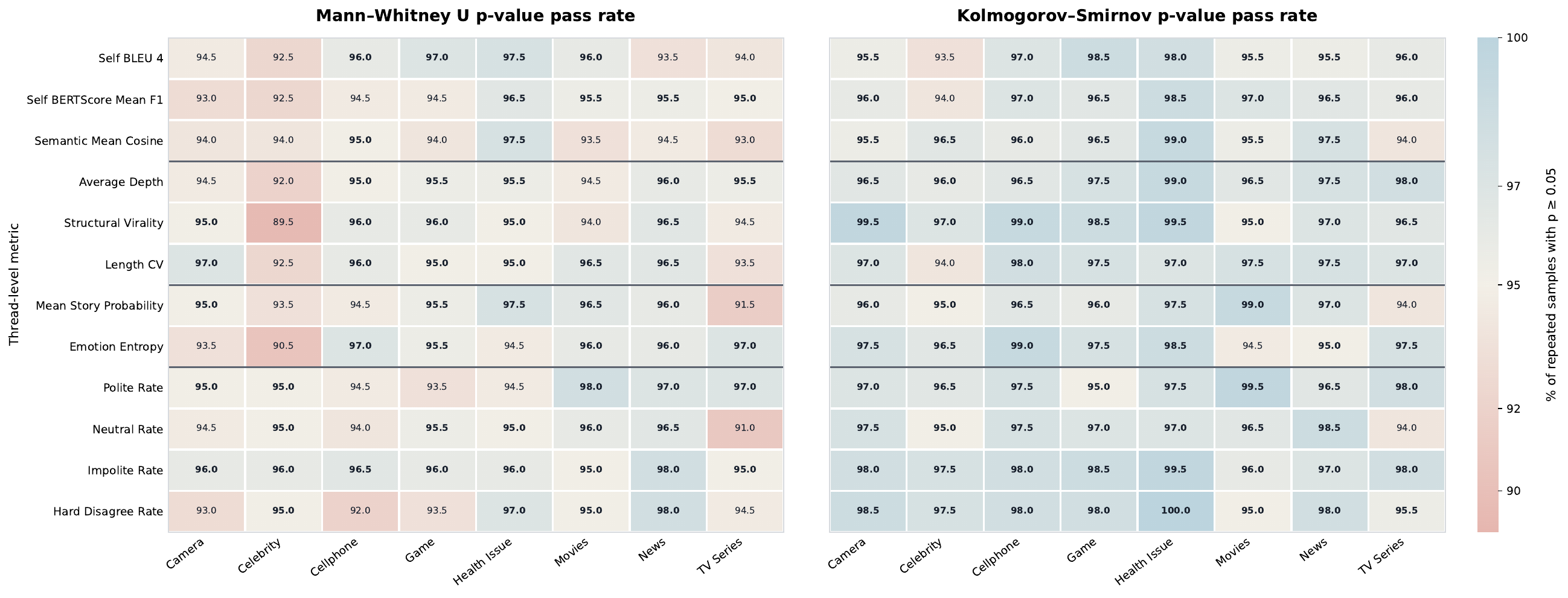}
    \vspace{-8pt}
    \caption{
    Real-vs.-real $p$-value pass rates across 200 repeated comparisons.
    Each cell reports the percentage of comparisons with $p\geq0.05$
    for the corresponding metric and domain, using the MWU test
    (left) and KS test (right).
    }
    \vspace{-16pt}
    \label{fig:repeated_sampling_pvalue}
\end{figure*}

\noindent\textbf{Real-vs.-Real Sanity Check.} We first check whether two randomly sampled groups of real discussions from
the same domain usually pass ThreadBench's MWU and KS tests, meaning that the
tests do not find a significant difference between the two real samples. For each domain, we randomly sample two groups of 150 real threads and compare
their metric distributions using the same MWU, KS, $W_1$, and
$\left|\delta_{\mathrm{Cliff}}\right|$ measures described above over 200
repeated comparisons. As shown in Figure~\ref{fig:repeated_sampling_pvalue},
94.99\% of MWU tests and 96.95\% of KS tests satisfy $p\geq0.05$, with both
tests passing simultaneously in 94.30\% of cases.
Figure~\ref{fig:real_real_distances} shows the corresponding real-vs.-real
distances across metric families, providing a reference for the differences
between two samples of real discussions.
Full results for all domains and metrics are reported in
Appendix~\ref{app:real_vs_real_complete_results}.

\subsection{Statistical Guarantee}
\label{sec:statistical_guarantee}

We provide a statistical guarantee for ThreadBench's metric-level pass criterion. For each metric $f\in\mathcal{F}$,
let $F_f^R$ and $F_f^G$ denote the population cumulative
distribution functions (CDFs) of real and generated
metric values. We test
\(
H_{0,f}: F_f^R=F_f^G.
\)
ThreadBench records a pass when neither the two-sided
MWU test nor the two-sample KS test rejects at
significance level $\alpha$:
\[
\mathsf{Pass}_f
=
\left\{
p_f^{\mathrm{MWU}}\geq\alpha,\;
p_f^{\mathrm{KS}}\geq\alpha
\right\}.
\]

\textbf{Assumptions.}
We consider the following assumptions.

\noindent\textbf{A1 (Independent samples).}
The real and generated metric samples contain i.i.d.
observations from $F_f^R$ and $F_f^G$, respectively,
and the two samples are mutually independent.

\noindent\textbf{A2 (Balanced sample sizes).}
As $n_R,n_G\to\infty$,
\(
\frac{n_R}{n_R+n_G}
\longrightarrow \rho\in(0,1).
\)
We use tie-corrected normal calibration for MWU
and continuous-null calibration for KS, as specified
in Appendix~\ref{app:statistical_guarantee}.

\begin{theorem}[Validity and consistency of ThreadBench decisions]
\label{thm:statistical_guarantee}
Under Assumptions A1--A2 and the stated test
calibrations, let $N=n_R+n_G$. For any fixed metric
$f\in\mathcal{F}$ and $\alpha\in(0,1/2)$,
the ThreadBench decision satisfies:

\begin{enumerate}
    \item \textbf{Validity.}
    If $F_f^R=F_f^G$, then
    \(
    \liminf_{N\to\infty}
    \Pr(\mathsf{Pass}_f)
    \geq 1-2\alpha.
    \)

    \item \textbf{Consistency.}
    If $F_f^R\neq F_f^G$, then
    \(
    \lim_{N\to\infty}
    \Pr(\mathsf{Pass}_f)
    =0.
    \)
\end{enumerate}
\end{theorem}

\textbf{Implications for ThreadBench.} Theorem~\ref{thm:statistical_guarantee}indicates that if the real and generated
metric distributions are truly the same, ThreadBench should not systematically
report a difference because of sampling noise alone. If the two distributions
are different, the tests detect the mismatch with probability approaching one
as the sample size grows. In other words, passing a metric means that ThreadBench
does not find sufficient evidence to distinguish the generated distribution
from the real one. It does not, however, prove that the two distributions are
identical, especially at a finite sample size. We provide the formal proof in
Appendix~\ref{app:statistical_guarantee}.


\section{ThreadPDG: Planner--Writer Discussion Generation}
\label{sec:miropdg}

We introduce ThreadPDG, a case-study of \textbf{P}rompt-based \textbf{D}iscussion
\textbf{G}eneration with a Planner--Writer framework. ThreadPDG aims to
model how human discussions develop rather than reproduce the wording or
content of a particular real thread. Given a real seed discussion, the Planner
analyzes how branches evolve, how conversational directions change, and how
each comment relates to the preceding discussion. It then converts these
patterns into thread- and comment-level plans for a new discussion. The Writer
generates new comments according to these plans. We generate $K$ candidate
threads for each seed and combine their best branches to form the final thread.

\subsection{Planner, Writer, and Composition}
\label{sec:planner_writer}

\textbf{Planner.}
Let $r_i$ denote the complete real seed thread, including its root post, comment text, and reply relationships. We extract a thread-level structural template $\tau_i$ that specifies the comment slots, their comments and sub-comments reply structure, and length targets. The Planner reads $r_i$ and $\tau_i$ to learn how discussion formed. It first identifies the main branches(a branch consists a parent comment with all sub-comments) and plans a overall conversational direction for each branch in a thread. It then produces a plan for every comment slot in the branch, specifying what the comment should contribute and how it relates to the preceding discussion. A domain profile $P_d$, constructed from the real seed discussion pool for domain $d$, provides additional guidance on discussion patterns. For draft $k$, we write the set of branch that forms a thread and comment plans as
\(
\pi_i^{(k)}=\operatorname{Planner}(r_i,\tau_i,P_d).
\)

\textbf{Writer.}
Given $\pi_i^{(k)}$, the Writer generates comments in the slots of template $\tau_i$, based on plans, assigned personas that randomly picked from MatrAIx \citep{li2026matraix}, and previously generated discussions. We denote the resulting draft by
\(
G_i^{(k)}=\operatorname{Writer}(\tau_i,\pi_i^{(k)},P_d).
\)
Detailed implementation can be found in Appendix~\ref{app:miropdg}, and the prompt templates of the Planner and the Writer in Appendix~\ref{app:prompts}

\subsection{Multi-Draft Composition}
\label{sec:multi_draft}

A generated thread may contain both strong and weak branches. We therefore run the Planner-Writer framework $K$ times based on the same real root post, the framework stops either when the pass rate is 100\% or budget $K$ is reached. We then combine the best branches from $K$ drafts of the same real seed post $G_i^{(1)},\ldots,G_i^{(K)}$. Let $B_i$ be the number of branches in the thread template $\tau_i$. We define the branch assignment as $\mathbf z_i=(z_{i1},\ldots,z_{iB_i})$, where $z_{ib}\in\{1,\ldots,K\}$ indicates which draft supplies branch $b$. We denote the resulting thread by $\widehat G_i(\mathbf z_i)$ and its score on metric $f_m$ by
\[
x_{im}(\mathbf z_i)=f_m(\widehat G_i(\mathbf z_i)).
\]
For each seed, we independently draw a reference thread $c_i$ from the reference pool $C_d$ and set $t_{im}=f_m(c_i)$. We use the same reference thread for all metrics, so the targets retain the relationships between metrics observed in that thread. We then search for a branch assignment that reduces
\[
J_i(\mathbf z_i)=\sum_{m\in\mathcal M_i}w_m
\left(\frac{x_{im}(\mathbf z_i)-t_{im}}{\operatorname{IQR}_m}\right)^2.
\]
Here, $\mathcal M_i$ is the set of metrics, $w_m$ is a fixed weight, and $\operatorname{IQR}_m$ is the interquartile range computed from $C_d$. We select branches using only the drafts and target for the current seed, together with fixed statistics from $C_d$ without other held-out test data. Appendix~\ref{app:miropdg} provides the target sampling and search details.
\section{Experiments and Results}
\label{sec:experiments_results}

\subsection{Experimental Setup}
\label{sec:setup}

\textbf{Baseline.}
We compare ThreadPDG with OASIS~\citep{yang2411oasis}, which models social-media interactions using LLM agents and dynamic social networks, and SynthPAI~\citep{yukhymenko2024synthetic}, which generates Reddit discussions using synthetic personal profiles. We evaluate all these method with \texttt{gpt-5.4-mini}, \texttt{gpt-4o-mini}, and \texttt{DeepSeek-V4.1-Flash}.

\begin{wraptable}{r}{0.27\textwidth}
\vspace{-12pt}
\centering
\scriptsize
\setlength{\tabcolsep}{4pt}
\caption{Human turing test results from 436 annotators using pair-level majority vote.}
\label{tab:turing-pair-majority}
\begin{tabular}{lcc}
\toprule
Method & Not dist. & Dist. \\
\midrule
\textbf{ThreadPDG }& \textbf{78.7\%} & \textbf{21.3\%} \\
OASIS      & 61.1\% & 38.9\% \\
SynthPAI   & 58.7\% & 41.3\% \\
\bottomrule
\end{tabular}
\vspace{-1pt}
\end{wraptable}

\textbf{Implementation.}
For every 8 domains in ThreadBench, we divide the real threads into three non-overlapping sets: a seed pool $S_d$, a reference pool $C_d$, and a test pool $T_d$. The seed pool provides the real discussions used for all three methods generation. Specifically for ThreadPDG, we use $C_d$ to build the domain profile, collect reference comments, and compute metric targets statistics. In every domain, we generate 150 threads for all three methods. Particularly for ThreadPDG, we set $K=3$ to generate up to $K$ drafts for each seed post and combine branches from these drafts to form the final thread. We then independently sample 150 real threads from $T_d$ to compare with the generated distributions using twelve thread-level metrics and two-sided Mann--Whitney U (MWU) and Kolmogorov--Smirnov (KS) tests.

\begin{table*}[!t]
\centering
\scriptsize
\large\setlength{\tabcolsep}{4.5pt}\renewcommand{\arraystretch}{0.90}
\caption{Simulated-vs.-real evaluation on ThreadBench,
over a held-out real population.
Each cell gives exact MWU $p$-value/KS $p$-value for corresponding domain, method, and LLM model with related metrics;
a metric passes when both $p$-values exceed
$\alpha=0.05$ (in bold). Pass rate is the percentage of the twelve metrics that pass; the highest in each domain is in bold.}
\label{tab:holdout_pvalues}
\resizebox{\textwidth}{!}{%
\begin{tabular}{@{}ccc|ccc|ccc|cc|cccc|c@{}}
\toprule
\multirow{2}{*}[-2.5em]{\textbf{Domain}} & \multirow{2}{*}[-2.5em]{\textbf{Method}} & \multirow{2}{*}[-2.5em]{\textbf{Model}} & \multicolumn{3}{c}{\textit{Uniformity}} & \multicolumn{3}{c}{\textit{Interaction}} & \multicolumn{2}{c}{\textit{Expression}} & \multicolumn{4}{c}{\textit{Tone}} & \multirow{2}{*}[-2.5em]{\textbf{Pass rate}} \\
\cmidrule(lr){4-6}\cmidrule(lr){7-9}\cmidrule(lr){10-11}\cmidrule(lr){12-15}
& & & \rotatebox{90}{Self-BLEU} & \rotatebox{90}{Self-BERT} & \rotatebox{90}{Sem.Cos} & \rotatebox{90}{Depth} & \rotatebox{90}{Virality} & \rotatebox{90}{Len.CV} & \rotatebox{90}{Story} & \rotatebox{90}{Emotion} & \rotatebox{90}{Polite} & \rotatebox{90}{Neutral} & \rotatebox{90}{Impolite} & \rotatebox{90}{Disagr.} & \\
\midrule
\multirow{9}{*}{\rotatebox[origin=c]{90}{\textbf{Camera}}} & \multirow{3}{*}{OASIS} & \texttt{DeepSeek-V4.1-Flash} & \textcolor{black!30}{$<$.01/$<$.01} & \textcolor{black!30}{$<$.01/$<$.01} & \textcolor{black!30}{$<$.01/$<$.01} & \textcolor{black!30}{$<$.01/$<$.01} & \textcolor{black!30}{$<$.01/$<$.01} & \textcolor{black!30}{$<$.01/$<$.01} & \textcolor{black!30}{$<$.01/$<$.01} & \textcolor{black!30}{$<$.01/$<$.01} & \textcolor{black!30}{$<$.01/$<$.01} & \textcolor{black!30}{$<$.01/$<$.01} & \textcolor{black!45}{.08/.03} & \textbf{.46/.06} & \textcolor{black!55}{8\% (1/12)} \\
 &  & \texttt{gpt-5.4-mini} & \textcolor{black!30}{$<$.01/$<$.01} & \textcolor{black!30}{$<$.01/$<$.01} & \textcolor{black!30}{$<$.01/$<$.01} & \textcolor{black!30}{$<$.01/$<$.01} & \textcolor{black!30}{$<$.01/$<$.01} & \textcolor{black!30}{$<$.01/$<$.01} & \textcolor{black!30}{$<$.01/$<$.01} & \textcolor{black!30}{$<$.01/$<$.01} & \textcolor{black!30}{$<$.01/$<$.01} & \textcolor{black!30}{$<$.01/$<$.01} & \textcolor{black!30}{$<$.01/$<$.01} & \textcolor{black!30}{$<$.01/$<$.01} & \textcolor{black!55}{0\% (0/12)} \\
 &  & \texttt{gpt-4o-mini} & \textcolor{black!30}{$<$.01/$<$.01} & \textcolor{black!30}{$<$.01/$<$.01} & \textcolor{black!30}{$<$.01/$<$.01} & \textcolor{black!30}{$<$.01/$<$.01} & \textcolor{black!30}{$<$.01/$<$.01} & \textcolor{black!30}{$<$.01/$<$.01} & \textcolor{black!30}{$<$.01/$<$.01} & \textcolor{black!30}{$<$.01/$<$.01} & \textcolor{black!30}{$<$.01/$<$.01} & \textcolor{black!30}{$<$.01/$<$.01} & \textcolor{black!30}{$<$.01/$<$.01} & \textcolor{black!30}{$<$.01/$<$.01} & \textcolor{black!55}{0\% (0/12)} \\
\cmidrule(l){2-16}
 & \multirow{3}{*}{SynthPAI} & \texttt{DeepSeek-V4.1-Flash} & \textcolor{black!30}{$<$.01/$<$.01} & \textcolor{black!30}{$<$.01/$<$.01} & \textbf{.08/.11} & \textcolor{black!30}{.01/$<$.01} & \textcolor{black!45}{.39/.03} & \textcolor{black!30}{$<$.01/$<$.01} & \textcolor{black!30}{$<$.01/$<$.01} & \textcolor{black!30}{$<$.01/$<$.01} & \textcolor{black!30}{$<$.01/$<$.01} & \textcolor{black!30}{$<$.01/$<$.01} & \textcolor{black!30}{$<$.01/$<$.01} & \textcolor{black!30}{$<$.01/$<$.01} & \textcolor{black!55}{8\% (1/12)} \\
 &  & \texttt{gpt-5.4-mini} & \textcolor{black!30}{$<$.01/$<$.01} & \textcolor{black!30}{$<$.01/$<$.01} & \textbf{.55/.14} & \textcolor{black!30}{$<$.01/$<$.01} & \textcolor{black!30}{$<$.01/$<$.01} & \textcolor{black!30}{.46/$<$.01} & \textcolor{black!30}{$<$.01/$<$.01} & \textcolor{black!30}{$<$.01/$<$.01} & \textcolor{black!30}{$<$.01/$<$.01} & \textcolor{black!45}{.03/.07} & \textcolor{black!30}{$<$.01/$<$.01} & \textcolor{black!30}{$<$.01/$<$.01} & \textcolor{black!55}{8\% (1/12)} \\
 &  & \texttt{gpt-4o-mini} & \textcolor{black!30}{$<$.01/$<$.01} & \textcolor{black!30}{$<$.01/$<$.01} & \textcolor{black!30}{$<$.01/$<$.01} & \textcolor{black!30}{.05/$<$.01} & \textbf{.85/.22} & \textcolor{black!30}{$<$.01/$<$.01} & \textcolor{black!30}{$<$.01/$<$.01} & \textcolor{black!30}{$<$.01/$<$.01} & \textbf{.07/.17} & \textcolor{black!30}{$<$.01/$<$.01} & \textcolor{black!30}{$<$.01/$<$.01} & \textcolor{black!30}{$<$.01/$<$.01} & \textcolor{black!55}{17\% (2/12)} \\
\cmidrule(l){2-16}
 & \multirow{3}{*}{\textbf{ThreadPDG}} & \texttt{DeepSeek-V4.1-Flash} & \textbf{.44/.71} & \textbf{.92/.28} & \textbf{.68/.44} & \textbf{.86/.98} & \textbf{.81/.85} & \textbf{.21/.44} & \textbf{.15/.16} & \textbf{.50/.79} & \textbf{.76/.33} & \textbf{.15/.07} & \textcolor{black!30}{$<$.01/.05} & \textbf{.10/.09} & \textbf{92\% (11/12)} \\
 &  & \texttt{gpt-5.4-mini} & \textcolor{black!45}{.12/.01} & \textbf{.27/.19} & \textbf{.72/.61} & \textbf{.98/.92} & \textbf{.70/.57} & \textbf{.70/.34} & \textbf{.17/.51} & \textbf{.65/.65} & \textcolor{black!30}{.02/$<$.01} & \textbf{.06/.20} & \textbf{.83/.90} & \textbf{.19/.37} & \textcolor{black!55}{83\% (10/12)} \\
 &  & \texttt{gpt-4o-mini} & \textcolor{black!45}{.03/.02} & \textcolor{black!30}{$<$.01/$<$.01} & \textcolor{black!30}{$<$.01/$<$.01} & \textbf{.41/.37} & \textbf{.23/.23} & \textcolor{black!30}{$<$.01/$<$.01} & \textcolor{black!30}{$<$.01/$<$.01} & \textcolor{black!30}{$<$.01/$<$.01} & \textcolor{black!30}{$<$.01/$<$.01} & \textcolor{black!30}{$<$.01/$<$.01} & \textcolor{black!30}{$<$.01/$<$.01} & \textcolor{black!30}{$<$.01/$<$.01} & \textcolor{black!55}{17\% (2/12)} \\
\midrule
\multirow{9}{*}{\rotatebox[origin=c]{90}{\textbf{Celebrity}}} & \multirow{3}{*}{OASIS} & \texttt{DeepSeek-V4.1-Flash} & \textcolor{black!30}{$<$.01/$<$.01} & \textcolor{black!30}{$<$.01/$<$.01} & \textcolor{black!30}{$<$.01/$<$.01} & \textcolor{black!30}{$<$.01/$<$.01} & \textcolor{black!30}{$<$.01/$<$.01} & \textcolor{black!30}{$<$.01/$<$.01} & \textcolor{black!30}{$<$.01/$<$.01} & \textcolor{black!30}{$<$.01/$<$.01} & \textcolor{black!30}{$<$.01/$<$.01} & \textcolor{black!30}{$<$.01/$<$.01} & \textbf{.67/.15} & \textcolor{black!30}{$<$.01/$<$.01} & \textcolor{black!55}{8\% (1/12)} \\
 &  & \texttt{gpt-5.4-mini} & \textcolor{black!30}{$<$.01/$<$.01} & \textcolor{black!30}{$<$.01/$<$.01} & \textcolor{black!30}{$<$.01/$<$.01} & \textcolor{black!30}{$<$.01/$<$.01} & \textcolor{black!30}{$<$.01/$<$.01} & \textcolor{black!30}{$<$.01/$<$.01} & \textcolor{black!30}{$<$.01/$<$.01} & \textcolor{black!30}{$<$.01/$<$.01} & \textcolor{black!30}{$<$.01/$<$.01} & \textcolor{black!30}{$<$.01/$<$.01} & \textcolor{black!30}{$<$.01/$<$.01} & \textbf{.37/.27} & \textcolor{black!55}{8\% (1/12)} \\
 &  & \texttt{gpt-4o-mini} & \textcolor{black!30}{$<$.01/$<$.01} & \textcolor{black!30}{$<$.01/$<$.01} & \textcolor{black!30}{$<$.01/$<$.01} & \textcolor{black!30}{$<$.01/$<$.01} & \textcolor{black!30}{$<$.01/$<$.01} & \textcolor{black!30}{$<$.01/$<$.01} & \textcolor{black!30}{$<$.01/$<$.01} & \textcolor{black!30}{$<$.01/$<$.01} & \textcolor{black!30}{$<$.01/$<$.01} & \textcolor{black!30}{$<$.01/$<$.01} & \textcolor{black!30}{$<$.01/$<$.01} & \textcolor{black!30}{.15/$<$.01} & \textcolor{black!55}{0\% (0/12)} \\
\cmidrule(l){2-16}
 & \multirow{3}{*}{SynthPAI} & \texttt{DeepSeek-V4.1-Flash} & \textcolor{black!30}{$<$.01/$<$.01} & \textcolor{black!30}{$<$.01/$<$.01} & \textcolor{black!30}{$<$.01/$<$.01} & \textcolor{black!30}{$<$.01/$<$.01} & \textcolor{black!30}{$<$.01/$<$.01} & \textcolor{black!30}{$<$.01/$<$.01} & \textcolor{black!30}{$<$.01/$<$.01} & \textcolor{black!30}{$<$.01/$<$.01} & \textcolor{black!30}{$<$.01/$<$.01} & \textcolor{black!45}{.02/.02} & \textcolor{black!30}{$<$.01/$<$.01} & \textcolor{black!30}{$<$.01/$<$.01} & \textcolor{black!55}{0\% (0/12)} \\
 &  & \texttt{gpt-5.4-mini} & \textcolor{black!30}{$<$.01/$<$.01} & \textcolor{black!30}{$<$.01/$<$.01} & \textcolor{black!30}{$<$.01/$<$.01} & \textcolor{black!30}{$<$.01/$<$.01} & \textcolor{black!30}{$<$.01/$<$.01} & \textcolor{black!30}{.11/$<$.01} & \textcolor{black!30}{$<$.01/$<$.01} & \textcolor{black!30}{$<$.01/$<$.01} & \textcolor{black!30}{$<$.01/$<$.01} & \textcolor{black!30}{.02/$<$.01} & \textcolor{black!30}{$<$.01/$<$.01} & \textcolor{black!45}{.03/.02} & \textcolor{black!55}{0\% (0/12)} \\
 &  & \texttt{gpt-4o-mini} & \textcolor{black!30}{$<$.01/$<$.01} & \textcolor{black!30}{$<$.01/$<$.01} & \textcolor{black!30}{$<$.01/$<$.01} & \textcolor{black!30}{$<$.01/$<$.01} & \textcolor{black!30}{$<$.01/$<$.01} & \textcolor{black!30}{$<$.01/$<$.01} & \textcolor{black!30}{$<$.01/$<$.01} & \textcolor{black!30}{$<$.01/$<$.01} & \textcolor{black!30}{$<$.01/$<$.01} & \textcolor{black!30}{$<$.01/$<$.01} & \textcolor{black!30}{$<$.01/$<$.01} & \textcolor{black!30}{$<$.01/$<$.01} & \textcolor{black!55}{0\% (0/12)} \\
\cmidrule(l){2-16}
 & \multirow{3}{*}{\textbf{ThreadPDG}} & \texttt{DeepSeek-V4.1-Flash} & \textbf{.77/.71} & \textbf{.77/.88} & \textbf{.90/.88} & \textbf{.58/.56} & \textbf{.30/.33} & \textbf{.05/.20} & \textbf{.56/.67} & \textbf{.67/.80} & \textbf{.72/.55} & \textbf{.68/.77} & \textbf{.69/.66} & \textbf{.78/.64} & \textbf{100\% (12/12)} \\
 &  & \texttt{gpt-5.4-mini} & \textbf{.12/.12} & \textbf{.29/.70} & \textbf{.91/.79} & \textbf{.56/.50} & \textbf{.26/.22} & \textbf{.17/.11} & \textbf{.24/.53} & \textbf{.49/.52} & \textcolor{black!45}{.05/.06} & \textbf{.58/.45} & \textbf{.31/.43} & \textbf{.41/.62} & \textcolor{black!55}{92\% (11/12)} \\
 &  & \texttt{gpt-4o-mini} & \textcolor{black!30}{$<$.01/$<$.01} & \textcolor{black!30}{$<$.01/$<$.01} & \textcolor{black!30}{$<$.01/$<$.01} & \textbf{.77/.71} & \textbf{.91/.83} & \textcolor{black!45}{.04/.05} & \textcolor{black!30}{$<$.01/.02} & \textcolor{black!30}{$<$.01/$<$.01} & \textcolor{black!30}{$<$.01/$<$.01} & \textcolor{black!30}{$<$.01/.01} & \textbf{.51/.92} & \textcolor{black!30}{$<$.01/$<$.01} & \textcolor{black!55}{25\% (3/12)} \\
\midrule
\multirow{9}{*}{\rotatebox[origin=c]{90}{\textbf{Cellphone}}} & \multirow{3}{*}{OASIS} & \texttt{DeepSeek-V4.1-Flash} & \textcolor{black!30}{$<$.01/$<$.01} & \textcolor{black!30}{$<$.01/$<$.01} & \textcolor{black!30}{$<$.01/$<$.01} & \textcolor{black!30}{$<$.01/$<$.01} & \textcolor{black!30}{$<$.01/$<$.01} & \textcolor{black!30}{$<$.01/$<$.01} & \textcolor{black!30}{$<$.01/$<$.01} & \textcolor{black!30}{$<$.01/$<$.01} & \textcolor{black!30}{$<$.01/$<$.01} & \textcolor{black!30}{$<$.01/$<$.01} & \textcolor{black!30}{$<$.01/$<$.01} & \textbf{.87/.90} & \textcolor{black!55}{8\% (1/12)} \\
 &  & \texttt{gpt-5.4-mini} & \textcolor{black!30}{$<$.01/$<$.01} & \textcolor{black!30}{$<$.01/$<$.01} & \textcolor{black!30}{$<$.01/$<$.01} & \textcolor{black!30}{$<$.01/$<$.01} & \textcolor{black!30}{$<$.01/$<$.01} & \textcolor{black!30}{$<$.01/$<$.01} & \textcolor{black!30}{$<$.01/$<$.01} & \textcolor{black!30}{$<$.01/$<$.01} & \textcolor{black!30}{$<$.01/$<$.01} & \textcolor{black!30}{$<$.01/$<$.01} & \textcolor{black!30}{$<$.01/$<$.01} & \textbf{.93/.20} & \textcolor{black!55}{8\% (1/12)} \\
 &  & \texttt{gpt-4o-mini} & \textcolor{black!30}{$<$.01/$<$.01} & \textcolor{black!30}{$<$.01/$<$.01} & \textcolor{black!30}{$<$.01/$<$.01} & \textcolor{black!30}{$<$.01/$<$.01} & \textcolor{black!30}{$<$.01/$<$.01} & \textcolor{black!30}{$<$.01/$<$.01} & \textcolor{black!30}{$<$.01/$<$.01} & \textcolor{black!30}{.05/$<$.01} & \textcolor{black!30}{$<$.01/$<$.01} & \textcolor{black!30}{$<$.01/$<$.01} & \textcolor{black!30}{$<$.01/$<$.01} & \textcolor{black!45}{.05/.01} & \textcolor{black!55}{0\% (0/12)} \\
\cmidrule(l){2-16}
 & \multirow{3}{*}{SynthPAI} & \texttt{DeepSeek-V4.1-Flash} & \textcolor{black!30}{$<$.01/$<$.01} & \textcolor{black!30}{$<$.01/$<$.01} & \textbf{.17/.12} & \textcolor{black!30}{$<$.01/$<$.01} & \textcolor{black!30}{$<$.01/$<$.01} & \textcolor{black!30}{$<$.01/$<$.01} & \textcolor{black!30}{$<$.01/$<$.01} & \textcolor{black!45}{.17/.03} & \textcolor{black!30}{$<$.01/$<$.01} & \textcolor{black!30}{$<$.01/$<$.01} & \textcolor{black!30}{$<$.01/$<$.01} & \textcolor{black!30}{$<$.01/$<$.01} & \textcolor{black!55}{8\% (1/12)} \\
 &  & \texttt{gpt-5.4-mini} & \textcolor{black!30}{$<$.01/$<$.01} & \textcolor{black!30}{$<$.01/$<$.01} & \textbf{.64/.06} & \textcolor{black!30}{$<$.01/$<$.01} & \textcolor{black!30}{$<$.01/$<$.01} & \textcolor{black!30}{.07/$<$.01} & \textcolor{black!30}{$<$.01/$<$.01} & \textcolor{black!30}{$<$.01/$<$.01} & \textcolor{black!30}{$<$.01/$<$.01} & \textcolor{black!30}{$<$.01/$<$.01} & \textcolor{black!30}{$<$.01/$<$.01} & \textcolor{black!30}{$<$.01/$<$.01} & \textcolor{black!55}{8\% (1/12)} \\
 &  & \texttt{gpt-4o-mini} & \textcolor{black!30}{$<$.01/$<$.01} & \textcolor{black!30}{$<$.01/$<$.01} & \textcolor{black!30}{$<$.01/$<$.01} & \textcolor{black!30}{$<$.01/$<$.01} & \textbf{.63/.26} & \textcolor{black!30}{$<$.01/$<$.01} & \textcolor{black!30}{$<$.01/$<$.01} & \textcolor{black!30}{.11/$<$.01} & \textcolor{black!30}{$<$.01/$<$.01} & \textcolor{black!30}{$<$.01/$<$.01} & \textcolor{black!30}{$<$.01/$<$.01} & \textcolor{black!30}{$<$.01/$<$.01} & \textcolor{black!55}{8\% (1/12)} \\
\cmidrule(l){2-16}
 & \multirow{3}{*}{\textbf{ThreadPDG}} & \texttt{DeepSeek-V4.1-Flash} & \textbf{.09/.42} & \textbf{.23/.33} & \textbf{.90/.35} & \textbf{.43/.41} & \textbf{.20/.43} & \textcolor{black!30}{.07/$<$.01} & \textbf{.78/.42} & \textbf{.12/.08} & \textbf{.18/.15} & \textbf{.97/1.0} & \textbf{.05/.13} & \textcolor{black!30}{$<$.01/.01} & \textbf{83\% (10/12)} \\
 &  & \texttt{gpt-5.4-mini} & \textcolor{black!30}{$<$.01/.04} & \textbf{.95/.70} & \textbf{.13/.41} & \textbf{.76/.70} & \textbf{.91/.93} & \textbf{.31/.49} & \textbf{.92/.88} & \textbf{.84/.55} & \textcolor{black!30}{$<$.01/$<$.01} & \textbf{.26/.12} & \textbf{.17/.33} & \textbf{.17/.27} & \textbf{83\% (10/12)} \\
 &  & \texttt{gpt-4o-mini} & \textbf{.55/.72} & \textcolor{black!30}{$<$.01/$<$.01} & \textcolor{black!30}{$<$.01/$<$.01} & \textbf{.07/.16} & \textcolor{black!45}{.01/.04} & \textcolor{black!30}{$<$.01/$<$.01} & \textcolor{black!30}{$<$.01/.06} & \textcolor{black!30}{$<$.01/$<$.01} & \textbf{.46/.29} & \textbf{.20/.15} & \textcolor{black!45}{.02/.02} & \textbf{.10/.07} & \textcolor{black!55}{42\% (5/12)} \\
\midrule
\multirow{9}{*}{\rotatebox[origin=c]{90}{\textbf{Game}}} & \multirow{3}{*}{OASIS} & \texttt{DeepSeek-V4.1-Flash} & \textbf{.07/.18} & \textcolor{black!30}{$<$.01/$<$.01} & \textcolor{black!30}{$<$.01/$<$.01} & \textcolor{black!30}{$<$.01/$<$.01} & \textcolor{black!30}{$<$.01/$<$.01} & \textcolor{black!30}{$<$.01/$<$.01} & \textcolor{black!30}{$<$.01/$<$.01} & \textcolor{black!30}{$<$.01/$<$.01} & \textcolor{black!30}{$<$.01/$<$.01} & \textcolor{black!30}{$<$.01/$<$.01} & \textcolor{black!30}{$<$.01/$<$.01} & \textbf{.08/.08} & \textcolor{black!55}{17\% (2/12)} \\
 &  & \texttt{gpt-5.4-mini} & \textcolor{black!30}{$<$.01/$<$.01} & \textcolor{black!30}{$<$.01/$<$.01} & \textcolor{black!30}{$<$.01/$<$.01} & \textcolor{black!30}{$<$.01/$<$.01} & \textcolor{black!30}{$<$.01/$<$.01} & \textcolor{black!30}{$<$.01/$<$.01} & \textcolor{black!30}{$<$.01/$<$.01} & \textcolor{black!30}{$<$.01/$<$.01} & \textcolor{black!30}{$<$.01/$<$.01} & \textcolor{black!30}{$<$.01/$<$.01} & \textcolor{black!30}{$<$.01/$<$.01} & \textcolor{black!45}{.20/.02} & \textcolor{black!55}{0\% (0/12)} \\
 &  & \texttt{gpt-4o-mini} & \textcolor{black!30}{$<$.01/$<$.01} & \textcolor{black!30}{$<$.01/$<$.01} & \textcolor{black!30}{$<$.01/$<$.01} & \textcolor{black!30}{$<$.01/$<$.01} & \textcolor{black!30}{$<$.01/$<$.01} & \textcolor{black!30}{$<$.01/$<$.01} & \textcolor{black!30}{$<$.01/$<$.01} & \textcolor{black!30}{$<$.01/$<$.01} & \textcolor{black!30}{$<$.01/$<$.01} & \textcolor{black!30}{$<$.01/$<$.01} & \textcolor{black!30}{$<$.01/$<$.01} & \textcolor{black!30}{$<$.01/$<$.01} & \textcolor{black!55}{0\% (0/12)} \\
\cmidrule(l){2-16}
 & \multirow{3}{*}{SynthPAI} & \texttt{DeepSeek-V4.1-Flash} & \textcolor{black!30}{$<$.01/$<$.01} & \textcolor{black!30}{$<$.01/$<$.01} & \textcolor{black!30}{$<$.01/$<$.01} & \textcolor{black!30}{$<$.01/$<$.01} & \textcolor{black!30}{$<$.01/$<$.01} & \textcolor{black!30}{$<$.01/$<$.01} & \textcolor{black!30}{$<$.01/$<$.01} & \textcolor{black!30}{$<$.01/$<$.01} & \textcolor{black!30}{$<$.01/$<$.01} & \textcolor{black!30}{$<$.01/$<$.01} & \textcolor{black!30}{$<$.01/$<$.01} & \textbf{.08/.07} & \textcolor{black!55}{8\% (1/12)} \\
 &  & \texttt{gpt-5.4-mini} & \textcolor{black!30}{$<$.01/$<$.01} & \textcolor{black!30}{$<$.01/$<$.01} & \textbf{.07/.07} & \textcolor{black!30}{$<$.01/$<$.01} & \textcolor{black!30}{$<$.01/$<$.01} & \textcolor{black!30}{.04/$<$.01} & \textcolor{black!30}{$<$.01/$<$.01} & \textcolor{black!30}{$<$.01/$<$.01} & \textcolor{black!30}{$<$.01/$<$.01} & \textbf{.37/.63} & \textcolor{black!30}{$<$.01/$<$.01} & \textcolor{black!30}{.02/$<$.01} & \textcolor{black!55}{17\% (2/12)} \\
 &  & \texttt{gpt-4o-mini} & \textcolor{black!30}{$<$.01/$<$.01} & \textcolor{black!30}{$<$.01/$<$.01} & \textcolor{black!30}{$<$.01/$<$.01} & \textcolor{black!30}{$<$.01/$<$.01} & \textcolor{black!45}{.48/.01} & \textcolor{black!30}{$<$.01/$<$.01} & \textcolor{black!30}{$<$.01/$<$.01} & \textcolor{black!30}{$<$.01/$<$.01} & \textcolor{black!30}{$<$.01/$<$.01} & \textcolor{black!30}{$<$.01/$<$.01} & \textcolor{black!30}{$<$.01/$<$.01} & \textcolor{black!30}{$<$.01/$<$.01} & \textcolor{black!55}{0\% (0/12)} \\
\cmidrule(l){2-16}
 & \multirow{3}{*}{\textbf{ThreadPDG}} & \texttt{DeepSeek-V4.1-Flash} & \textbf{.75/.51} & \textbf{.11/.08} & \textbf{.15/.45} & \textbf{.18/.28} & \textcolor{black!45}{.03/.05} & \textbf{.71/.69} & \textbf{.71/.91} & \textbf{.30/.22} & \textbf{.17/.05} & \textbf{.60/.63} & \textbf{.88/.36} & \textcolor{black!30}{$<$.01/$<$.01} & \textbf{83\% (10/12)} \\
 &  & \texttt{gpt-5.4-mini} & \textbf{.36/.72} & \textcolor{black!45}{.03/.02} & \textbf{.29/.63} & \textbf{.19/.29} & \textcolor{black!45}{.07/.03} & \textbf{.22/.29} & \textbf{.93/1.0} & \textbf{.70/.53} & \textcolor{black!45}{.03/.02} & \textbf{.63/.89} & \textbf{.46/.53} & \textcolor{black!30}{$<$.01/$<$.01} & \textcolor{black!55}{67\% (8/12)} \\
 &  & \texttt{gpt-4o-mini} & \textcolor{black!30}{$<$.01/$<$.01} & \textcolor{black!30}{$<$.01/$<$.01} & \textcolor{black!30}{$<$.01/$<$.01} & \textbf{.06/.07} & \textcolor{black!45}{.01/.03} & \textcolor{black!30}{$<$.01/$<$.01} & \textbf{.69/.84} & \textcolor{black!30}{$<$.01/$<$.01} & \textcolor{black!30}{$<$.01/$<$.01} & \textcolor{black!45}{.02/.04} & \textcolor{black!30}{$<$.01/$<$.01} & \textbf{.67/.53} & \textcolor{black!55}{25\% (3/12)} \\
\midrule
\multirow{9}{*}{\rotatebox[origin=c]{90}{\textbf{Health Issue}}} & \multirow{3}{*}{OASIS} & \texttt{DeepSeek-V4.1-Flash} & \textcolor{black!30}{$<$.01/$<$.01} & \textcolor{black!30}{$<$.01/$<$.01} & \textcolor{black!30}{$<$.01/$<$.01} & \textcolor{black!30}{$<$.01/$<$.01} & \textcolor{black!30}{$<$.01/$<$.01} & \textcolor{black!30}{$<$.01/$<$.01} & \textcolor{black!45}{.02/.15} & \textcolor{black!30}{$<$.01/$<$.01} & \textbf{.08/.09} & \textcolor{black!30}{$<$.01/$<$.01} & \textcolor{black!30}{$<$.01/$<$.01} & \textcolor{black!30}{$<$.01/$<$.01} & \textcolor{black!55}{8\% (1/12)} \\
 &  & \texttt{gpt-5.4-mini} & \textcolor{black!30}{$<$.01/$<$.01} & \textcolor{black!30}{$<$.01/$<$.01} & \textcolor{black!30}{$<$.01/$<$.01} & \textcolor{black!30}{$<$.01/$<$.01} & \textcolor{black!30}{$<$.01/$<$.01} & \textcolor{black!30}{$<$.01/$<$.01} & \textcolor{black!30}{$<$.01/$<$.01} & \textcolor{black!30}{$<$.01/$<$.01} & \textcolor{black!30}{$<$.01/$<$.01} & \textcolor{black!30}{$<$.01/$<$.01} & \textcolor{black!30}{$<$.01/$<$.01} & \textbf{.51/.10} & \textcolor{black!55}{8\% (1/12)} \\
 &  & \texttt{gpt-4o-mini} & \textcolor{black!30}{$<$.01/$<$.01} & \textcolor{black!30}{$<$.01/$<$.01} & \textcolor{black!30}{$<$.01/$<$.01} & \textcolor{black!30}{$<$.01/$<$.01} & \textcolor{black!30}{$<$.01/$<$.01} & \textcolor{black!30}{$<$.01/$<$.01} & \textcolor{black!30}{$<$.01/$<$.01} & \textcolor{black!30}{$<$.01/$<$.01} & \textcolor{black!30}{$<$.01/$<$.01} & \textcolor{black!30}{$<$.01/$<$.01} & \textcolor{black!30}{$<$.01/$<$.01} & \textbf{.79/.66} & \textcolor{black!55}{8\% (1/12)} \\
\cmidrule(l){2-16}
 & \multirow{3}{*}{SynthPAI} & \texttt{DeepSeek-V4.1-Flash} & \textcolor{black!30}{$<$.01/$<$.01} & \textcolor{black!30}{$<$.01/$<$.01} & \textbf{.47/.57} & \textcolor{black!30}{$<$.01/$<$.01} & \textcolor{black!30}{$<$.01/$<$.01} & \textcolor{black!30}{$<$.01/$<$.01} & \textcolor{black!30}{$<$.01/$<$.01} & \textbf{.49/.19} & \textcolor{black!30}{$<$.01/$<$.01} & \textcolor{black!30}{$<$.01/$<$.01} & \textcolor{black!30}{$<$.01/$<$.01} & \textcolor{black!30}{$<$.01/$<$.01} & \textcolor{black!55}{17\% (2/12)} \\
 &  & \texttt{gpt-5.4-mini} & \textcolor{black!30}{$<$.01/$<$.01} & \textcolor{black!30}{$<$.01/$<$.01} & \textbf{.43/.57} & \textcolor{black!30}{$<$.01/$<$.01} & \textcolor{black!30}{$<$.01/$<$.01} & \textcolor{black!30}{.66/$<$.01} & \textcolor{black!30}{$<$.01/$<$.01} & \textcolor{black!30}{$<$.01/$<$.01} & \textcolor{black!30}{$<$.01/$<$.01} & \textcolor{black!45}{.13/.02} & \textcolor{black!30}{$<$.01/$<$.01} & \textcolor{black!30}{$<$.01/$<$.01} & \textcolor{black!55}{8\% (1/12)} \\
 &  & \texttt{gpt-4o-mini} & \textcolor{black!30}{$<$.01/$<$.01} & \textcolor{black!30}{$<$.01/$<$.01} & \textcolor{black!30}{$<$.01/$<$.01} & \textcolor{black!30}{$<$.01/$<$.01} & \textcolor{black!30}{$<$.01/$<$.01} & \textcolor{black!30}{$<$.01/$<$.01} & \textcolor{black!30}{$<$.01/$<$.01} & \textcolor{black!30}{$<$.01/$<$.01} & \textcolor{black!30}{$<$.01/$<$.01} & \textcolor{black!30}{$<$.01/$<$.01} & \textcolor{black!30}{$<$.01/$<$.01} & \textcolor{black!30}{$<$.01/$<$.01} & \textcolor{black!55}{0\% (0/12)} \\
\cmidrule(l){2-16}
 & \multirow{3}{*}{\textbf{ThreadPDG}} & \texttt{DeepSeek-V4.1-Flash} & \textcolor{black!45}{.02/.02} & \textcolor{black!45}{.02/.14} & \textbf{.40/.53} & \textbf{.48/.23} & \textbf{.39/.72} & \textbf{.74/.36} & \textbf{.37/.72} & \textbf{.35/.36} & \textcolor{black!30}{$<$.01/$<$.01} & \textcolor{black!45}{.03/.11} & \textcolor{black!30}{$<$.01/$<$.01} & \textbf{.29/.23} & \textcolor{black!55}{58\% (7/12)} \\
 &  & \texttt{gpt-5.4-mini} & \textbf{.44/.11} & \textbf{.92/.96} & \textbf{.21/.25} & \textbf{.47/.23} & \textbf{.41/.70} & \textcolor{black!45}{.22/.04} & \textbf{.76/.82} & \textbf{.06/.09} & \textcolor{black!30}{$<$.01/$<$.01} & \textcolor{black!45}{.02/.08} & \textcolor{black!45}{.05/.16} & \textbf{.34/.16} & \textbf{67\% (8/12)} \\
 &  & \texttt{gpt-4o-mini} & \textcolor{black!30}{.01/$<$.01} & \textcolor{black!30}{$<$.01/$<$.01} & \textcolor{black!30}{$<$.01/$<$.01} & \textbf{.42/.26} & \textbf{.41/.67} & \textcolor{black!30}{$<$.01/$<$.01} & \textcolor{black!30}{$<$.01/.04} & \textcolor{black!30}{$<$.01/$<$.01} & \textcolor{black!30}{$<$.01/$<$.01} & \textbf{.27/.36} & \textcolor{black!30}{$<$.01/$<$.01} & \textcolor{black!45}{.01/.02} & \textcolor{black!55}{25\% (3/12)} \\
\midrule
\multirow{9}{*}{\rotatebox[origin=c]{90}{\textbf{Movies}}} & \multirow{3}{*}{OASIS} & \texttt{DeepSeek-V4.1-Flash} & \textcolor{black!30}{$<$.01/$<$.01} & \textcolor{black!30}{$<$.01/$<$.01} & \textcolor{black!30}{$<$.01/$<$.01} & \textcolor{black!30}{$<$.01/$<$.01} & \textcolor{black!30}{$<$.01/$<$.01} & \textcolor{black!30}{$<$.01/$<$.01} & \textbf{.12/.13} & \textcolor{black!30}{$<$.01/$<$.01} & \textcolor{black!30}{$<$.01/$<$.01} & \textcolor{black!30}{$<$.01/$<$.01} & \textbf{.34/.70} & \textcolor{black!45}{.02/.07} & \textcolor{black!55}{17\% (2/12)} \\
 &  & \texttt{gpt-5.4-mini} & \textcolor{black!30}{$<$.01/$<$.01} & \textcolor{black!30}{$<$.01/$<$.01} & \textcolor{black!30}{$<$.01/$<$.01} & \textcolor{black!30}{$<$.01/$<$.01} & \textcolor{black!30}{$<$.01/$<$.01} & \textcolor{black!30}{$<$.01/$<$.01} & \textcolor{black!30}{$<$.01/$<$.01} & \textcolor{black!30}{$<$.01/$<$.01} & \textcolor{black!30}{$<$.01/$<$.01} & \textcolor{black!30}{$<$.01/$<$.01} & \textcolor{black!30}{$<$.01/$<$.01} & \textcolor{black!30}{.02/$<$.01} & \textcolor{black!55}{0\% (0/12)} \\
 &  & \texttt{gpt-4o-mini} & \textcolor{black!30}{$<$.01/$<$.01} & \textcolor{black!30}{$<$.01/$<$.01} & \textcolor{black!30}{$<$.01/$<$.01} & \textcolor{black!30}{$<$.01/$<$.01} & \textcolor{black!30}{$<$.01/$<$.01} & \textcolor{black!30}{$<$.01/$<$.01} & \textcolor{black!30}{$<$.01/$<$.01} & \textcolor{black!30}{$<$.01/$<$.01} & \textcolor{black!30}{$<$.01/$<$.01} & \textcolor{black!30}{$<$.01/$<$.01} & \textcolor{black!30}{$<$.01/$<$.01} & \textcolor{black!30}{.08/$<$.01} & \textcolor{black!55}{0\% (0/12)} \\
\cmidrule(l){2-16}
 & \multirow{3}{*}{SynthPAI} & \texttt{DeepSeek-V4.1-Flash} & \textcolor{black!30}{$<$.01/$<$.01} & \textcolor{black!30}{$<$.01/$<$.01} & \textcolor{black!45}{.03/.04} & \textcolor{black!30}{$<$.01/$<$.01} & \textcolor{black!30}{$<$.01/$<$.01} & \textcolor{black!30}{$<$.01/$<$.01} & \textcolor{black!30}{$<$.01/$<$.01} & \textbf{.55/.72} & \textcolor{black!30}{$<$.01/$<$.01} & \textcolor{black!30}{$<$.01/$<$.01} & \textcolor{black!30}{$<$.01/$<$.01} & \textcolor{black!30}{$<$.01/$<$.01} & \textcolor{black!55}{8\% (1/12)} \\
 &  & \texttt{gpt-5.4-mini} & \textcolor{black!30}{$<$.01/$<$.01} & \textcolor{black!30}{$<$.01/$<$.01} & \textbf{.07/.17} & \textcolor{black!30}{$<$.01/$<$.01} & \textcolor{black!30}{$<$.01/$<$.01} & \textcolor{black!30}{.01/$<$.01} & \textcolor{black!45}{.04/.13} & \textcolor{black!30}{$<$.01/$<$.01} & \textcolor{black!30}{$<$.01/$<$.01} & \textcolor{black!30}{$<$.01/$<$.01} & \textcolor{black!30}{$<$.01/$<$.01} & \textcolor{black!30}{$<$.01/$<$.01} & \textcolor{black!55}{8\% (1/12)} \\
 &  & \texttt{gpt-4o-mini} & \textcolor{black!30}{$<$.01/$<$.01} & \textcolor{black!30}{$<$.01/$<$.01} & \textcolor{black!30}{$<$.01/$<$.01} & \textcolor{black!30}{$<$.01/$<$.01} & \textcolor{black!30}{.08/$<$.01} & \textcolor{black!30}{$<$.01/$<$.01} & \textcolor{black!30}{$<$.01/$<$.01} & \textcolor{black!45}{.02/.06} & \textcolor{black!30}{$<$.01/$<$.01} & \textcolor{black!30}{$<$.01/$<$.01} & \textcolor{black!30}{$<$.01/$<$.01} & \textcolor{black!30}{$<$.01/$<$.01} & \textcolor{black!55}{0\% (0/12)} \\
\cmidrule(l){2-16}
 & \multirow{3}{*}{\textbf{ThreadPDG}} & \texttt{DeepSeek-V4.1-Flash} & \textbf{.97/.42} & \textbf{.78/.70} & \textbf{.82/.88} & \textbf{.72/.61} & \textbf{.28/.36} & \textcolor{black!30}{$<$.01/$<$.01} & \textbf{.47/.56} & \textbf{.80/.96} & \textbf{.58/.47} & \textbf{.77/.94} & \textcolor{black!45}{.02/.09} & \textbf{.36/.13} & \textcolor{black!55}{83\% (10/12)} \\
 &  & \texttt{gpt-5.4-mini} & \textbf{.13/.08} & \textbf{.31/.64} & \textbf{.38/.38} & \textbf{.61/.34} & \textbf{.19/.25} & \textbf{.07/.05} & \textbf{.67/.65} & \textbf{.73/.76} & \textcolor{black!45}{.01/.01} & \textbf{.33/.37} & \textbf{.09/.15} & \textbf{.50/.11} & \textbf{92\% (11/12)} \\
 &  & \texttt{gpt-4o-mini} & \textcolor{black!30}{$<$.01/$<$.01} & \textcolor{black!30}{$<$.01/$<$.01} & \textcolor{black!30}{$<$.01/$<$.01} & \textbf{.43/.36} & \textcolor{black!45}{.05/.14} & \textcolor{black!45}{.03/.04} & \textbf{.30/.15} & \textcolor{black!30}{$<$.01/.04} & \textcolor{black!30}{$<$.01/$<$.01} & \textcolor{black!30}{$<$.01/$<$.01} & \textbf{.08/.15} & \textcolor{black!30}{$<$.01/$<$.01} & \textcolor{black!55}{25\% (3/12)} \\
\midrule
\multirow{9}{*}{\rotatebox[origin=c]{90}{\textbf{News}}} & \multirow{3}{*}{OASIS} & \texttt{DeepSeek-V4.1-Flash} & \textcolor{black!30}{$<$.01/$<$.01} & \textcolor{black!30}{$<$.01/$<$.01} & \textcolor{black!30}{$<$.01/$<$.01} & \textcolor{black!30}{$<$.01/$<$.01} & \textcolor{black!30}{$<$.01/$<$.01} & \textcolor{black!30}{$<$.01/$<$.01} & \textcolor{black!30}{$<$.01/$<$.01} & \textcolor{black!30}{$<$.01/$<$.01} & \textcolor{black!30}{$<$.01/$<$.01} & \textcolor{black!30}{$<$.01/$<$.01} & \textcolor{black!30}{.43/$<$.01} & \textbf{.50/.09} & \textcolor{black!55}{8\% (1/12)} \\
 &  & \texttt{gpt-5.4-mini} & \textcolor{black!30}{$<$.01/$<$.01} & \textcolor{black!30}{$<$.01/$<$.01} & \textcolor{black!30}{$<$.01/$<$.01} & \textcolor{black!30}{$<$.01/$<$.01} & \textcolor{black!30}{$<$.01/$<$.01} & \textcolor{black!30}{$<$.01/$<$.01} & \textcolor{black!30}{$<$.01/$<$.01} & \textcolor{black!30}{$<$.01/$<$.01} & \textcolor{black!30}{$<$.01/$<$.01} & \textcolor{black!30}{$<$.01/$<$.01} & \textcolor{black!30}{$<$.01/$<$.01} & \textcolor{black!30}{$<$.01/$<$.01} & \textcolor{black!55}{0\% (0/12)} \\
 &  & \texttt{gpt-4o-mini} & \textcolor{black!30}{$<$.01/$<$.01} & \textcolor{black!30}{$<$.01/$<$.01} & \textcolor{black!30}{$<$.01/$<$.01} & \textcolor{black!30}{$<$.01/$<$.01} & \textcolor{black!30}{$<$.01/$<$.01} & \textcolor{black!30}{$<$.01/$<$.01} & \textcolor{black!30}{$<$.01/$<$.01} & \textcolor{black!30}{$<$.01/$<$.01} & \textcolor{black!30}{$<$.01/$<$.01} & \textcolor{black!30}{$<$.01/$<$.01} & \textcolor{black!30}{$<$.01/$<$.01} & \textcolor{black!30}{.02/$<$.01} & \textcolor{black!55}{0\% (0/12)} \\
\cmidrule(l){2-16}
 & \multirow{3}{*}{SynthPAI} & \texttt{DeepSeek-V4.1-Flash} & \textcolor{black!30}{$<$.01/$<$.01} & \textcolor{black!30}{$<$.01/$<$.01} & \textcolor{black!30}{$<$.01/$<$.01} & \textcolor{black!30}{$<$.01/$<$.01} & \textcolor{black!30}{$<$.01/$<$.01} & \textcolor{black!30}{$<$.01/$<$.01} & \textcolor{black!30}{$<$.01/$<$.01} & \textcolor{black!30}{$<$.01/$<$.01} & \textcolor{black!30}{$<$.01/$<$.01} & \textcolor{black!30}{$<$.01/$<$.01} & \textcolor{black!30}{$<$.01/$<$.01} & \textcolor{black!30}{$<$.01/$<$.01} & \textcolor{black!55}{0\% (0/12)} \\
 &  & \texttt{gpt-5.4-mini} & \textcolor{black!30}{$<$.01/$<$.01} & \textcolor{black!30}{$<$.01/$<$.01} & \textcolor{black!30}{.03/$<$.01} & \textcolor{black!30}{$<$.01/$<$.01} & \textcolor{black!30}{$<$.01/$<$.01} & \textcolor{black!30}{$<$.01/$<$.01} & \textcolor{black!45}{.01/.05} & \textcolor{black!30}{$<$.01/$<$.01} & \textcolor{black!30}{$<$.01/$<$.01} & \textcolor{black!30}{$<$.01/$<$.01} & \textcolor{black!30}{$<$.01/$<$.01} & \textcolor{black!45}{.05/.02} & \textcolor{black!55}{0\% (0/12)} \\
 &  & \texttt{gpt-4o-mini} & \textcolor{black!30}{$<$.01/$<$.01} & \textcolor{black!30}{$<$.01/$<$.01} & \textcolor{black!30}{$<$.01/$<$.01} & \textcolor{black!30}{$<$.01/$<$.01} & \textbf{.44/.37} & \textcolor{black!30}{$<$.01/$<$.01} & \textcolor{black!30}{$<$.01/$<$.01} & \textbf{.46/.54} & \textcolor{black!30}{$<$.01/$<$.01} & \textcolor{black!30}{$<$.01/$<$.01} & \textcolor{black!30}{$<$.01/$<$.01} & \textcolor{black!30}{$<$.01/$<$.01} & \textcolor{black!55}{17\% (2/12)} \\
\cmidrule(l){2-16}
 & \multirow{3}{*}{\textbf{ThreadPDG}} & \texttt{DeepSeek-V4.1-Flash} & \textbf{.13/.16} & \textbf{.63/.74} & \textcolor{black!45}{.01/.02} & \textbf{.38/.36} & \textbf{.78/.29} & \textcolor{black!30}{$<$.01/$<$.01} & \textbf{.14/.19} & \textbf{.55/.83} & \textbf{.09/.07} & \textcolor{black!45}{.02/.07} & \textcolor{black!30}{$<$.01/$<$.01} & \textcolor{black!30}{.01/$<$.01} & \textcolor{black!55}{58\% (7/12)} \\
 &  & \texttt{gpt-5.4-mini} & \textbf{.42/.65} & \textbf{.20/.38} & \textbf{.35/.53} & \textbf{.57/.59} & \textbf{.99/.57} & \textbf{.45/.57} & \textbf{.92/1.0} & \textbf{.21/.20} & \textcolor{black!30}{$<$.01/$<$.01} & \textcolor{black!30}{$<$.01/.03} & \textcolor{black!30}{$<$.01/$<$.01} & \textbf{.12/.06} & \textbf{75\% (9/12)} \\
 &  & \texttt{gpt-4o-mini} & \textcolor{black!30}{$<$.01/$<$.01} & \textcolor{black!30}{$<$.01/$<$.01} & \textcolor{black!30}{$<$.01/$<$.01} & \textbf{.45/.66} & \textbf{.94/.54} & \textcolor{black!30}{$<$.01/$<$.01} & \textbf{.69/.91} & \textcolor{black!30}{$<$.01/$<$.01} & \textcolor{black!30}{$<$.01/$<$.01} & \textcolor{black!30}{$<$.01/$<$.01} & \textbf{.54/.45} & \textcolor{black!30}{$<$.01/$<$.01} & \textcolor{black!55}{33\% (4/12)} \\
\midrule
\multirow{9}{*}{\rotatebox[origin=c]{90}{\textbf{TV Series}}} & \multirow{3}{*}{OASIS} & \texttt{DeepSeek-V4.1-Flash} & \textcolor{black!30}{$<$.01/$<$.01} & \textcolor{black!30}{$<$.01/$<$.01} & \textcolor{black!30}{$<$.01/$<$.01} & \textcolor{black!30}{$<$.01/$<$.01} & \textcolor{black!30}{$<$.01/$<$.01} & \textcolor{black!30}{$<$.01/$<$.01} & \textcolor{black!30}{$<$.01/$<$.01} & \textcolor{black!30}{$<$.01/$<$.01} & \textcolor{black!30}{$<$.01/$<$.01} & \textcolor{black!30}{$<$.01/$<$.01} & \textbf{.09/.33} & \textbf{.83/.82} & \textcolor{black!55}{17\% (2/12)} \\
 &  & \texttt{gpt-5.4-mini} & \textcolor{black!30}{$<$.01/$<$.01} & \textcolor{black!30}{$<$.01/$<$.01} & \textcolor{black!30}{$<$.01/$<$.01} & \textcolor{black!30}{$<$.01/$<$.01} & \textcolor{black!30}{$<$.01/$<$.01} & \textcolor{black!30}{$<$.01/$<$.01} & \textcolor{black!30}{$<$.01/$<$.01} & \textcolor{black!30}{$<$.01/$<$.01} & \textcolor{black!30}{$<$.01/$<$.01} & \textcolor{black!30}{$<$.01/$<$.01} & \textcolor{black!30}{$<$.01/$<$.01} & \textcolor{black!30}{$<$.01/$<$.01} & \textcolor{black!55}{0\% (0/12)} \\
 &  & \texttt{gpt-4o-mini} & \textcolor{black!30}{$<$.01/$<$.01} & \textcolor{black!30}{$<$.01/$<$.01} & \textcolor{black!30}{$<$.01/$<$.01} & \textcolor{black!30}{$<$.01/$<$.01} & \textcolor{black!30}{$<$.01/$<$.01} & \textcolor{black!30}{$<$.01/$<$.01} & \textcolor{black!30}{$<$.01/$<$.01} & \textcolor{black!30}{$<$.01/$<$.01} & \textcolor{black!30}{$<$.01/$<$.01} & \textcolor{black!30}{$<$.01/$<$.01} & \textcolor{black!30}{$<$.01/$<$.01} & \textcolor{black!30}{$<$.01/$<$.01} & \textcolor{black!55}{0\% (0/12)} \\
\cmidrule(l){2-16}
 & \multirow{3}{*}{SynthPAI} & \texttt{DeepSeek-V4.1-Flash} & \textcolor{black!30}{$<$.01/$<$.01} & \textcolor{black!30}{$<$.01/$<$.01} & \textcolor{black!30}{$<$.01/$<$.01} & \textcolor{black!30}{$<$.01/$<$.01} & \textcolor{black!30}{$<$.01/$<$.01} & \textcolor{black!30}{$<$.01/$<$.01} & \textcolor{black!30}{$<$.01/$<$.01} & \textbf{.87/.09} & \textcolor{black!30}{$<$.01/$<$.01} & \textcolor{black!30}{$<$.01/$<$.01} & \textcolor{black!30}{$<$.01/$<$.01} & \textcolor{black!30}{$<$.01/$<$.01} & \textcolor{black!55}{8\% (1/12)} \\
 &  & \texttt{gpt-5.4-mini} & \textcolor{black!30}{$<$.01/$<$.01} & \textcolor{black!30}{$<$.01/$<$.01} & \textcolor{black!30}{$<$.01/$<$.01} & \textcolor{black!30}{$<$.01/$<$.01} & \textcolor{black!30}{$<$.01/$<$.01} & \textcolor{black!30}{.07/$<$.01} & \textcolor{black!30}{$<$.01/$<$.01} & \textcolor{black!30}{$<$.01/$<$.01} & \textcolor{black!30}{$<$.01/$<$.01} & \textcolor{black!30}{$<$.01/$<$.01} & \textcolor{black!30}{$<$.01/$<$.01} & \textcolor{black!30}{$<$.01/$<$.01} & \textcolor{black!55}{0\% (0/12)} \\
 &  & \texttt{gpt-4o-mini} & \textcolor{black!30}{$<$.01/$<$.01} & \textcolor{black!30}{$<$.01/$<$.01} & \textcolor{black!30}{$<$.01/$<$.01} & \textcolor{black!30}{$<$.01/$<$.01} & \textcolor{black!30}{$<$.01/$<$.01} & \textcolor{black!30}{$<$.01/$<$.01} & \textcolor{black!30}{$<$.01/$<$.01} & \textcolor{black!45}{.79/.03} & \textcolor{black!30}{$<$.01/$<$.01} & \textcolor{black!30}{$<$.01/$<$.01} & \textcolor{black!30}{$<$.01/$<$.01} & \textcolor{black!30}{$<$.01/$<$.01} & \textcolor{black!55}{0\% (0/12)} \\
\cmidrule(l){2-16}
 & \multirow{3}{*}{\textbf{ThreadPDG}} & \texttt{DeepSeek-V4.1-Flash} & \textbf{.27/.31} & \textbf{.08/.09} & \textbf{.46/.44} & \textbf{.51/.25} & \textbf{.63/.76} & \textcolor{black!30}{$<$.01/$<$.01} & \textbf{.58/.46} & \textbf{.96/.21} & \textbf{.61/.72} & \textbf{.20/.34} & \textbf{.08/.41} & \textbf{.65/.90} & \textbf{92\% (11/12)} \\
 &  & \texttt{gpt-5.4-mini} & \textcolor{black!30}{$<$.01/$<$.01} & \textbf{.24/.36} & \textbf{1.0/.70} & \textbf{.46/.35} & \textbf{.58/.74} & \textbf{.06/.16} & \textbf{.74/.77} & \textbf{.38/.06} & \textbf{.19/.14} & \textbf{.43/.30} & \textbf{.70/.86} & \textbf{.26/.42} & \textbf{92\% (11/12)} \\
 &  & \texttt{gpt-4o-mini} & \textcolor{black!30}{$<$.01/$<$.01} & \textcolor{black!30}{$<$.01/$<$.01} & \textcolor{black!30}{$<$.01/$<$.01} & \textbf{.61/.33} & \textbf{.82/.89} & \textbf{.05/.11} & \textcolor{black!45}{.08/.04} & \textcolor{black!30}{$<$.01/$<$.01} & \textcolor{black!30}{$<$.01/$<$.01} & \textcolor{black!30}{$<$.01/.03} & \textbf{.18/.15} & \textcolor{black!30}{$<$.01/$<$.01} & \textcolor{black!55}{33\% (4/12)} \\
\bottomrule
\end{tabular}
}
\vspace{-10pt}
\end{table*}


\subsection{Results on ThreadBench}
\label{sec:baseline_results}
As Table~\ref{tab:holdout_pvalues} shows with \texttt{gpt-5.4-mini}, ThreadPDG achieves pass rates ranging
from 67\% (8/12) to 92\% (11/12) across 8 domains, compared with
0\%-8\% for OASIS and 0\%-17\% for SynthPAI.
With \texttt{DeepSeek-V4.1-Flash}, ThreadPDG achieves pass rates ranging
from 58\% (7/12) to 100\% (12/12) across 8 domains, compared with
8\%-17\% for OASIS and 0\%-17\% for SynthPAI. SynthPAI and OASIS'
metric distributions differ from real discussions on nearly every
metric. ThreadPDG reduces these differences, but its results remain least consistent in politeness, disagreement, and length. Appendix~\ref{app:distance_measures} gives the Wasserstein distance and Cliff's delta for the same comparisons. Appendix~\ref{app:examples} shows one post discussed by real users and by each method.

\paragraph{Human evaluation.}
We conduct a turing test with 436 annotators on 240 matched discussion
pairs across eight domains. Each pair contains one real Reddit discussion and
one discussion generated by ThreadPDG, SynthPAI, or OASIS from the same post,
and annotators choose which is the real thread or cannot distinguish. We aggregate
responses by pair-level majority vote, with full survey and filtering details
provided in Appendix~\ref{app:turing}. As shown in
Table~\ref{tab:turing-pair-majority}, ThreadPDG is not distinguished in
78.7\% of pairs, compared with 61.1\% for OASIS and 58.7\% for SynthPAI.

\subsection{Effect of the Number of Drafts}
\label{sec:ablation}

\begin{table}[!t]
\centering
\scriptsize
\setlength{\tabcolsep}{3pt}
\renewcommand{\arraystretch}{1.15}

\caption{ThreadBench pass rates (\%) across draft budgets.
Parentheses show relative improvements over the prior budget.}
\label{tab:ablation_k}

\resizebox{\linewidth}{!}{%
\begin{tabular}{@{}lccccccccc@{}}
\toprule
& \multicolumn{3}{c}{\texttt{gpt-5.4-mini}}
& \multicolumn{3}{c}{\texttt{DeepSeek-V4.1-Flash}}
& \multicolumn{3}{c}{\texttt{gpt-4o-mini}} \\
\cmidrule(lr){2-4}
\cmidrule(lr){5-7}
\cmidrule(l){8-10}

\textbf{Domain}
& $K=1$ & $K=2$ & $K=3$
& $K=1$ & $K=2$ & $K=3$
& $K=1$ & $K=2$ & $K=3$ \\
\midrule

Camera
& 42\% & 75\% ($\uparrow$\,80\%) & 83\% ($\uparrow$\,11\%)
& 33\% & 75\% ($\uparrow$\,125\%) & 92\% ($\uparrow$\,22\%)
& 17\% & 17\% ($\uparrow$\,0\%) & 17\% ($\uparrow$\,0\%) \\

Celebrity
& 33\% & 75\% ($\uparrow$\,125\%) & 92\% ($\uparrow$\,22\%)
& 67\% & 92\% ($\uparrow$\,38\%) & 100\% ($\uparrow$\,9\%)
& 25\% & 25\% ($\uparrow$\,0\%) & 25\% ($\uparrow$\,0\%) \\

Cellphone
& 50\% & 67\% ($\uparrow$\,33\%) & 83\% ($\uparrow$\,25\%)
& 50\% & 75\% ($\uparrow$\,50\%) & 83\% ($\uparrow$\,11\%)
& 33\% & 33\% ($\uparrow$\,0\%) & 42\% ($\uparrow$\,27\%) \\

Game
& 33\% & 67\% ($\uparrow$\,100\%) & 67\% ($\uparrow$\,0\%)
& 58\% & 83\% ($\uparrow$\,43\%) & 83\% ($\uparrow$\,0\%)
& 17\% & 17\% ($\uparrow$\,0\%) & 25\% ($\uparrow$\,50\%) \\

Health
& 50\% & 67\% ($\uparrow$\,33\%) & 67\% ($\uparrow$\,0\%)
& 33\% & 58\% ($\uparrow$\,75\%) & 58\% ($\uparrow$\,0\%)
& 17\% & 25\% ($\uparrow$\,50\%) & 25\% ($\uparrow$\,0\%) \\

Movies
& 50\% & 67\% ($\uparrow$\,33\%) & 92\% ($\uparrow$\,38\%)
& 50\% & 83\% ($\uparrow$\,67\%) & 83\% ($\uparrow$\,0\%)
& 17\% & 17\% ($\uparrow$\,0\%) & 25\% ($\uparrow$\,50\%) \\

News
& 58\% & 67\% ($\uparrow$\,14\%) & 75\% ($\uparrow$\,13\%)
& 25\% & 50\% ($\uparrow$\,100\%) & 58\% ($\uparrow$\,17\%)
& 33\% & 33\% ($\uparrow$\,0\%) & 33\% ($\uparrow$\,0\%) \\

TV Series
& 50\% & 75\% ($\uparrow$\,50\%) & 92\% ($\uparrow$\,22\%)
& 67\% & 75\% ($\uparrow$\,13\%) & 92\% ($\uparrow$\,22\%)
& 25\% & 25\% ($\uparrow$\,0\%) & 33\% ($\uparrow$\,33\%) \\

\bottomrule
\end{tabular}%
}
\end{table}

We study how the draft budget $K$ affects ThreadPDG performance while keeping the
Planner--Writer framework and branch-selection pipeline fixed. Table~\ref{tab:ablation_k} shows that increasing the draft
budget from $K=1$ to $K=3$ improves the pass rate in every domain: by 29\% to 179\%
for \texttt{gpt-5.4-mini} and by 37\% to 179\% for \texttt{DeepSeek-V4.1-Flash},
over the same eight domains. \texttt{gpt-4o-mini} gains at most 47\% and does not
improve in three domains.

\subsection{Qualitative Observations}
\label{sec:observations}

\begin{wrapfigure}{r}{0.5\textwidth}
    \centering
    \vspace{-12pt}
    \includegraphics[width=\linewidth]{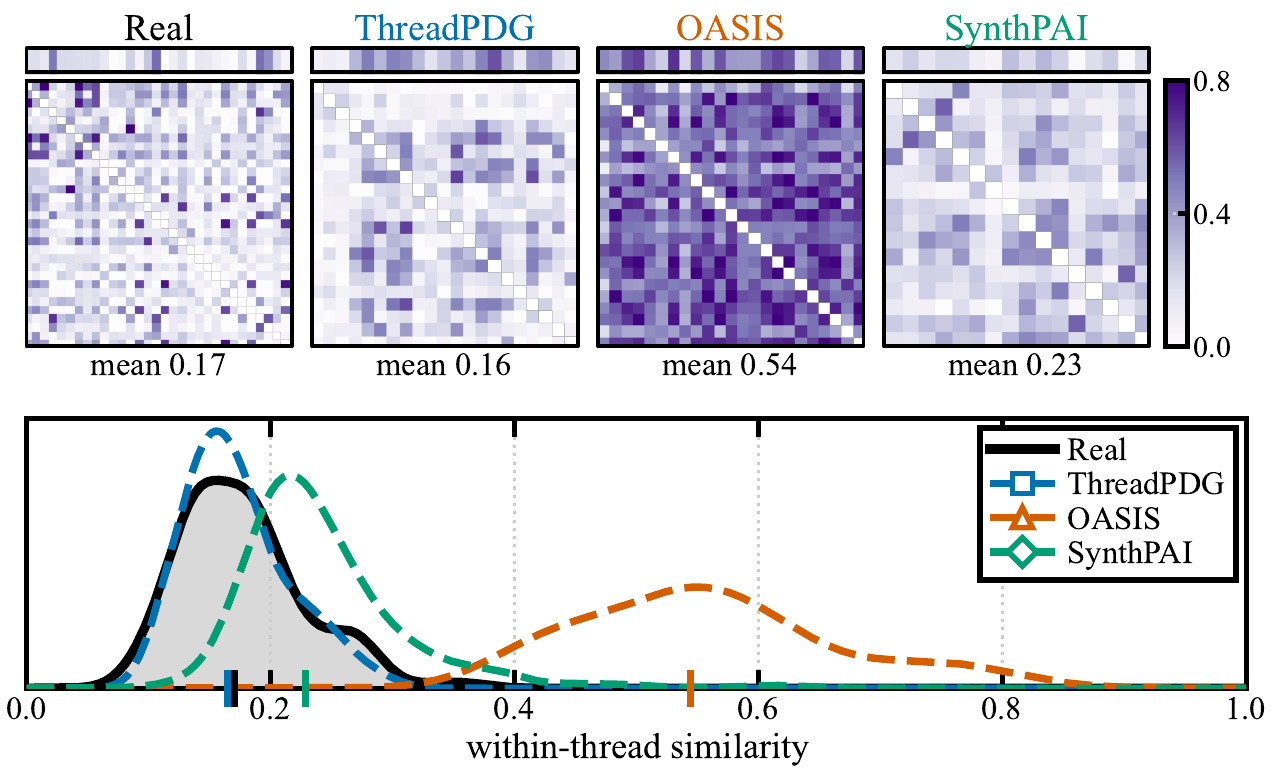}
    \caption{Comment's semantic similarity within-thread. Top: A representative Celebrity thread for each method, shown as a pairwise comment-similarity matrix. Darker cells indicate more similar comments. Bottom: The distribution of mean within-thread similarity across Celebrity threads.}
    \label{fig:qual_semantic}
    \vspace{-20pt}
\end{wrapfigure}

\textbf{Semantic diversity.} Figure~\ref{fig:qual_semantic} compares the similarity between comments within a thread, using the same embedding and cosine similarity as \texttt{semantic\_mean\_cosine}. Top Figure~\ref{fig:qual_semantic}shows, for each method, the thread closest to its median similarity. Real and ThreadPDG threads contain comments that cover more varied directions, while OASIS comments are much more similar to one another. 
At the distribution level, bottom Figure~\ref{fig:qual_semantic} shows that OASIS is clearly shifted toward higher within-thread similarity, while SynthPAI lies between OASIS and real discussions. Additional results and analysis for other LLMs and domains are provided in Appendix~\ref{app:qualitative}.

\textbf{Comment length.}
Figure~\ref{fig:qual_length} illustrates real discussions that combine short reactions with longer replies, whereas OASIS
produces mostly longer comments and SynthPAI produces mostly short ones. ThreadPDG closely matches the real median but still differs from real discussions in some domains, as detailed in Appendix~\ref{app:qualitative} with all eight
domains and each LLMs.

\textbf{Tone and structure.}
Figure~\ref{fig:qual_tone} shows that both OASIS and SynthPAI begin comments with
agreement more often than real users do. OASIS also offers advice and hedges
more often, while it begins with direct disagreement less often. ThreadPDG
matches the observed rates more closely, but its politeness and disagreement
metrics still away from real comments. The reply structures also differ:
Table~\ref{tab:fingerprint} in Appendix \ref{app:qualitative} shows that 46\% of real comments reply to another
comment, compared with 0\% for OASIS, 85\% for SynthPAI, and 47\% for
ThreadPDG. 

\section{Conclusion}
\label{sec:conclusion}

We introduced ThreadBench, a benchmark for evaluating Reddit discussion simulation at the thread level. ThreadBench contains 9{,}817 real Reddit threads across 8 domains and 116 subdomains, and evaluates simulated discussions using 12 metrics covering multiple aspects of discussion behavior. Using ThreadBench, we show that existing simulators still differ substantially from real discussions, while ThreadPDG more closely matches real thread-level behavior across domains and LLM backbones. ThreadPDG improves realism by learning patterns of how real discussions develop rather than reproducing the content of individual threads, although important differences from human discussions still remain. We view ThreadBench as a starting point for the community to measure these gaps more systematically, compare future simulation methods on a common basis, and develop discussion simulators that more faithfully reflect real human behavior.


\newpage

\bibliography{paper}
\bibliographystyle{iclr2027_conference}

\clearpage
\appendix


\section{ThreadBench Benchmark Inventory}
\label{app:benchmark_inventory}

This appendix provides the complete inventory and sampling configuration
of ThreadBench, including the domain-level sampling threshold, the number
of subdomains represented within each domain, and the full
subdomain-level composition of the benchmark.

\paragraph{Data and ethics.}
The corpus was collected by the authors from publicly visible Reddit threads and is
used for research only. Before scoring we remove usernames, permalinks, and thread
identifiers, and we release derived thread-level metric values rather than raw post
text. A thread enters the benchmark only if it passes the content and structural
checks described below, and every measurement in this paper is computed over threads
with at least five comments. We make no attempt to identify, profile, or contact any
user, and the corpus is not used for any purpose beyond the evaluation reported here.

\subsection{Sampling Configuration}
\label{app:benchmark_sampling}

To maintain balanced coverage across high-level discussion domains,
ThreadBench draws each domain from a fixed set of source subreddits under a
per-domain crawl quota. The retained discussions preserve
the observed coverage of the domain's constituent subdomains rather than
forcing an artificial uniform distribution across subdomains.

This procedure yields a high-level inventory of 9,817 discussion
threads across 8 domains. The domain-level thread count is not uniform, and
the underlying subdomain structure is intentionally not forced to be uniform
either. The number of represented subdomains therefore varies by domain. In
total, ThreadBench contains 116 distinct subdomains.










\subsection{Subdomain-level Inventory}
\label{app:benchmark_subdomain_inventory}

Table~\ref{tab:complete_subdomain_inventory} reports the complete
subdomain-level composition of ThreadBench. For each subdomain, we report
the number of retained discussion threads, its share of the corresponding
domain (\textit{Domain \%}), and its share of the complete benchmark
(\textit{ThreadBench \%}). \textit{Domain \%} is computed relative to that
domain's own thread count, while \textit{ThreadBench \%} is computed relative
to all 9,817 benchmark threads.

\newpage
\begin{longtable}{
    @{}
    llrrr
    @{}
}
\caption{
Complete subdomain-level inventory of ThreadBench.
\textit{Domain \%} denotes the fraction of threads within the corresponding
domain inventory, while \textit{ThreadBench \%} denotes the fraction
of all 9,817 benchmark threads.
}
\label{tab:complete_subdomain_inventory} \\

\toprule
\textbf{Domain}
& \textbf{Subdomain}
& \textbf{\# Threads}
& \textbf{Domain \%}
& \textbf{ThreadBench \%} \\
\midrule
\endfirsthead

\multicolumn{5}{c}{
\tablename\ \thetable{} -- continued from previous page
} \\

\toprule
\textbf{Domain}
& \textbf{Subdomain}
& \textbf{\# Threads}
& \textbf{Domain \%}
& \textbf{ThreadBench \%} \\
\midrule
\endhead

\midrule
\multicolumn{5}{r}{Continued on next page} \\
\endfoot

\bottomrule
\endlastfoot

\textbf{Camera}      & photography              &   557 & 29.0\% & 5.67\% \\
                     & Cameras                  &   411 & 21.4\% & 4.19\% \\
                     & AskPhotography           &   302 & 15.7\% & 3.08\% \\
                     & CameraRestocks           &   270 & 14.0\% & 2.75\% \\
                     & analog                   &   200 & 10.4\% & 2.04\% \\
                     & SonyAlpha                &   110 &  5.7\% & 1.12\% \\
                     & hasselblad               &    16 &  0.8\% & 0.16\% \\
                     & bandccamera              &     6 &  0.3\% & 0.06\% \\
                     & Leica                    &     4 &  0.2\% & 0.04\% \\
                     & CameraDeals              &     3 &  0.2\% & 0.03\% \\
                     & fujifilm                 &     3 &  0.2\% & 0.03\% \\
                     & mediumformat             &     3 &  0.2\% & 0.03\% \\
                     & photomarket              &     3 &  0.2\% & 0.03\% \\
                     & AllElectronicsDeals      &     2 &  0.1\% & 0.02\% \\
                     & OzBargainNew             &     2 &  0.1\% & 0.02\% \\
                     & nelsonphotovideo         &     2 &  0.1\% & 0.02\% \\
                     & AnalogCommunity          &     1 &  0.1\% & 0.01\% \\
                     & BirdPhotography          &     1 &  0.1\% & 0.01\% \\
                     & CalPoly                  &     1 &  0.1\% & 0.01\% \\
                     & DealForDays              &     1 &  0.1\% & 0.01\% \\
                     & India\_Today             &     1 &  0.1\% & 0.01\% \\
                     & Instagram                &     1 &  0.1\% & 0.01\% \\
                     & M43                      &     1 &  0.1\% & 0.01\% \\
                     & Nikon                    &     1 &  0.1\% & 0.01\% \\
                     & PowerlessGamingNews      &     1 &  0.1\% & 0.01\% \\
                     & SkydTech                 &     1 &  0.1\% & 0.01\% \\
                     & Sports\_Videography      &     1 &  0.1\% & 0.01\% \\
                     & Tech2019                 &     1 &  0.1\% & 0.01\% \\
                     & astrophotography         &     1 &  0.1\% & 0.01\% \\
                     & burlington               &     1 &  0.1\% & 0.01\% \\
                     & canon                    &     1 &  0.1\% & 0.01\% \\
                     & captureone               &     1 &  0.1\% & 0.01\% \\
                     & classifiedsph            &     1 &  0.1\% & 0.01\% \\
                     & gadgets                  &     1 &  0.1\% & 0.01\% \\
                     & hifi\_fan                &     1 &  0.1\% & 0.01\% \\
                     & masters                  &     1 &  0.1\% & 0.01\% \\
                     & phclassifieds            &     1 &  0.1\% & 0.01\% \\
                     & pointandshoot            &     1 &  0.1\% & 0.01\% \\
                     & portraits                &     1 &  0.1\% & 0.01\% \\
                     & sonya7iv                 &     1 &  0.1\% & 0.01\% \\
                     & techcrunch               &     1 &  0.1\% & 0.01\% \\
                     & uAlberta                 &     1 &  0.1\% & 0.01\% \\
                     & underwaterphotography    &     1 &  0.1\% & 0.01\% \\
                     & vermont                  &     1 &  0.1\% & 0.01\% \\
\addlinespace[3pt]
\textbf{Celebrity}   & Fauxmoi                  &   200 & 28.6\% & 2.04\% \\
                     & entertainment            &   200 & 28.6\% & 2.04\% \\
                     & popculturechat           &   200 & 28.6\% & 2.04\% \\
                     & Celebs                   &   100 & 14.3\% & 1.02\% \\
\addlinespace[3pt]
\textbf{Cellphone}   & Android                  &   400 & 28.7\% & 4.07\% \\
                     & iphone                   &   301 & 21.6\% & 3.07\% \\
                     & Smartphones              &   200 & 14.3\% & 2.04\% \\
                     & GooglePixel              &   101 &  7.2\% & 1.03\% \\
                     & samsung                  &   100 &  7.2\% & 1.02\% \\
                     & Visible                  &    88 &  6.3\% & 0.90\% \\
                     & mintmobile               &    57 &  4.1\% & 0.58\% \\
                     & NoContract               &    50 &  3.6\% & 0.51\% \\
                     & hardwareswap             &    32 &  2.3\% & 0.33\% \\
                     & androiddeals             &     6 &  0.4\% & 0.06\% \\
                     & dealsonamazon            &     5 &  0.4\% & 0.05\% \\
                     & pixel\_phones            &     5 &  0.4\% & 0.05\% \\
                     & AmazonofDeals            &     4 &  0.3\% & 0.04\% \\
                     & DripDeals                &     4 &  0.3\% & 0.04\% \\
                     & appleswap                &     4 &  0.3\% & 0.04\% \\
                     & samsunggalaxy            &     4 &  0.3\% & 0.04\% \\
                     & dealsInAmazon            &     3 &  0.2\% & 0.03\% \\
                     & DealForDays              &     2 &  0.1\% & 0.02\% \\
                     & GoogleStoreReferrals     &     2 &  0.1\% & 0.02\% \\
                     & dumbphones               &     2 &  0.1\% & 0.02\% \\
                     & pixelreferral            &     2 &  0.1\% & 0.02\% \\
                     & AmazonVine               &     1 &  0.1\% & 0.01\% \\
                     & BangaloreMarketplace     &     1 &  0.1\% & 0.01\% \\
                     & BestOfPrimeDay           &     1 &  0.1\% & 0.01\% \\
                     & Bestbuy                  &     1 &  0.1\% & 0.01\% \\
                     & ChicagoList              &     1 &  0.1\% & 0.01\% \\
                     & DealsRUs                 &     1 &  0.1\% & 0.01\% \\
                     & Deals\_Central           &     1 &  0.1\% & 0.01\% \\
                     & ItaliaBox                &     1 &  0.1\% & 0.01\% \\
                     & JBBuddy                  &     1 &  0.1\% & 0.01\% \\
                     & MotoG                    &     1 &  0.1\% & 0.01\% \\
                     & PickAnAndroidForMe       &     1 &  0.1\% & 0.01\% \\
                     & Pixel9Pro                &     1 &  0.1\% & 0.01\% \\
                     & PixelUsersIndia          &     1 &  0.1\% & 0.01\% \\
                     & TopDealsUsa              &     1 &  0.1\% & 0.01\% \\
                     & asmr                     &     1 &  0.1\% & 0.01\% \\
                     & blackfriday              &     1 &  0.1\% & 0.01\% \\
                     & dailydeals               &     1 &  0.1\% & 0.01\% \\
                     & flipphones               &     1 &  0.1\% & 0.01\% \\
                     & gtamarketplace           &     1 &  0.1\% & 0.01\% \\
                     & homelabsales             &     1 &  0.1\% & 0.01\% \\
                     & iPhone17Pro              &     1 &  0.1\% & 0.01\% \\
                     & phoneswap                &     1 &  0.1\% & 0.01\% \\
                     & u\_arongae               &     1 &  0.1\% & 0.01\% \\
\addlinespace[3pt]
\textbf{Game}        & Games                    &   300 & 30.0\% & 3.06\% \\
                     & gaming                   &   300 & 30.0\% & 3.06\% \\
                     & pcgaming                 &   200 & 20.0\% & 2.04\% \\
                     & GamingLeaksAndRumours    &   100 & 10.0\% & 1.02\% \\
                     & patientgamers            &   100 & 10.0\% & 1.02\% \\
\addlinespace[3pt]
\textbf{Health Issue} & AskDocs                 &   600 & 28.6\% & 6.11\% \\
                     & medical                  &   500 & 23.8\% & 5.09\% \\
                     & HealthAnxiety            &   400 & 19.0\% & 4.07\% \\
                     & ChronicIllness           &   300 & 14.3\% & 3.06\% \\
                     & Health                   &   300 & 14.3\% & 3.06\% \\
\addlinespace[3pt]
\textbf{Movies}      & MovieDetails             &   200 & 28.6\% & 2.04\% \\
                     & movies                   &   200 & 28.6\% & 2.04\% \\
                     & TrueFilm                 &   100 & 14.3\% & 1.02\% \\
                     & boxoffice                &   100 & 14.3\% & 1.02\% \\
                     & flicks                   &   100 & 14.3\% & 1.02\% \\
                     & TrueFilm                 &   100 & 14.3\% & 1.02\% \\
                     & boxoffice                &   100 & 14.3\% & 1.02\% \\
                     & flicks                   &   100 & 14.3\% & 1.02\% \\
\addlinespace[3pt]
\textbf{News}        & news                     &   300 & 30.0\% & 3.06\% \\
                     & worldnews                &   300 & 30.0\% & 3.06\% \\
                     & TrueReddit               &   200 & 20.0\% & 2.04\% \\
                     & UpliftingNews            &   100 & 10.0\% & 1.02\% \\
                     & nottheonion              &   100 & 10.0\% & 1.02\% \\
\addlinespace[3pt]
\textbf{TV Series}   & netflix                  &   300 & 30.0\% & 3.06\% \\
                     & television               &   300 & 30.0\% & 3.06\% \\
                     & TVDetails                &   200 & 20.0\% & 2.04\% \\
                     & HBOMAX                   &   100 & 10.0\% & 1.02\% \\
                     & Hulu                     &   100 & 10.0\% & 1.02\% \\

\midrule

\textbf{Total}
& \textbf{116 subdomains}
& \textbf{9,5817}
& \textbf{100.0\%}
& \textbf{100.0\%} \\

\end{longtable}

\newpage
\section{Metrics Calculation}
\label{app:metric_calculation}
For a thread $t$, let
$C_t=\{c_1,\ldots,c_{n_t}\}$ denote its comments.
Let $\mathcal{P}(t)=\{(i,j):1\leq i<j\leq n_t\}$ be the set of
unordered comment pairs, and let $M_t=|\mathcal{P}(t)|$.
Metrics with no valid comments or pairs are assigned a value of zero.

\begin{table}[H]
\centering
\scriptsize
\setlength{\tabcolsep}{2.6pt}
\renewcommand{\arraystretch}{1.06}
\caption{Thread-level formulas for the 12 core metrics in ThreadBench}
\vspace{4pt}
\label{tab:metric_formula_table}
\resizebox{\textwidth}{!}{%
\begin{tabular}{
  >{\raggedright\arraybackslash}p{2.25cm}
  >{\raggedright\arraybackslash}p{3.45cm}
  >{\raggedright\arraybackslash}p{8.85cm}
}
\toprule
Family & Metric & Formula \\
\midrule

\multirow[t]{3}{2.25cm}{\raggedright Uniformity}
& \texttt{self\_bleu\_4}
& $\displaystyle
\frac{1}{M_t}
\sum_{(i,j)\in\mathcal{P}(t)}
\widetilde{\mathrm{BLEU}}_4(c_i,c_j)$
\\

& \texttt{self\_bertscore\_mean\_f1}
& $\displaystyle
\frac{1}{M_t}
\sum_{(i,j)\in\mathcal{P}(t)}
\mathrm{BERTScore}_{F1}(c_i,c_j)$
\\

& \texttt{semantic\_mean\_cosine}
& $\displaystyle
\frac{1}{M_t}
\sum_{(i,j)\in\mathcal{P}(t)}
\frac{e_i^\top e_j}{\lVert e_i\rVert_2\lVert e_j\rVert_2}$
\\

\midrule

\multirow[t]{3}{2.25cm}{\raggedright Interaction}
& \texttt{avg\_depth}
& $\displaystyle
\frac{1}{n_t}\sum_{i=1}^{n_t}d_i$
\\

& \texttt{structural\_virality}
& $\displaystyle
\frac{1}{R_t}
\sum_{(i,j)\in\mathcal{R}(t)}
\operatorname{dist}_{G_t}(i,j),
\qquad
R_t=|\mathcal{R}(t)|$
\\

& \texttt{length\_cv}
& $\displaystyle
\frac{
\sqrt{
\frac{1}{n_t}
\sum_{i=1}^{n_t}
(\ell_i-\bar{\ell}_t)^2
}
}{
\bar{\ell}_t
}$
\\

\midrule

\multirow[t]{2}{2.25cm}{\raggedright Expression}
& \texttt{mean\_story\_probability}
& $\displaystyle
\frac{1}{n_t}\sum_{i=1}^{n_t}s_i$
\\

& \texttt{emotion\_entropy}
& $\displaystyle
-\sum_{a\in\mathcal{A}:\,\pi_{t,a}>0}
\pi_{t,a}\log \pi_{t,a},
\qquad
\pi_{t,a}
=
\frac{1}{n_t}
\sum_{i=1}^{n_t}
\mathbf{1}[\hat{a}_i=a]$
\\

\midrule

\multirow[t]{4}{2.25cm}{\raggedright Tone}
& \texttt{polite\_rate}
& $\displaystyle
\frac{1}{n_t}
\sum_{i=1}^{n_t}
\mathbf{1}[\hat{y}_i=\mathrm{polite}]$
\\

& \texttt{neutral\_rate}
& $\displaystyle
\frac{1}{n_t}
\sum_{i=1}^{n_t}
\mathbf{1}[\hat{y}_i=\mathrm{neutral}]$
\\

& \texttt{impolite\_rate}
& $\displaystyle
\frac{1}{n_t}
\sum_{i=1}^{n_t}
\mathbf{1}[\hat{y}_i=\mathrm{impolite}]$
\\

& \texttt{hard\_disagree\_rate}
& $\displaystyle
\frac{1}{|\mathcal{E}(t)|}
\sum_{(u,v)\in\mathcal{E}(t)}
\mathbf{1}[\hat{z}_{uv}=\mathrm{disagree}]$
\\

\bottomrule
\end{tabular}%
}
\end{table}

Here, $\widetilde{\mathrm{BLEU}}_4$ is the average of the two
directional BLEU-4 scores for a comment pair, with add-one smoothing.
The vector $e_i$ is the embedding of comment $c_i$, and $s_i$ is its
StorySeeker story probability. The label $\hat{a}_i$ is the dominant
emotion assigned to comment $c_i$, and $\mathcal{A}$ is the set of
emotion labels.

The label $\hat{y}_i$ is the politeness label assigned to comment
$c_i$. The set $\mathcal{E}(t)$ contains the valid parent--reply pairs
in thread $t$, and $\hat{z}_{uv}$ is the predicted stance label for
pair $(u,v)$. Top-level comments have depth $d_i=1$.

The graph $G_t$ contains comments connected by reply links.
The set $\mathcal{R}(t)$ contains the unordered comment pairs connected
by a path in $G_t$. Comment length $\ell_i$ is measured by the number
of whitespace-separated tokens.

\newpage
\section{Detailed Real-vs-Real Sanity-Check Statistics}
\label{app:real_vs_real_complete_results}

\begin{figure}[h]
    \centering
    \includegraphics[width=0.7\linewidth]{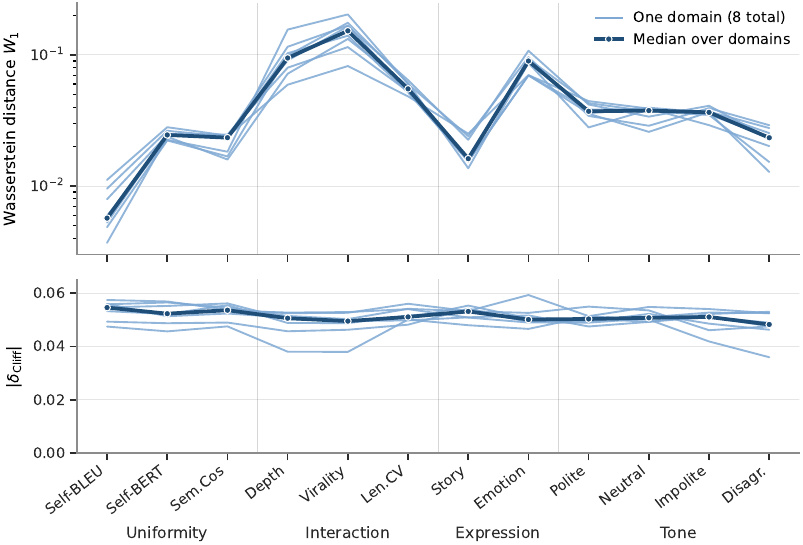}
    \caption{Real-vs.-real distance profiles over the twelve metrics, with one
    line per domain and each point the mean over 200 repeated draws of two
    150-thread samples. The domains trace the same profile: the pairwise Spearman
    correlation between two domains' $\log W_1$ profiles is at least $0.87$, with
    a median of $0.96$, and a metric's mean $W_1$ varies by at most a factor of
    $3$ across domains. Absolute Cliff's delta is flat as well, between $0.036$
    and $0.059$ for every metric in every domain, so rank separation between two
    real samples does not depend on which metric or which domain is measured.
    $W_1$ is shown on a logarithmic axis because each metric carries its own
    scale.}
    \label{fig:real_real_distances}
\end{figure}

For every domain and metric, we draw two $n=150$ samples from the real-thread pool 200 times and compare their thread-level score distributions. The tables report the median and mean $p$-values, the count and percentage of repeats with $p\geq0.05$, and the mean $W_1$ and mean $|\delta_{\mathrm{Cliff}}|$. Lower distance values indicate smaller differences. Distance values should be compared within the same metric, rather than across unrelated metrics.

\subsection{Camera}
\label{app:real_vs_real_camera}

\begin{table*}[h]
\centering
\scriptsize
\setlength{\tabcolsep}{2.9pt}
\renewcommand{\arraystretch}{1.06}
\caption{Repeated real-vs-real statistics for the camera domain. Each row summarizes 200 $n=150$-vs.-$n=150$ comparisons.}
\label{tab:real_vs_real_camera}
\resizebox{\textwidth}{!}{%
\begin{tabular}{lrrrrrrrr}
\toprule
Metric & Med. MWU $p$ & Mean MWU $p$ & MWU $p\geq0.05$ & Med. KS $p$ & Mean KS $p$ & KS $p\geq0.05$ & Mean $W_1$ & Mean $|\delta_{\mathrm{Cliff}}|$ \\
\midrule
\texttt{self\_bleu\_4} & 0.458 & 0.480 & 189/200 (94.5\%) & 0.595 & 0.555 & 191/200 (95.5\%) & 0.0055 & 0.0558 \\
\texttt{self\_bertscore\_mean\_f1} & 0.443 & 0.477 & 186/200 (93.0\%) & 0.501 & 0.549 & 192/200 (96.0\%) & 0.0249 & 0.0565 \\
\texttt{semantic\_mean\_cosine} & 0.504 & 0.496 & 188/200 (94.0\%) & 0.595 & 0.554 & 191/200 (95.5\%) & 0.0247 & 0.0546 \\
\texttt{avg\_depth} & 0.483 & 0.479 & 189/200 (94.5\%) & 0.787 & 0.684 & 193/200 (96.5\%) & 0.0800 & 0.0514 \\
\texttt{structural\_virality} & 0.485 & 0.483 & 190/200 (95.0\%) & 0.871 & 0.742 & 199/200 (99.5\%) & 0.1148 & 0.0502 \\
\texttt{length\_cv} & 0.452 & 0.484 & 194/200 (97.0\%) & 0.595 & 0.550 & 194/200 (97.0\%) & 0.0527 & 0.0541 \\
\texttt{mean\_story\_probability} & 0.464 & 0.497 & 190/200 (95.0\%) & 0.595 & 0.533 & 192/200 (96.0\%) & 0.0154 & 0.0532 \\
\texttt{emotion\_entropy} & 0.492 & 0.496 & 187/200 (93.5\%) & 0.693 & 0.652 & 195/200 (97.5\%) & 0.0699 & 0.0526 \\
\texttt{polite\_rate} & 0.482 & 0.483 & 190/200 (95.0\%) & 0.693 & 0.616 & 194/200 (97.0\%) & 0.0384 & 0.0549 \\
\texttt{neutral\_rate} & 0.483 & 0.488 & 189/200 (94.5\%) & 0.693 & 0.641 & 195/200 (97.5\%) & 0.0386 & 0.0535 \\
\texttt{impolite\_rate} & 0.572 & 0.531 & 192/200 (96.0\%) & 0.787 & 0.695 & 196/200 (98.0\%) & 0.0352 & 0.0461 \\
\texttt{hard\_disagree\_rate} & 0.468 & 0.480 & 186/200 (93.0\%) & 0.913 & 0.766 & 197/200 (98.5\%) & 0.0236 & 0.0475 \\
\bottomrule
\end{tabular}%
}
\end{table*}

\newpage
\subsection{Celebrity}
\label{app:real_vs_real_celebrity}

\begin{table*}[h]
\centering
\scriptsize
\setlength{\tabcolsep}{2.9pt}
\renewcommand{\arraystretch}{1.06}
\caption{Repeated real-vs-real statistics for the celebrity domain. Each row summarizes 200 $n=150$-vs.-$n=150$ comparisons.}
\label{tab:real_vs_real_celebrity}
\resizebox{\textwidth}{!}{%
\begin{tabular}{lrrrrrrrr}
\toprule
Metric & Med. MWU $p$ & Mean MWU $p$ & MWU $p\geq0.05$ & Med. KS $p$ & Mean KS $p$ & KS $p\geq0.05$ & Mean $W_1$ & Mean $|\delta_{\mathrm{Cliff}}|$ \\
\midrule
\texttt{self\_bleu\_4} & 0.436 & 0.471 & 185/200 (92.5\%) & 0.595 & 0.525 & 187/200 (93.5\%) & 0.0112 & 0.0574 \\
\texttt{self\_bertscore\_mean\_f1} & 0.490 & 0.479 & 185/200 (92.5\%) & 0.501 & 0.528 & 188/200 (94.0\%) & 0.0281 & 0.0568 \\
\texttt{semantic\_mean\_cosine} & 0.510 & 0.501 & 188/200 (94.0\%) & 0.595 & 0.557 & 193/200 (96.5\%) & 0.0243 & 0.0537 \\
\texttt{avg\_depth} & 0.514 & 0.493 & 184/200 (92.0\%) & 0.787 & 0.650 & 192/200 (96.0\%) & 0.0907 & 0.0524 \\
\texttt{structural\_virality} & 0.490 & 0.484 & 179/200 (89.5\%) & 0.693 & 0.644 & 194/200 (97.0\%) & 0.1653 & 0.0526 \\
\texttt{length\_cv} & 0.494 & 0.485 & 185/200 (92.5\%) & 0.595 & 0.560 & 188/200 (94.0\%) & 0.0643 & 0.0560 \\
\texttt{mean\_story\_probability} & 0.480 & 0.503 & 187/200 (93.5\%) & 0.595 & 0.533 & 190/200 (95.0\%) & 0.0226 & 0.0533 \\
\texttt{emotion\_entropy} & 0.458 & 0.453 & 181/200 (90.5\%) & 0.595 & 0.567 & 193/200 (96.5\%) & 0.1076 & 0.0593 \\
\texttt{polite\_rate} & 0.559 & 0.515 & 190/200 (95.0\%) & 0.595 & 0.584 & 193/200 (96.5\%) & 0.0419 & 0.0513 \\
\texttt{neutral\_rate} & 0.478 & 0.480 & 190/200 (95.0\%) & 0.693 & 0.606 & 190/200 (95.0\%) & 0.0339 & 0.0548 \\
\texttt{impolite\_rate} & 0.466 & 0.476 & 192/200 (96.0\%) & 0.644 & 0.594 & 195/200 (97.5\%) & 0.0410 & 0.0540 \\
\texttt{hard\_disagree\_rate} & 0.512 & 0.504 & 190/200 (95.0\%) & 0.732 & 0.651 & 195/200 (97.5\%) & 0.0232 & 0.0525 \\
\bottomrule
\end{tabular}%
}
\end{table*}

\subsection{Cellphone}
\label{app:real_vs_real_cellphone}

\begin{table*}[h]
\centering
\scriptsize
\setlength{\tabcolsep}{2.9pt}
\renewcommand{\arraystretch}{1.06}
\caption{Repeated real-vs-real statistics for the cellphone domain. Each row summarizes 200 $n=150$-vs.-$n=150$ comparisons.}
\label{tab:real_vs_real_cellphone}
\resizebox{\textwidth}{!}{%
\begin{tabular}{lrrrrrrrr}
\toprule
Metric & Med. MWU $p$ & Mean MWU $p$ & MWU $p\geq0.05$ & Med. KS $p$ & Mean KS $p$ & KS $p\geq0.05$ & Mean $W_1$ & Mean $|\delta_{\mathrm{Cliff}}|$ \\
\midrule
\texttt{self\_bleu\_4} & 0.502 & 0.495 & 192/200 (96.0\%) & 0.595 & 0.559 & 194/200 (97.0\%) & 0.0053 & 0.0532 \\
\texttt{self\_bertscore\_mean\_f1} & 0.499 & 0.501 & 189/200 (94.5\%) & 0.693 & 0.595 & 194/200 (97.0\%) & 0.0245 & 0.0521 \\
\texttt{semantic\_mean\_cosine} & 0.498 & 0.481 & 190/200 (95.0\%) & 0.595 & 0.565 & 192/200 (96.0\%) & 0.0232 & 0.0555 \\
\texttt{avg\_depth} & 0.517 & 0.511 & 190/200 (95.0\%) & 0.787 & 0.669 & 193/200 (96.5\%) & 0.1031 & 0.0488 \\
\texttt{structural\_virality} & 0.515 & 0.508 & 192/200 (96.0\%) & 0.787 & 0.711 & 198/200 (99.0\%) & 0.1401 & 0.0487 \\
\texttt{length\_cv} & 0.490 & 0.502 & 192/200 (96.0\%) & 0.595 & 0.575 & 196/200 (98.0\%) & 0.0518 & 0.0516 \\
\texttt{mean\_story\_probability} & 0.476 & 0.506 & 189/200 (94.5\%) & 0.501 & 0.517 & 193/200 (96.5\%) & 0.0155 & 0.0530 \\
\texttt{emotion\_entropy} & 0.519 & 0.501 & 194/200 (97.0\%) & 0.693 & 0.662 & 198/200 (99.0\%) & 0.0698 & 0.0496 \\
\texttt{polite\_rate} & 0.517 & 0.513 & 189/200 (94.5\%) & 0.787 & 0.665 & 195/200 (97.5\%) & 0.0360 & 0.0493 \\
\texttt{neutral\_rate} & 0.499 & 0.502 & 188/200 (94.0\%) & 0.693 & 0.613 & 195/200 (97.5\%) & 0.0396 & 0.0522 \\
\texttt{impolite\_rate} & 0.528 & 0.525 & 193/200 (96.5\%) & 0.595 & 0.628 & 196/200 (98.0\%) & 0.0363 & 0.0485 \\
\texttt{hard\_disagree\_rate} & 0.545 & 0.509 & 184/200 (92.0\%) & 0.928 & 0.757 & 196/200 (98.0\%) & 0.0128 & 0.0462 \\
\bottomrule
\end{tabular}%
}
\end{table*}

\subsection{Game}
\label{app:real_vs_real_game}

\begin{table*}[h]
\centering
\scriptsize
\setlength{\tabcolsep}{2.9pt}
\renewcommand{\arraystretch}{1.06}
\caption{Repeated real-vs-real statistics for the game domain. Each row summarizes 200 $n=150$-vs.-$n=150$ comparisons.}
\label{tab:real_vs_real_game}
\resizebox{\textwidth}{!}{%
\begin{tabular}{lrrrrrrrr}
\toprule
Metric & Med. MWU $p$ & Mean MWU $p$ & MWU $p\geq0.05$ & Med. KS $p$ & Mean KS $p$ & KS $p\geq0.05$ & Mean $W_1$ & Mean $|\delta_{\mathrm{Cliff}}|$ \\
\midrule
\texttt{self\_bleu\_4} & 0.473 & 0.497 & 194/200 (97.0\%) & 0.787 & 0.674 & 197/200 (98.5\%) & 0.0037 & 0.0493 \\
\texttt{self\_bertscore\_mean\_f1} & 0.521 & 0.512 & 189/200 (94.5\%) & 0.787 & 0.679 & 193/200 (96.5\%) & 0.0238 & 0.0487 \\
\texttt{semantic\_mean\_cosine} & 0.507 & 0.512 & 188/200 (94.0\%) & 0.787 & 0.685 & 193/200 (96.5\%) & 0.0160 & 0.0489 \\
\texttt{avg\_depth} & 0.543 & 0.526 & 191/200 (95.5\%) & 0.787 & 0.721 & 195/200 (97.5\%) & 0.0719 & 0.0457 \\
\texttt{structural\_virality} & 0.507 & 0.518 & 192/200 (96.0\%) & 0.829 & 0.731 & 197/200 (98.5\%) & 0.1325 & 0.0462 \\
\texttt{length\_cv} & 0.527 & 0.504 & 190/200 (95.0\%) & 0.787 & 0.683 & 195/200 (97.5\%) & 0.0576 & 0.0481 \\
\texttt{mean\_story\_probability} & 0.459 & 0.482 & 191/200 (95.5\%) & 0.501 & 0.533 & 192/200 (96.0\%) & 0.0160 & 0.0541 \\
\texttt{emotion\_entropy} & 0.526 & 0.502 & 191/200 (95.5\%) & 0.787 & 0.690 & 195/200 (97.5\%) & 0.0931 & 0.0491 \\
\texttt{polite\_rate} & 0.537 & 0.521 & 187/200 (93.5\%) & 0.693 & 0.628 & 190/200 (95.0\%) & 0.0429 & 0.0504 \\
\texttt{neutral\_rate} & 0.516 & 0.518 & 191/200 (95.5\%) & 0.693 & 0.621 & 194/200 (97.0\%) & 0.0371 & 0.0492 \\
\texttt{impolite\_rate} & 0.472 & 0.484 & 192/200 (96.0\%) & 0.787 & 0.678 & 197/200 (98.5\%) & 0.0349 & 0.0510 \\
\texttt{hard\_disagree\_rate} & 0.525 & 0.500 & 187/200 (93.5\%) & 0.866 & 0.730 & 196/200 (98.0\%) & 0.0153 & 0.0476 \\
\bottomrule
\end{tabular}%
}
\end{table*}

\newpage

\subsection{Health Issue}
\label{app:real_vs_real_health_issue}

\begin{table*}[h]
\centering
\scriptsize
\setlength{\tabcolsep}{2.9pt}
\renewcommand{\arraystretch}{1.06}
\caption{Repeated real-vs-real statistics for the health issue domain. Each row summarizes 200 $n=150$-vs.-$n=150$ comparisons.}
\label{tab:real_vs_real_health_issue}
\resizebox{\textwidth}{!}{%
\begin{tabular}{lrrrrrrrr}
\toprule
Metric & Med. MWU $p$ & Mean MWU $p$ & MWU $p\geq0.05$ & Med. KS $p$ & Mean KS $p$ & KS $p\geq0.05$ & Mean $W_1$ & Mean $|\delta_{\mathrm{Cliff}}|$ \\
\midrule
\texttt{self\_bleu\_4} & 0.520 & 0.515 & 195/200 (97.5\%) & 0.787 & 0.671 & 196/200 (98.0\%) & 0.0096 & 0.0474 \\
\texttt{self\_bertscore\_mean\_f1} & 0.524 & 0.531 & 193/200 (96.5\%) & 0.787 & 0.697 & 197/200 (98.5\%) & 0.0264 & 0.0456 \\
\texttt{semantic\_mean\_cosine} & 0.544 & 0.521 & 195/200 (97.5\%) & 0.787 & 0.672 & 198/200 (99.0\%) & 0.0239 & 0.0475 \\
\texttt{avg\_depth} & 0.605 & 0.554 & 191/200 (95.5\%) & 0.976 & 0.860 & 198/200 (99.0\%) & 0.0593 & 0.0380 \\
\texttt{structural\_virality} & 0.593 & 0.547 & 190/200 (95.0\%) & 0.994 & 0.884 & 199/200 (99.5\%) & 0.0823 & 0.0379 \\
\texttt{length\_cv} & 0.502 & 0.497 & 190/200 (95.0\%) & 0.693 & 0.647 & 194/200 (97.0\%) & 0.0483 & 0.0501 \\
\texttt{mean\_story\_probability} & 0.575 & 0.542 & 195/200 (97.5\%) & 0.595 & 0.606 & 195/200 (97.5\%) & 0.0251 & 0.0479 \\
\texttt{emotion\_entropy} & 0.569 & 0.523 & 189/200 (94.5\%) & 0.935 & 0.773 & 197/200 (98.5\%) & 0.0701 & 0.0466 \\
\texttt{polite\_rate} & 0.495 & 0.504 & 189/200 (94.5\%) & 0.787 & 0.675 & 195/200 (97.5\%) & 0.0445 & 0.0513 \\
\texttt{neutral\_rate} & 0.473 & 0.485 & 190/200 (95.0\%) & 0.787 & 0.726 & 194/200 (97.0\%) & 0.0390 & 0.0500 \\
\texttt{impolite\_rate} & 0.542 & 0.528 & 192/200 (96.0\%) & 0.935 & 0.796 & 199/200 (99.5\%) & 0.0292 & 0.0418 \\
\texttt{hard\_disagree\_rate} & 0.507 & 0.510 & 194/200 (97.0\%) & 0.987 & 0.902 & 200/200 (100.0\%) & 0.0202 & 0.0359 \\
\bottomrule
\end{tabular}%
}
\end{table*}

\subsection{Movies}
\label{app:real_vs_real_movies}

\begin{table*}[h]
\centering
\scriptsize
\setlength{\tabcolsep}{2.9pt}
\renewcommand{\arraystretch}{1.06}
\caption{Repeated real-vs-real statistics for the movies domain. Each row summarizes 200 $n=150$-vs.-$n=150$ comparisons.}
\label{tab:real_vs_real_movies}
\resizebox{\textwidth}{!}{%
\begin{tabular}{lrrrrrrrr}
\toprule
Metric & Med. MWU $p$ & Mean MWU $p$ & MWU $p\geq0.05$ & Med. KS $p$ & Mean KS $p$ & KS $p\geq0.05$ & Mean $W_1$ & Mean $|\delta_{\mathrm{Cliff}}|$ \\
\midrule
\texttt{self\_bleu\_4} & 0.458 & 0.476 & 192/200 (96.0\%) & 0.595 & 0.544 & 191/200 (95.5\%) & 0.0079 & 0.0559 \\
\texttt{self\_bertscore\_mean\_f1} & 0.526 & 0.508 & 191/200 (95.5\%) & 0.644 & 0.591 & 194/200 (97.0\%) & 0.0246 & 0.0513 \\
\texttt{semantic\_mean\_cosine} & 0.512 & 0.513 & 187/200 (93.5\%) & 0.595 & 0.574 & 191/200 (95.5\%) & 0.0236 & 0.0523 \\
\texttt{avg\_depth} & 0.507 & 0.508 & 189/200 (94.5\%) & 0.693 & 0.633 & 193/200 (96.5\%) & 0.1155 & 0.0508 \\
\texttt{structural\_virality} & 0.515 & 0.520 & 188/200 (94.0\%) & 0.787 & 0.670 & 190/200 (95.0\%) & 0.1675 & 0.0495 \\
\texttt{length\_cv} & 0.531 & 0.519 & 193/200 (96.5\%) & 0.693 & 0.616 & 195/200 (97.5\%) & 0.0523 & 0.0497 \\
\texttt{mean\_story\_probability} & 0.483 & 0.507 & 193/200 (96.5\%) & 0.595 & 0.558 & 198/200 (99.0\%) & 0.0164 & 0.0509 \\
\texttt{emotion\_entropy} & 0.502 & 0.505 & 192/200 (96.0\%) & 0.595 & 0.600 & 189/200 (94.5\%) & 0.0929 & 0.0516 \\
\texttt{polite\_rate} & 0.533 & 0.530 & 196/200 (98.0\%) & 0.693 & 0.632 & 199/200 (99.5\%) & 0.0356 & 0.0475 \\
\texttt{neutral\_rate} & 0.513 & 0.519 & 192/200 (96.0\%) & 0.693 & 0.636 & 193/200 (96.5\%) & 0.0259 & 0.0492 \\
\texttt{impolite\_rate} & 0.463 & 0.490 & 190/200 (95.0\%) & 0.595 & 0.608 & 192/200 (96.0\%) & 0.0367 & 0.0521 \\
\texttt{hard\_disagree\_rate} & 0.486 & 0.489 & 190/200 (95.0\%) & 0.675 & 0.621 & 190/200 (95.0\%) & 0.0254 & 0.0530 \\
\bottomrule
\end{tabular}%
}
\end{table*}

\subsection{News}
\label{app:real_vs_real_news}

\begin{table*}[h]
\centering
\scriptsize
\setlength{\tabcolsep}{2.9pt}
\renewcommand{\arraystretch}{1.06}
\caption{Repeated real-vs-real statistics for the news domain. Each row summarizes 200 $n=150$-vs.-$n=150$ comparisons.}
\label{tab:real_vs_real_news}
\resizebox{\textwidth}{!}{%
\begin{tabular}{lrrrrrrrr}
\toprule
Metric & Med. MWU $p$ & Mean MWU $p$ & MWU $p\geq0.05$ & Med. KS $p$ & Mean KS $p$ & KS $p\geq0.05$ & Mean $W_1$ & Mean $|\delta_{\mathrm{Cliff}}|$ \\
\midrule
\texttt{self\_bleu\_4} & 0.475 & 0.492 & 187/200 (93.5\%) & 0.595 & 0.566 & 191/200 (95.5\%) & 0.0049 & 0.0544 \\
\texttt{self\_bertscore\_mean\_f1} & 0.532 & 0.502 & 191/200 (95.5\%) & 0.595 & 0.565 & 193/200 (96.5\%) & 0.0223 & 0.0524 \\
\texttt{semantic\_mean\_cosine} & 0.464 & 0.496 & 189/200 (94.5\%) & 0.595 & 0.567 & 195/200 (97.5\%) & 0.0168 & 0.0534 \\
\texttt{avg\_depth} & 0.479 & 0.490 & 192/200 (96.0\%) & 0.595 & 0.594 & 195/200 (97.5\%) & 0.0989 & 0.0527 \\
\texttt{structural\_virality} & 0.493 & 0.489 & 193/200 (96.5\%) & 0.693 & 0.619 & 194/200 (97.0\%) & 0.1760 & 0.0529 \\
\texttt{length\_cv} & 0.488 & 0.486 & 193/200 (96.5\%) & 0.501 & 0.546 & 195/200 (97.5\%) & 0.0611 & 0.0540 \\
\texttt{mean\_story\_probability} & 0.541 & 0.519 & 192/200 (96.0\%) & 0.595 & 0.558 & 194/200 (97.0\%) & 0.0137 & 0.0508 \\
\texttt{emotion\_entropy} & 0.522 & 0.521 & 192/200 (96.0\%) & 0.693 & 0.611 & 190/200 (95.0\%) & 0.0868 & 0.0489 \\
\texttt{polite\_rate} & 0.533 & 0.516 & 194/200 (97.0\%) & 0.693 & 0.633 & 193/200 (96.5\%) & 0.0280 & 0.0489 \\
\texttt{neutral\_rate} & 0.486 & 0.510 & 193/200 (96.5\%) & 0.595 & 0.578 & 197/200 (98.5\%) & 0.0382 & 0.0505 \\
\texttt{impolite\_rate} & 0.454 & 0.492 & 196/200 (98.0\%) & 0.595 & 0.569 & 194/200 (97.0\%) & 0.0370 & 0.0510 \\
\texttt{hard\_disagree\_rate} & 0.540 & 0.515 & 196/200 (98.0\%) & 0.691 & 0.615 & 196/200 (98.0\%) & 0.0277 & 0.0489 \\
\bottomrule
\end{tabular}%
}
\end{table*}

\newpage

\subsection{TV Series}
\label{app:real_vs_real_tv_series}

\begin{table*}[h]
\centering
\scriptsize
\setlength{\tabcolsep}{2.9pt}
\renewcommand{\arraystretch}{1.06}
\caption{Repeated real-vs-real statistics for the tv series domain. Each row summarizes 200 $n=150$-vs.-$n=150$ comparisons.}
\label{tab:real_vs_real_tv_series}
\resizebox{\textwidth}{!}{%
\begin{tabular}{lrrrrrrrr}
\toprule
Metric & Med. MWU $p$ & Mean MWU $p$ & MWU $p\geq0.05$ & Med. KS $p$ & Mean KS $p$ & KS $p\geq0.05$ & Mean $W_1$ & Mean $|\delta_{\mathrm{Cliff}}|$ \\
\midrule
\texttt{self\_bleu\_4} & 0.505 & 0.501 & 188/200 (94.0\%) & 0.595 & 0.570 & 192/200 (96.0\%) & 0.0059 & 0.0547 \\
\texttt{self\_bertscore\_mean\_f1} & 0.458 & 0.480 & 190/200 (95.0\%) & 0.501 & 0.531 & 192/200 (96.0\%) & 0.0225 & 0.0552 \\
\texttt{semantic\_mean\_cosine} & 0.479 & 0.487 & 186/200 (93.0\%) & 0.595 & 0.569 & 188/200 (94.0\%) & 0.0183 & 0.0561 \\
\texttt{avg\_depth} & 0.508 & 0.517 & 191/200 (95.5\%) & 0.693 & 0.632 & 196/200 (98.0\%) & 0.1561 & 0.0503 \\
\texttt{structural\_virality} & 0.555 & 0.518 & 189/200 (94.5\%) & 0.787 & 0.652 & 193/200 (96.5\%) & 0.2037 & 0.0495 \\
\texttt{length\_cv} & 0.566 & 0.528 & 187/200 (93.5\%) & 0.595 & 0.588 & 194/200 (97.0\%) & 0.0600 & 0.0505 \\
\texttt{mean\_story\_probability} & 0.503 & 0.496 & 183/200 (91.5\%) & 0.595 & 0.536 & 188/200 (94.0\%) & 0.0241 & 0.0553 \\
\texttt{emotion\_entropy} & 0.507 & 0.521 & 194/200 (97.0\%) & 0.693 & 0.599 & 195/200 (97.5\%) & 0.0963 & 0.0505 \\
\texttt{polite\_rate} & 0.513 & 0.502 & 194/200 (97.0\%) & 0.693 & 0.619 & 196/200 (98.0\%) & 0.0343 & 0.0502 \\
\texttt{neutral\_rate} & 0.563 & 0.532 & 182/200 (91.0\%) & 0.693 & 0.635 & 188/200 (94.0\%) & 0.0288 & 0.0510 \\
\texttt{impolite\_rate} & 0.501 & 0.498 & 190/200 (95.0\%) & 0.595 & 0.574 & 196/200 (98.0\%) & 0.0393 & 0.0527 \\
\texttt{hard\_disagree\_rate} & 0.497 & 0.504 & 189/200 (94.5\%) & 0.702 & 0.639 & 191/200 (95.5\%) & 0.0291 & 0.0525 \\
\bottomrule
\end{tabular}%
}
\end{table*}

\newpage
\section{Proof of the Statistical Guarantee}
\label{app:statistical_guarantee}

We prove Theorem~\ref{thm:statistical_guarantee} for an
arbitrary fixed metric $f\in\mathcal F$.
For simplicity, we omit the subscript $f$ and write
$F_R=F_f^R$, $F_G=F_f^G$, $n=n_R$, $m=n_G$, and $N=n+m$.

Under A1, $X_1,\ldots,X_n$ and $Y_1,\ldots,Y_m$
are mutually independent samples from $F_R$ and
$F_G$, respectively. A2 requires
$n/N\to\rho\in(0,1)$.

Define
\[
\theta=\Pr(X<Y)+\tfrac12\Pr(X=Y),
\qquad
D=\sup_x|F_R(x)-F_G(x)|,
\]
where $X\sim F_R$ and $Y\sim F_G$ are independent. We use tie-corrected normal calibration for the
two-sided MWU test, with an optional continuity
correction, and exact or asymptotic continuous-null
calibration for the two-sample KS test.

\subsection{Auxiliary Results}

\begin{lemma}[MWU validity and rank consistency]
\label{lem:mwu}
Let
\[
h(x,y)=\mathbf 1\{x<y\}+\tfrac12\mathbf 1\{x=y\},\qquad
\widehat\theta=\frac{U}{nm}
=\frac1{nm}\sum_{i=1}^n\sum_{j=1}^m h(X_i,Y_j).
\]
Under A1--A2, $\widehat\theta\to\theta$ almost surely. Under
$H_0:F_R=F_G$, the tie-corrected two-sided MWU test has
asymptotic Type-I error at most $\alpha$. If
$\theta\neq\tfrac12$, its rejection probability tends to one.
\end{lemma}

\begin{proof}
The kernel $h$ is bounded. The strong law for two-sample
U-statistics therefore gives $\widehat\theta\to\theta$ almost
surely. Under $H_0$, exchangeability of independent $X,Y$ gives
$\Pr(X<Y)=\Pr(Y<X)$; together with the tie probability, this
yields $\theta=\tfrac12$.

For null calibration, denote the common CDF by $H$ and its
mid-distribution function by
\[
H_{\mathrm{mid}}(z)=\tfrac12\{H(z-)+H(z)\},\qquad
\sigma_H^2=\operatorname{Var}\{H_{\mathrm{mid}}(Z)\},
\quad Z\sim H.
\]
If $H$ is not a point mass, $\sigma_H^2>0$. The first-order
Hoeffding decomposition is
\[
\widehat\theta-\tfrac12
=-\frac1n\sum_{i=1}^n
 \bigl\{H_{\mathrm{mid}}(X_i)-\tfrac12\bigr\}
 +\frac1m\sum_{j=1}^m
 \bigl\{H_{\mathrm{mid}}(Y_j)-\tfrac12\bigr\}
 +O_p\bigl((nm)^{-1/2}\bigr).
\]
The summands are bounded, centered, and have variance
$\sigma_H^2$. By the independent-sample CLT and A2,
\[
\frac{\widehat\theta-1/2}
{\sigma_H\sqrt{1/n+1/m}}
\;\Rightarrow\;\mathcal N(0,1).
\]

It remains to justify the variance used by the test when ties
occur. Let $t_1,\ldots,t_J$ be the sizes of the distinct tie
groups in the pooled sample. The usual midrank variance is
\[
V_{n,m}=\frac{nm}{12}
\left[N+1-\frac{\sum_{k=1}^J(t_k^3-t_k)}{N(N-1)}\right].
\]
Writing $R_1,\ldots,R_N$ for the pooled midranks, the
midrank-sum identity gives
\[
V_{n,m}=\frac{nm}{N(N-1)}
\sum_{\ell=1}^N\left(R_\ell-\frac{N+1}{2}\right)^2.
\]
Under $H_0$, the pooled observations are i.i.d. from $H$.
Uniform convergence of their empirical CDF and the strong law
for the bounded squared mid-distribution values imply
\[
\frac{V_{n,m}}{nmN}\xrightarrow{\mathrm{a.s.}}\sigma_H^2.
\]
Consequently, the ratio of $V_{n,m}/(nm)^2$ to
$\sigma_H^2(1/n+1/m)$ converges to one, and Slutsky's theorem yields
\[
\frac{U-nm/2}{\sqrt{V_{n,m}}}
\;\Rightarrow\;\mathcal N(0,1).
\]
The optional half-unit continuity correction is asymptotically
negligible. Hence the two-sided normal-calibrated $p$-value is
asymptotically uniform and its rejection probability tends to
$\alpha$. If $H$ is a point mass, all pooled observations tie,
$U=nm/2$ and $V_{n,m}=0$; using the non-rejection convention
$p^{\mathrm{MWU}}=1$ gives Type-I error zero. Thus the null
Type-I error is asymptotically at most $\alpha$ for any $H$.

For consistency, if $\theta\neq\tfrac12$, the strong law gives
$|\widehat\theta-1/2|\to|\theta-1/2|>0$. The tie correction
satisfies $V_{n,m}\le nm(N+1)/12$, so the critical deviation for
$\widehat\theta$ on the normal scale is at most
$O(\sqrt{N/(nm)})=O(N^{-1/2})$ by A2. Therefore the standardized
statistic diverges in absolute value and the two-sided rejection
probability tends to one. Replacing $U$ by its complementary
orientation $nm-U$ does not change the two-sided $p$-value.
\end{proof}

\begin{lemma}[KS validity and distribution consistency]
\label{lem:ks}
Let
\[
\widehat D_{n,m}
=\sup_x|\widehat F_{R,n}(x)-\widehat F_{G,m}(x)|.
\]
Under A1--A2, $\widehat D_{n,m}\to D$ almost surely, and
$D=0$ if and only if $F_R=F_G$. Under the null, calibration
against the continuous-null KS distribution has asymptotic
Type-I error at most $\alpha$, even when the common CDF has
jumps. If $D>0$, the rejection probability tends to one.
\end{lemma}

\begin{proof}
The reverse triangle inequality gives
\[
|\widehat D_{n,m}-D|
\le\sup_x|\widehat F_{R,n}(x)-F_R(x)|
 +\sup_x|\widehat F_{G,m}(x)-F_G(x)|.
\]
Both terms vanish almost surely by Glivenko--Cantelli. Thus
$\widehat D_{n,m}\to D$ almost surely; by the definition of
$D$, $D=0$ is equivalent to $F_R=F_G$.

For null calibration, write $F_R=F_G=H$, without assuming that
$H$ is continuous. Couple both samples using independent
uniform random variables $A_1,\ldots,A_n$ and
$B_1,\ldots,B_m$ and the generalized inverse
$Q_H(u)=\inf\{x:H(x)\ge u\}$. With
$X_i=Q_H(A_i)$ and $Y_j=Q_H(B_j)$, their empirical CDFs obey
\[
\widehat F_{R,n}(x)=\widehat F_{A,n}\{H(x)\},\qquad
\widehat F_{G,m}(x)=\widehat F_{B,m}\{H(x)\}
\quad\text{almost surely}.
\]
Therefore, on this coupling,
\[
\widehat D_{n,m}
=\sup_x|\widehat F_{A,n}\{H(x)\}
           -\widehat F_{B,m}\{H(x)\}|
\le\sup_{0\le u\le1}
     |\widehat F_{A,n}(u)-\widehat F_{B,m}(u)|
=:D^{\mathrm{unif}}_{n,m}.
\]
Thus the statistic for any common $H$ is stochastically bounded
by the statistic under the continuous uniform null. An exact
continuous-null KS $p$-value is consequently conservative for
arbitrary $H$. The asymptotic calibration has the same limiting
bound because
\[
\sqrt{\frac{nm}{N}}\,D^{\mathrm{unif}}_{n,m}
\Rightarrow \sup_{0\le u\le1}|\mathbb B(u)|,
\]
whose CDF is continuous. In particular,
$\limsup_{N\to\infty}\Pr_{H_0}(p^{\mathrm{KS}}<\alpha)
\le\alpha$.

Finally, if $D>0$, the strong convergence above gives
$\widehat D_{n,m}\to D>0$, whereas the continuous-null KS
critical value is $O(\sqrt{N/(nm)})=O(N^{-1/2})$ under A2.
Consequently, the rejection probability tends to one.
\end{proof}

\subsection{Proof of Theorem~\ref{thm:statistical_guarantee}}

\begin{proof}
\textbf{Validity.}
Under $H_{0,f}$, Lemmas~\ref{lem:mwu}
and~\ref{lem:ks} establish that
\[
\limsup_{N\to\infty}
\Pr(p_f^{(q)}<\alpha)\leq\alpha,
\qquad q\in\{\mathrm{MWU},\mathrm{KS}\}.
\]

Since a metric fails when either test rejects,
the union bound gives
\[
\begin{aligned}
\Pr(\mathsf{Pass}_f^c)
&=
\Pr\bigl(
p_f^{\mathrm{MWU}}<\alpha
\;\text{or}\;
p_f^{\mathrm{KS}}<\alpha
\bigr)\\
&\leq
\Pr(p_f^{\mathrm{MWU}}<\alpha)
+
\Pr(p_f^{\mathrm{KS}}<\alpha).
\end{aligned}
\]

Taking the limit superior, we obtain
\[
\limsup_{N\to\infty}
\Pr(\mathsf{Pass}_f^c)
\leq 2\alpha.
\]
Equivalently,
\[
\liminf_{N\to\infty}
\Pr(\mathsf{Pass}_f)
\geq 1-2\alpha.
\]

\textbf{Consistency.}
If $F_f^R\neq F_f^G$, then
\[
D_f=\sup_x|F_f^R(x)-F_f^G(x)|>0.
\]
By Lemma~\ref{lem:ks}, the KS test is consistent,
so
\[
\Pr(p_f^{\mathrm{KS}}<\alpha)
\longrightarrow 1.
\]

Since ThreadBench requires both tests to pass,
\[
\Pr(\mathsf{Pass}_f)
\leq
\Pr(p_f^{\mathrm{KS}}\geq\alpha)
\longrightarrow 0.
\]
This completes the proof.
\end{proof}



\newpage
\section{ThreadPDG Implementation Details}
\label{app:miropdg}

\subsection{Data Isolation and Sampling}
\label{app:miropdg_sampling}

For each domain $d$, the eligible source threads are assigned to three disjoint sets by source-thread identifier: the seed pool $S_d$, the reference pool $C_d$, and the held-out test pool $T_d$. We construct every domain profile and all reference-derived conditioning data exclusively from $C_d$. This includes measured writing patterns, reference comment examples, metric-based thread templates, and the reference metric distributions used by the composer. No thread in $T_d$ enters these objects, the prompts, draft selection, or budget selection.

We treat the development pools and their derived configurations as fixed before the final evaluation. For each domain we draw 150 output indices; each independently samples a seed from $S_d$ with replacement and a reference target thread from $C_d$ with replacement. Generation uses independent random draws across output indices; it shares only the fixed model configuration, domain profile, and reference statistics. We independently draw 150 real evaluation observations with replacement from $T_d$ after generation choices are fixed and use the same drawn reference across methods and draft budgets. This sampling protocol defines the evaluation distributions conditional on the fixed data partitions and configurations; a finite sampled corpus does not by itself establish population-level sampling assumptions.

\subsection{Thread Shape}
\label{app:shape}

For a sampled seed thread, we retain its root post and construct a template with one slot per comment, the original parent--reply links, and a coarse word-count bucket for each slot. Each draft fills the same slots and therefore inherits the seed's comment count and reply-tree topology. Sampling seed contexts with replacement does not force the output cohort to reproduce the fixed seed pool's size histogram exactly.

\subsection{Planner Conditioning}
\label{app:planner_writer}

The branch Planner receives the seed discussion, its structural template, and a domain profile computed from $C_d$. It prepares a plan for each top-level branch. The comment Planner then provides one plan per slot, specifying \texttt{semantic\_move}, \texttt{comment\_function}, \texttt{content\_angle}, \texttt{detail\_focus}, \texttt{claim\_key}, and \texttt{tone\_target}. Comment plans are produced in batches of eight.

The branch Planner and the comment Planner both read the seed discussion together with its structural template and the domain profile computed from $C_d$. These are development-side inputs drawn from $S_d$; the held-out pool $T_d$ is never read at any stage of generation.

\subsection{Writer and Output Guards}
\label{app:writer}

The Writer fills comment slots using their plans, the preceding generated discussion context, and assigned personas. We check each output for excessive reuse of Planner wording, lack of concrete discussion detail, and generic long-form filler. A rejected comment is regenerated with a reduced prompt; an entire draft can be retried if it continues to fail the checks. These checks operate within a draft and do not compare it with other generated threads or with $T_d$. Each completed draft is scored before composition.

\subsection{Persona Conditioning}
\label{app:persona}

Personas are drawn from MatrAIx~\citep{li2026matraix} and assigned to comment slots. The assignment for a sampled seed is fixed across its drafts so that selecting a branch does not change the persona attached to a slot. Persona conditioning applies to the Writer.

\subsection{Composition Targets and Search}
\label{app:composition}

Before generation, we compute each metric's reference interquartile range $\operatorname{IQR}_m$ from $C_d$ and fix the metric weights $w_m$. For output index $i$, we sample one reference thread $c_i$ independently from $C_d$ with replacement. Its metric vector supplies the targets
\[
t_{im}=f_m(c_i).
\]
Using a single reference thread retains the observed within-thread relationships among target metrics. The draw is separate for each output index, and the target does not depend on the ranks, scores, or branch assignments of any other generated thread.

Let $B_i$ be the number of top-level branches in the seed template. An assignment $\mathbf z_i=(z_{i1},\ldots,z_{iB_i})$ has $z_{ib}\in\{1,\ldots,K\}$, indicating the draft supplying branch $b$. We evaluate its objective $J_i$ as defined in Section~\ref{sec:multi_draft}. The eligible metric set $\mathcal M_i$ includes metrics with a defined target and a strictly positive reference scale. Missing targets and zero scales are excluded according to a rule fixed before the final evaluation; if no metric is eligible, the first complete draft is retained without further selection.

We minimize $J_i$ by multi-start coordinate descent, initializing from the $K$ whole-draft assignments. Each sweep tries replacing a branch with the corresponding branch from another draft and retains a replacement only when it reduces the objective. Search stops when no single-branch replacement improves the current assignment. It uses only the current seed's drafts, target vector, and frozen reference statistics. Neither cohort-wide rankings nor held-out evaluation scores are part of this search. The main comparison fixes $K=3$ in advance; $K=1$ and $K=2$ are pre-specified draft-budget conditions rather than stopping outcomes selected using test results. The budget study reuses the same sampled seeds, reference targets, and common first $K$ draft realizations, changing only the number of available drafts.




\subsection{Why Branch-Level Composition}
\label{app:granularity}

Selecting one whole draft offers only $K$ alternatives for a discussion. Selecting each comment independently can break the connection between a reply and the text it answers. We instead take a top-level comment together with its entire reply subtree from one draft, preserving its internal parent--reply context. All drafts share the same template, so selected branches occupy the same top-level slots. Any fallback for an incomplete branch must be handled within the same draft or recorded as a separate deviation; the protocol does not silently mix replies from different drafts. Comparisons with other composition granularities require results obtained under the same revised data protocol.

\newpage
\section{Limitations and Future Work}
\label{app:limitations}

ThreadPDG is intended as a case study of how ThreadBench can guide the design of more realistic discussion simulators, rather than as a general solution to discussion simulation. Motivated by the gaps identified by ThreadBench, ThreadPDG learns discussion patterns from real threads and uses them to generate new discussions. This dependence on real examples is therefore part of the current case study design. One direction for future work is to develop simulators that can generate realistic threads independently at inference time, without requiring a corresponding real discussion example.

\newpage
\section{Prompt Templates}
\label{app:prompts}

ThreadPDG makes three kinds of LLM calls: a thread planner that divides the
thread into branches, a comment planner that writes one plan per comment slot,
and a Writer that turns one plan into one comment. Below are abridged versions
of the three prompts. We keep each system prompt, the order of the inputs, a few
of the rules in their original wording, and the output format. We drop the
domain vocabularies, the quotas and the repair messages. Text in
$\langle\cdot\rangle$ is filled in at run time. Only the two planners see the
real seed thread; the Writer never does.

\newcommand{\pf}[1]{{\rmfamily$\langle$\textit{#1}$\rangle$}}
\newcommand{\pl}[1]{{\bfseries #1}}
\newcommand{\promptbox}[3]{%
  \par\noindent
  \begin{tabular}{@{}>{\raggedright\arraybackslash}p{\linewidth}@{}}
  \toprule
  \rowcolor{black!6}[0pt][0pt]\rule{0pt}{2.3ex}\,\textbf{#1}\hfill{\footnotesize #2}\,\\
  \midrule
  \hspace{0.35em}\begin{minipage}[t]{\dimexpr\linewidth-0.7em\relax}\vspace{0pt}
  \scriptsize\sffamily\raggedright\setlength{\parskip}{2pt}#3\par\vspace{2pt}
  \end{minipage}\\
  \bottomrule
  \end{tabular}\par\vspace{8pt}}

\promptbox{Thread planner}{one call per thread}{%
\pl{System.} You are a sampling planner for realistic Reddit discussions.
Return strict JSON only. Do not write comments.

\pl{User.} Plan the semantic axes of one Reddit discussion in
\pf{community}. The planner produces private controls. It does not write
comments.\\
Objectives:\\
- Preserve the matched real thread's comment-count scale, reply-tree shape,
local branch behavior, length variation, and mix of social roles.\\
- Ground every branch in the seed post or in a parent-local issue.\\
- Allow realistic low-information turns, questions, corrections, jokes,
acknowledgements, side tangents, and personal datapoints.

Domain: \pf{name, topic facets, perspective lenses from the profile $P_d$}\\
Reference comments: \pf{comments from the reference pool $C_d$}\\
Seed post: \pf{title and body}\\
Thread target: \pf{comments, top-level comments, maximum depth, length mix}\\
Matched real thread: \pf{one line per slot: depth, parent, words, text}\\
Return at least \pf{B} genuinely distinct semantic branches.

\pl{Returns.} \{"branches": [\{branch\_id, anchor\_quote, branch\_goal,
perspective\_id, owned\_decision\_subject, decision\_boundary,
branch\_exclusion\}]\}}

\promptbox{Comment planner}{one call per batch of eight slots}{%
\pl{System.} You extract abstract semantic writing plans from real Reddit
comments. Return strict JSON only. Do not write final comments.

\pl{User.} Assign per-comment semantic and social controls to matched-real
structural slots.\\
Objectives:\\
- Produce a reusable instruction, not a comment.\\
- Preserve weak comments as weak comments. Do not upgrade reactions,
fragments, jokes, or questions into advice.

The post: \pf{title and body}\\
Branches: \pf{output of the thread planner}\\
Plans so far: \pf{plans for the earlier slots of this thread}\\
Slots: \pf{for each of the next eight slots: depth, parent, words, text}\\
Equivalent local claims should share claim\_key; different claims must not.
The slot text is displayed to you and withheld from the Writer.

\pl{Returns.} \{"comment\_plans": [\{sample\_id, branch\_id,
comment\_function, semantic\_move, content\_angle, detail\_focus, stance,
tone\_class, reply\_relation, claim\_key, avoid\_repeating, \ldots\}]\}}

\promptbox{Writer}{one call per comment}{%
\pl{System.} \pf{persona of the speaker, from MatrAIx} You are writing one
human Reddit comment in \pf{community}. Follow the sampled role, length, tone,
and local context exactly, but do not sound like an assistant, customer
support, or a polished explainer. Real Reddit comments can be partial,
thankful, uncertain, neutral, mildly skeptical, messy, or low-information.
\ldots{} Do not make comments unrelated just to be diverse. \ldots{} Return
only the comment.

\pl{User.} Write exactly one Reddit comment in \pf{community}. Output only the
comment body.\\
What this comment says (highest priority): The point this comment makes, in
your own words: \pf{semantic\_move}\\
What kind of turn this is: \pf{comment\_function, content\_angle,
detail\_focus, stance, reply\_relation}\\
Visible discussion: \pf{the parent comment, or the post for a top-level
comment}\\
How to write it: \pf{target length, tone, opening style}\\
Already used in this thread, so do not repeat them: \pf{openings and points
of the comments written so far}\\
Rules: Aim criticism at the product, claim, or process, never at a person.
\ldots

\pl{Returns.} The comment text only.}

\newpage
\section{Human Turing Test}
\label{app:turing}

\paragraph{Pairs.}
The test has 240 pairs: eight domains, three systems, and ten pairs per system in each domain. A
pair shows a real Reddit thread and a thread generated by one system for the same seed post, and no
seed post appears in more than one pair. Usernames are replaced by User 1, User 2 and so on
within each thread, and links and images are removed from both threads. The real thread is shown
on the left in 123 pairs and on the right in 117 (ThreadPDG 40/40, SynthPAI 39/41, OASIS 44/36).

\paragraph{Survey.}
We ran the survey on Qualtrics through an anonymous link. Each rater reads eight pairs drawn at
random from the 240, with Qualtrics balancing how often each pair is shown, and answers Thread
A, Thread B or Cannot distinguish for each pair. Raters are told that one thread in each
pair is real and the other was written by a computer, that there is no time limit, and that
cannot distinguish is also a valid answer.

\paragraph{Participants and filtering.}
Qualtrics records only completed responses, and 542 people completed the survey. We remove invalid
responses: responses with a reCAPTCHA score below 0.5, raters who gave the same answer to all eight
pairs, and repeated responses from the same IP address (we keep the first). We then remove raters
who finished in under two minutes. Table~\ref{tab:turing-filter} gives the counts. The 436
raters we keep made 3{,}484 judgments, and every pair was judged between 7 and 22 times
(median 15).

\begin{table}[h]
\centering
\small
\caption{Rater filtering.}
\label{tab:turing-filter}
\begin{tabular}{lc}
\toprule
 & Raters \\
\midrule
Completed responses                     & 542 \\
\quad reCAPTCHA score $<0.5$            & $-6$ \\
Valid responses                         & 515 \\
\quad Completed in under two minutes    & $-79$ \\
Raters kept                             & 436 \\
\bottomrule
\end{tabular}
\end{table}

\paragraph{Majority vote.}
A pair is \emph{not distinguished} when the raters who chose the generated thread or cannot
distinguish outnumber those who chose the real thread, and \emph{distinguished} in the reverse
case; tied pairs are left out.



\newpage
\section{Example Discussions}
\label{app:examples}

\newcommand{\excomment}[3]{\par{\leftskip=#1\relax\noindent#2~#3\par}\vspace{1.4pt}}
\providecommand{\mrk}[1]{\textcolor{black!45}{#1}}
\providecommand{\mkreuse}[1]{{\sethlcolor{mireuse}\hl{#1}}}
\providecommand{\mkpara}[1]{{\sethlcolor{miepara}\hl{#1}}}
\providecommand{\mkstory}[1]{{\sethlcolor{miestory}\hl{#1}}}
\providecommand{\mktone}[1]{{\sethlcolor{mietone}\hl{#1}}}
\newcommand{\exwc}[1]{\mrk{\mbox{\scriptsize(#1)}}}
\newcommand{\exre}{$\hookrightarrow$\,}
\newcommand{\excap}[2]{{\setlength{\fboxsep}{0.4pt}\colorbox{#1}{\vphantom{Ag}#2}}}

Figure~\ref{fig:app_examples} shows one post from the Game domain,
\emph{Is a monitor better than a TV?}, with its real discussion and the thread
each method generated for it with \texttt{gpt-5.4-mini}. Real users answer in
many ways: a short question back to the poster, technical answers, a joke about
the jargon, and a budget story from someone who games on a TV. OASIS writes 14
comments, all at the top level and most around 70 words long. All 14 mention a
desk, three open with the same sentence, and five say they are late to the
debate, so the mean similarity between two of its comments is 0.84, against
0.34 for the real thread. SynthPAI goes the other way: 92\% of its 80 comments
are replies, its median comment has 8.5 words, 31 comments open with
\emph{yep} or \emph{yeah}, and seven repeat the verdict \emph{monitor wins}.
ThreadPDG follows the real thread's shape and mix of lengths, because it inherits
the reply tree, but not every reply follows from its parent: \emph{Thanks, that
clears it up!} answers a question.

\begin{figure}[H]
\centering
\scriptsize
\setlength{\tabcolsep}{0pt}
\begin{tabular}{@{}p{0.48\textwidth}@{\hspace{0.8em}}|@{\hspace{0.8em}}p{0.48\textwidth}@{}}
\toprule
\multicolumn{2}{@{}l@{}}{\footnotesize\textbf{Post} (r/gaming):
  \emph{Is a monitor better than a TV?}\ For gaming, duh} \\
\midrule
\noindent\hfil\textbf{Real}\hfil & \noindent\hfil\textbf{ThreadPDG}\hfil \\
\addlinespace[2pt]
\begin{minipage}[t]{0.48\textwidth}\raggedright
\excomment{0em}{\textbf{User~1:}}{If you want low latency and high refresh rate, probably. \exwc{10 words}}
\excomment{1.1em}{\exre\textbf{User~2:}}{Don't make me google those words, please \exwc{7 words}}
\excomment{2.2em}{\exre\textbf{User~3:}}{For real? \exwc{2 words}}
\excomment{0em}{\textbf{User~4:}}{Console or PC? \exwc{3 words}}
\excomment{0em}{\textbf{User~5:}}{Imo a good quality monitor will always be better than a TV. But if you're not willing to spend the money, stick to the TV. Does the job just fine with most games/graphics. \exwc{33 words}}
\excomment{1.1em}{\exre\textbf{User~6:}}{This. \mkstory{I'm using a 4k tv for my gaming PC because I only had 200 bucks to spend at the time} and a good monitor was twice that. \exwc{28 words}}
\excomment{0em}{\textbf{User~7:}}{In every technical way, yes. In terms of comfort and where you're playing, it's up for debate. A 24" monitor is going to suck on your Playstation in the living room. \exwc{31 words}}
\end{minipage} &
\begin{minipage}[t]{0.48\textwidth}\raggedright
\excomment{0em}{\textbf{User~1:}}{Monitor or TV, taste matters \exwc{5 words}}
\excomment{0em}{\textbf{User~2:}}{Low latency and high refresh make the TV call pretty solid for gaming. \exwc{13 words}}
\excomment{1.1em}{\exre\textbf{User~3:}}{What jargon exactly? \exwc{3 words}}
\excomment{2.2em}{\exre\textbf{User~4:}}{\mkstory{I've got the monitor in the desk setup}, the TV only if it's just okay \exwc{15 words}}
\excomment{2.2em}{\exre\textbf{User~5:}}{\textgreater{} What jargon exactly? Refresh rate is how many times the image updates each second, and latency is the delay between your input and the screen reacting. \ldots{} \exwc{48 words}}
\excomment{0em}{\textbf{User~6:}}{Platform split first? huh! \exwc{4 words}}
\excomment{1.1em}{\exre\textbf{User~7:}}{Thanks, that clears it up! \exwc{5 words}}
\end{minipage} \\
\addlinespace[2pt]
\mrk{31 comments $\cdot$ median 14 words $\cdot$ 35\% replies $\cdot$ similarity 0.34} &
\mrk{37 comments $\cdot$ median 13 words $\cdot$ 41\% replies $\cdot$ similarity 0.32} \\
\midrule
\noindent\hfil\textbf{OASIS}\hfil & \noindent\hfil\textbf{SynthPAI}\hfil \\
\addlinespace[2pt]
\begin{minipage}[t]{0.48\textwidth}\raggedright
\excomment{0em}{\textbf{User~1:}}{\mkreuse{Kinda depends on the setup, honestly.} \mkpara{For pure desk gaming a monitor usually wins} on latency and convenience, but I've definitely used a TV when I just wanted to kick back from the couch \ldots{} \exwc{73 words}}
\excomment{0em}{\textbf{User~2:}}{Maybe I missed the moment, but the monitor vs TV war always felt way bigger online than in actual use. \mkpara{For a desk setup, sure, a monitor is usually the no-drama answer.} \ldots{} \exwc{58 words}}
\excomment{0em}{\textbf{User~3:}}{\mkreuse{Kinda depends on the setup, honestly.} \mkpara{For desk gaming a monitor usually makes more sense}, but I've also spent plenty of evenings on a TV because I wanted to sit back \ldots{} \exwc{66 words}}
\excomment{0em}{\textbf{User~4:}}{\mkreuse{Kinda depends on the setup, honestly.} \mkpara{For desk gaming I still lean monitor}, mostly because it's easier on the eyes and you're not fighting input lag. \ldots{} \exwc{70 words}}
\excomment{0em}{\textbf{User~5:}}{Yeah, ``better'' is doing a lot of work there. \mkpara{For desk gaming a monitor is usually the obvious pick}: lower latency, proper refresh rates, less input lag nonsense. \ldots{} \exwc{68 words}}
\end{minipage} &
\begin{minipage}[t]{0.48\textwidth}\raggedright
\excomment{0em}{\textbf{User~1:}}{depends on size and distance, \mkreuse{mate}. \exwc{6 words}}
\excomment{1.1em}{\exre\textbf{User~2:}}{nah, for most desk setups \mkreuse{monitor wins} hard. tv is fine if you sit far back, but on a small table the text looks worse and input lag can be annoying. \ldots{} \exwc{46 words}}
\excomment{2.2em}{\exre\textbf{User~3:}}{cheap monitor \textgreater{} tv tbh, \mkreuse{mate} \exwc{6 words}}
\excomment{3.3em}{\exre\textbf{User~1:}}{\mktone{yeah,} plus desk glare is such a pain \mkreuse{mate} \exwc{9 words}}
\excomment{2.2em}{\exre\textbf{User~4:}}{\mktone{yep,} desk life = \mkreuse{monitor wins} \exwc{6 words}}
\excomment{3.3em}{\exre\textbf{User~5:}}{desk setup, monitor every time lol \exwc{6 words}}
\excomment{3.3em}{\exre\textbf{User~6:}}{Desk gang, TV gets lazy. \exwc{5 words}}
\end{minipage} \\
\addlinespace[2pt]
\mrk{14 comments $\cdot$ median 69 words $\cdot$ 0\% replies $\cdot$ similarity 0.84} &
\mrk{80 comments $\cdot$ median 8.5 words $\cdot$ 92\% replies $\cdot$ similarity 0.37} \\
\bottomrule
\end{tabular}
\caption{One post and four discussions, from the Game domain on
\texttt{gpt-5.4-mini}. Each panel shows five to seven comments of one thread in
thread order. Indentation marks a reply, \ldots{} marks a cut, and the count
gives the full length. Highlights mark \excap{mireuse}{reused wording},
\excap{miepara}{paraphrases}, \excap{miestory}{personal experience} and
\excap{mietone}{agreement}. The
line under each panel describes the whole thread; similarity is the mean cosine
similarity between two of its comments, as in Figure~\ref{fig:qual_semantic}.}
\label{fig:app_examples}
\end{figure}

\newpage
\section{Qualitative Measurements per Domain and LLM}
\label{app:qualitative}

Section~\ref{sec:observations} shows one domain on \texttt{gpt-5.4-mini}. This
appendix gives every domain and every LLM. All measurements use the held-out
real threads, ThreadPDG at $K{=}3$, and threads with at least five comments.
The qualitative panels below use \texttt{gpt-5.4-mini}; the quantitative results
for every backbone are in Table~\ref{tab:holdout_pvalues}. Table~\ref{tab:fingerprint} summarizes the surface
properties on \texttt{gpt-5.4-mini}.

\begin{table}[H]
\centering
\caption{%
  Surface properties of a generated discussion, over the same held-out threads
  Table~\ref{tab:holdout_pvalues} tests, on GPT-5.4-mini. Every column is a rule
  a reader can check by opening a thread: \textsc{words} is the median comment
  length, \textsc{len.\,CV} the variation in comment length within a thread,
  \textsc{reply} the share of comments answering a comment rather than the post,
  \textsc{frag.}\ comments of at most twelve words with no terminal punctuation,
  \textsc{openers} the entropy in bits of a comment's first two words,
  \textsc{spread} the mean cosine distance between two comments of a thread and
  \textsc{on-topic} the mean similarity between a comment and its post. Real is
  the reference row and the closest simulator in each column is bold.}
\label{tab:fingerprint}
\resizebox{\textwidth}{!}{\begin{tabular}{lcccccccccccccc}
\toprule
 & \multicolumn{2}{c}{Length} & \multicolumn{2}{c}{Structure} & \multicolumn{4}{c}{Register} & \multicolumn{2}{c}{Stance} & \multicolumn{2}{c}{Variety} & \multicolumn{2}{c}{Semantics} \\
\cmidrule(lr){2-3}\cmidrule(lr){4-5}\cmidrule(lr){6-9}\cmidrule(lr){10-11}\cmidrule(lr){12-13}\cmidrule(lr){14-15}
 & words & len.\,CV & reply & depth & frag. & contr. & quest. & link & agree & hedge & openers & top 5 & spread & on-topic \\
\midrule
\textit{Real} & 19 & 0.95 & 46\% & 0.92 & 17\% & 47\% & 16\% & 2.8\% & 6\% & 5\% & 12.4 & 4\% & 0.776 & 0.271 \\
\midrule
ThreadPDG & \textbf{19} & \textbf{0.97} & \textbf{47\%} & \textbf{0.95} & 21\% & \textbf{48\%} & \textbf{15\%} & \textbf{2.4\%} & \textbf{6\%} & \textbf{5\%} & 13.0 & \textbf{4\%} & \textbf{0.777} & \textbf{0.277} \\
OASIS & 56 & 0.25 & 0\% & 0.00 & 0\% & 81\% & 13\% & 0.0\% & 35\% & 10\% & 8.6 & 24\% & 0.350 & 0.454 \\
SynthPAI & 7 & 0.93 & 85\% & 1.72 & \textbf{15\%} & 25\% & 1\% & 0.0\% & 30\% & 4\% & \textbf{12.5} & 9\% & 0.724 & 0.221 \\
\bottomrule
\end{tabular}}
\end{table}

\begin{figure}[H]
\scriptsize
\centering
\includegraphics[width=0.46\textwidth]{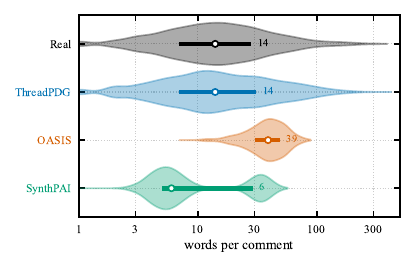}
\caption{Comment length on Celebrity by \texttt{gpt-5.4-mini}. Violins show the distribution on a logarithmic scale, the bar shows the interquartile
range, and the dot is the median. }
\label{fig:qual_length}
\vspace{-0.4\intextsep}
\end{figure}

\begin{figure}[H]
\centering
\includegraphics[width=\textwidth]{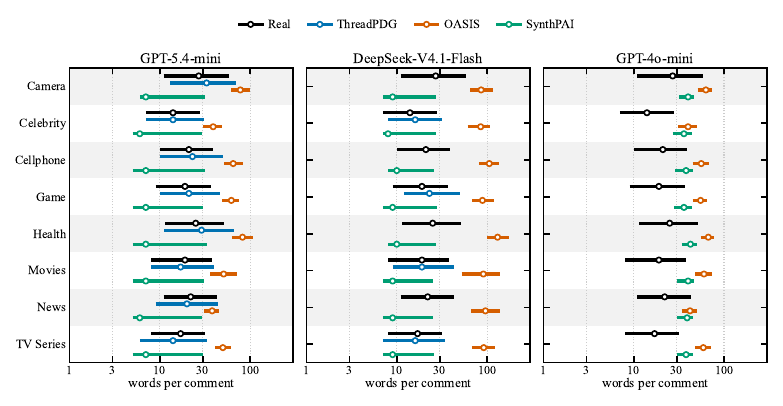}
\caption{Comment length for every domain and LLM. The dot is the median and the
bar the interquartile range, on a logarithmic axis.}
\label{fig:app_length}
\end{figure}

\textbf{Comment length.}
Figure~\ref{fig:app_length} shows the same pattern on every LLM. OASIS writes
the longest comments: its median is 38 to 82 words on \texttt{gpt-5.4-mini} and
85 to 131 words on \texttt{DeepSeek-V4.1-Flash}, against 14 to 27 words for
real. SynthPAI writes the shortest on these two LLMs, with a median of 6 to 10
words, but on \texttt{gpt-4o-mini} its median rises to 36 to 42 words and is
longer than real in every domain. ThreadPDG stays within six words of the real
median in every domain.

\begin{figure}[H]
\centering
\includegraphics[width=\textwidth]{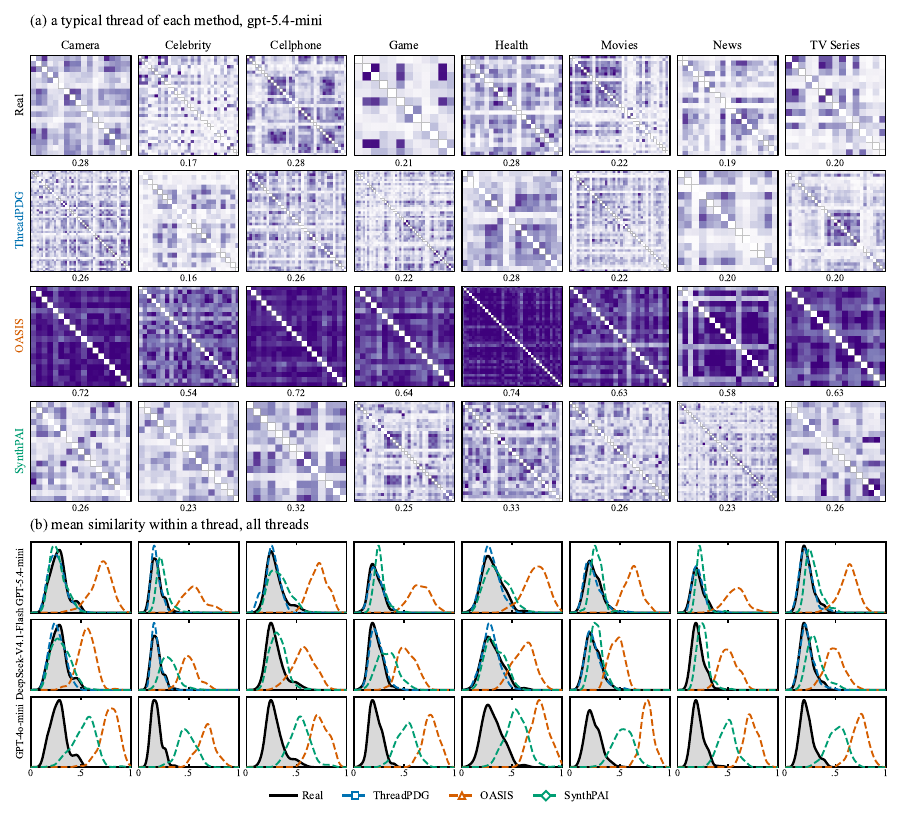}
\caption{Similarity between the comments of one thread, for every domain.
(a) One typical thread per method and domain on \texttt{gpt-5.4-mini}, chosen
and drawn as in Figure~\ref{fig:qual_semantic}a; the number is its mean
similarity. (b) The mean similarity within a thread over all threads, one row
per LLM.}
\label{fig:app_semantic}
\end{figure}

\textbf{Semantic diversity.}
Figure~\ref{fig:app_semantic} repeats Figure~\ref{fig:qual_semantic}a and b for
every domain. The order from the main text holds on every LLM and in every
domain. With \texttt{gpt-5.4-mini}, the mean similarity between two comments of
a thread is 0.18 to 0.30 for real, 0.17 to 0.29 for ThreadPDG, 0.23 to 0.35 for
SynthPAI and 0.56 to 0.73 for OASIS. With \texttt{DeepSeek-V4.1-Flash}, OASIS
is 2.0 to 2.9 times the real value, SynthPAI 1.1 to 1.7 times, and ThreadPDG 0.94
to 1.07 times in the eight domains where it has a run. With \texttt{gpt-4o-mini},
OASIS is 2.6 to 3.8 times the real value and SynthPAI 1.8 to 2.7 times. Short
comments do not cause the gap: on \texttt{gpt-5.4-mini}, when we keep only
comments of at least 15 words, the mean similarity is 0.23 to 0.36 for real,
0.25 to 0.36 for ThreadPDG, 0.40 to 0.56 for SynthPAI and 0.57 to 0.73 for OASIS.

\begin{figure}[H]
\centering
\includegraphics[width=\textwidth]{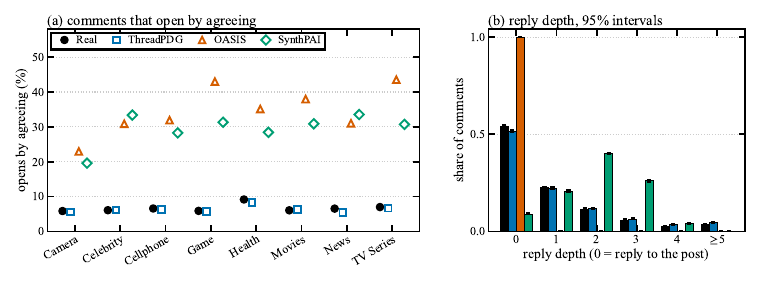}
\caption{Openings and reply depth on \texttt{gpt-5.4-mini}. (a) The share of
comments whose first word agrees, such as \emph{yeah}, \emph{exactly} or
\emph{same}. (b) The
share of comments at each reply depth, with 95\% Wilson intervals; depth 0 is
a reply to the post.}
\label{fig:app_turns}
\end{figure}

\textbf{Openings and replies.}
Figure~\ref{fig:app_turns}a shows that both baselines open far more comments by
agreeing than real users do, in every domain: 23\% to 44\% for OASIS and 20\%
to 34\% for SynthPAI, against 6\% to 9\% for real and 5\% to 8\% for ThreadPDG.
Figure~\ref{fig:app_turns}b shows the reply depth. OASIS places every comment
at depth 0, so it never answers another user. SynthPAI places most comments
two or more levels down, and ThreadPDG follows the real distribution at every
depth.

\begin{figure}[H]
\scriptsize
\vspace{-\intextsep}
\centering
\includegraphics[width=0.40\textwidth]{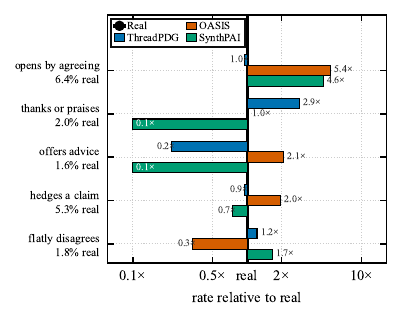}
\caption{Rates of five observable comment patterns relative to real
comments. Agreement and disagreement markers refer to how a comment begins;
advice and hedging markers indicate the corresponding expressions in the
comment.}
\label{fig:qual_tone}
\end{figure}

\newpage
\section{Additional Distance Measures}
\label{app:distance_measures}

In addition to the MWU and KS tests, we quantify the magnitude of the discrepancy between the real and generated metric samples, \(R_f\) and \(G_f\), using the Wasserstein distance and absolute Cliff's delta:
\[
\begin{aligned}
W_1(R_f,G_f)
&= \int_{-\infty}^{\infty}
\left|F_{R_f}(x)-F_{G_f}(x)\right|\,dx, \\
\left|\delta_{\mathrm{Cliff}}(R_f,G_f)\right|
&=
\left|
\frac{1}{n_Rn_G}
\sum_{i=1}^{n_R}\sum_{j=1}^{n_G}
\operatorname{sgn}(f(g_j)-f(r_i))
\right|.
\end{aligned}
\]
Here, \(F_{R_f}\) and \(F_{G_f}\) denote the corresponding empirical CDFs, and \(n_R\) and \(n_G\) are the sample sizes. The function \(\operatorname{sgn}\) returns \(1\), \(0\), or \(-1\) according to the sign of its argument.

The Wasserstein distance measures discrepancy on each metric's native numerical scale. Absolute Cliff's delta measures rank-level separation and is related to the MWU statistic. Smaller values indicate smaller discrepancies in these respective senses.

We use these measures to supplement the statistical tests. The MWU and KS tests provide the pass decisions, while the distance measures describe the magnitude of the observed differences. We do not use the distance measures to determine whether a metric passes. Table~\ref{tab:full_domain_wasserstein} and Table~\ref{tab:full_domain_cliffs_delta} report both measures for every arm, domain, and metric, computed on the same two samples the $p$-values in Table~\ref{tab:holdout_pvalues} come from.

\begin{table*}[htbp]
\centering
\footnotesize
\large\setlength{\tabcolsep}{13pt}\renewcommand{\arraystretch}{0.90}
\caption{Wasserstein distance $W_1$ between simulated and held-out real threads, on each metric's native scale, over the same comparisons as Table~\ref{tab:holdout_pvalues} and one arm more. Smaller is closer to real. \textbf{Mean} averages the twelve columns, which mixes scales and is a summary only.}
\label{tab:full_domain_wasserstein}
\resizebox{\textwidth}{!}{%
\begin{tabular}{@{}lll|ccc|ccc|cc|cccc|c@{}}
\toprule
& & & \multicolumn{3}{c}{\textit{Uniformity}} & \multicolumn{3}{c}{\textit{Interaction}} & \multicolumn{2}{c}{\textit{Expression}} & \multicolumn{4}{c}{\textit{Tone}} & \\
\cmidrule(lr){4-6}\cmidrule(lr){7-9}\cmidrule(lr){10-11}\cmidrule(lr){12-15}
\textbf{Domain} & \textbf{Method} & \textbf{Model} & \rotatebox{90}{Self-BLEU} & \rotatebox{90}{Self-BERT} & \rotatebox{90}{Sem.Cos} & \rotatebox{90}{Depth} & \rotatebox{90}{Virality} & \rotatebox{90}{Len.CV} & \rotatebox{90}{Story} & \rotatebox{90}{Emotion} & \rotatebox{90}{Polite} & \rotatebox{90}{Neutral} & \rotatebox{90}{Impolite} & \rotatebox{90}{Disagr.} & \textbf{Mean} \\
\midrule
\multirow{10}{*}{\rotatebox{90}{\textbf{Camera}}} & \multirow{3}{*}{OASIS} & \texttt{DeepSeek-V4.1-Flash} & .019 & .079 & .301 & 1.189 & 2.247 & .568 & .082 & .181 & .277 & .153 & .055 & .046 & .433 \\
 &  & \texttt{gpt-5.4-mini} & .094 & .162 & .435 & 1.189 & 2.247 & .706 & .119 & .821 & .074 & .124 & .224 & .082 & .523 \\
 &  & \texttt{gpt-4o-mini} & .076 & .207 & .511 & 1.189 & 2.247 & .730 & .116 & .302 & .614 & .146 & .372 & .068 & .548 \\
\cmidrule(l){2-16}
 & \multirow{4}{*}{SynthPAI} & \texttt{DeepSeek-V4.1-Flash} & .047 & .049 & .020 & .315 & .363 & .263 & .036 & .375 & .281 & .067 & .397 & .137 & .196 \\
 &  & \texttt{gpt-5.4-mini} & .062 & .060 & .016 & .569 & .791 & .142 & .035 & .565 & .261 & .029 & .303 & .092 & .244 \\
 &  & \texttt{gpt-4o-mini} & .021 & .124 & .176 & .268 & .324 & .691 & .097 & .200 & .052 & .131 & .138 & .129 & .196 \\
\cmidrule(l){2-16}
 & \multirow{3}{*}{\textbf{ThreadPDG}} & \texttt{DeepSeek-V4.1-Flash} & .002 & .005 & .011 & .090 & .139 & .047 & .017 & .068 & .041 & .017 & .064 & .022 & .044 \\
 &  & \texttt{gpt-5.4-mini} & .002 & .004 & .009 & .088 & .144 & .041 & .013 & .055 & .062 & .024 & .018 & .018 & .040 \\
 &  & \texttt{gpt-4o-mini} & .005 & .039 & .048 & .110 & .214 & .176 & .046 & .197 & .157 & .043 & .146 & .040 & .102 \\
\midrule
\multirow{10}{*}{\rotatebox{90}{\textbf{Celebrity}}} & \multirow{3}{*}{OASIS} & \texttt{DeepSeek-V4.1-Flash} & .015 & .059 & .346 & .728 & 1.946 & .608 & .160 & .558 & .140 & .085 & .053 & .144 & .404 \\
 &  & \texttt{gpt-5.4-mini} & .037 & .127 & .379 & .728 & 1.946 & .706 & .195 & .699 & .166 & .071 & .273 & .058 & .449 \\
 &  & \texttt{gpt-4o-mini} & .025 & .162 & .495 & .728 & 1.946 & .680 & .207 & .416 & .428 & .080 & .325 & .102 & .466 \\
\cmidrule(l){2-16}
 & \multirow{4}{*}{SynthPAI} & \texttt{DeepSeek-V4.1-Flash} & .029 & .063 & .058 & .414 & .797 & .304 & .101 & .165 & .161 & .025 & .215 & .108 & .203 \\
 &  & \texttt{gpt-5.4-mini} & .047 & .069 & .046 & .876 & .912 & .169 & .133 & .461 & .169 & .023 & .202 & .052 & .263 \\
 &  & \texttt{gpt-4o-mini} & .011 & .119 & .237 & .637 & .485 & .701 & .207 & .280 & .136 & .093 & .207 & .307 & .285 \\
\cmidrule(l){2-16}
 & \multirow{3}{*}{\textbf{ThreadPDG}} & \texttt{DeepSeek-V4.1-Flash} & .003 & .002 & .005 & .070 & .220 & .092 & .021 & .061 & .029 & .010 & .014 & .016 & .045 \\
 &  & \texttt{gpt-5.4-mini} & .004 & .003 & .006 & .070 & .221 & .078 & .024 & .079 & .045 & .011 & .021 & .019 & .048 \\
 &  & \texttt{gpt-4o-mini} & .011 & .023 & .056 & .090 & .176 & .118 & .043 & .221 & .057 & .033 & .026 & .092 & .079 \\
\midrule
\multirow{10}{*}{\rotatebox{90}{\textbf{Cellphone}}} & \multirow{3}{*}{OASIS} & \texttt{DeepSeek-V4.1-Flash} & .013 & .074 & .309 & 1.164 & 2.356 & .629 & .185 & .596 & .149 & .185 & .113 & .015 & .482 \\
 &  & \texttt{gpt-5.4-mini} & .092 & .166 & .431 & 1.164 & 2.356 & .720 & .228 & .576 & .097 & .193 & .285 & .061 & .531 \\
 &  & \texttt{gpt-4o-mini} & .053 & .178 & .445 & 1.164 & 2.356 & .720 & .206 & .172 & .481 & .189 & .260 & .094 & .527 \\
\cmidrule(l){2-16}
 & \multirow{4}{*}{SynthPAI} & \texttt{DeepSeek-V4.1-Flash} & .041 & .064 & .023 & .430 & .389 & .301 & .107 & .147 & .247 & .141 & .426 & .126 & .203 \\
 &  & \texttt{gpt-5.4-mini} & .061 & .068 & .016 & .558 & .636 & .088 & .118 & .389 & .236 & .063 & .293 & .130 & .221 \\
 &  & \texttt{gpt-4o-mini} & .018 & .121 & .192 & .285 & .290 & .683 & .190 & .140 & .153 & .208 & .290 & .176 & .229 \\
\cmidrule(l){2-16}
 & \multirow{3}{*}{\textbf{ThreadPDG}} & \texttt{DeepSeek-V4.1-Flash} & .010 & .008 & .018 & .122 & .264 & .131 & .018 & .124 & .038 & .013 & .048 & .038 & .069 \\
 &  & \texttt{gpt-5.4-mini} & .011 & .005 & .032 & .130 & .131 & .077 & .015 & .072 & .076 & .027 & .039 & .024 & .053 \\
 &  & \texttt{gpt-4o-mini} & .008 & .035 & .039 & .176 & .426 & .138 & .049 & .267 & .024 & .026 & .060 & .020 & .106 \\
\midrule
\multirow{10}{*}{\rotatebox{90}{\textbf{Game}}} & \multirow{3}{*}{OASIS} & \texttt{DeepSeek-V4.1-Flash} & .004 & .063 & .328 & .965 & 2.343 & .659 & .099 & .432 & .294 & .124 & .121 & .040 & .456 \\
 &  & \texttt{gpt-5.4-mini} & .053 & .145 & .436 & .965 & 2.343 & .769 & .145 & .635 & .080 & .128 & .189 & .092 & .498 \\
 &  & \texttt{gpt-4o-mini} & .063 & .208 & .544 & .965 & 2.343 & .795 & .144 & .597 & .761 & .139 & .561 & .114 & .603 \\
\cmidrule(l){2-16}
 & \multirow{4}{*}{SynthPAI} & \texttt{DeepSeek-V4.1-Flash} & .039 & .065 & .079 & .805 & 1.749 & .319 & .056 & .499 & .131 & .077 & .259 & .036 & .343 \\
 &  & \texttt{gpt-5.4-mini} & .054 & .056 & .017 & .786 & .712 & .152 & .069 & .465 & .131 & .019 & .183 & .036 & .223 \\
 &  & \texttt{gpt-4o-mini} & .014 & .115 & .237 & .458 & .441 & .733 & .134 & .198 & .073 & .126 & .119 & .217 & .239 \\
\cmidrule(l){2-16}
 & \multirow{3}{*}{\textbf{ThreadPDG}} & \texttt{DeepSeek-V4.1-Flash} & .002 & .004 & .013 & .172 & .378 & .032 & .009 & .071 & .025 & .012 & .028 & .061 & .067 \\
 &  & \texttt{gpt-5.4-mini} & .002 & .005 & .009 & .163 & .313 & .050 & .007 & .039 & .027 & .012 & .021 & .055 & .059 \\
 &  & \texttt{gpt-4o-mini} & .010 & .031 & .071 & .199 & .402 & .190 & .011 & .161 & .103 & .028 & .097 & .015 & .110 \\
\midrule
\multirow{10}{*}{\rotatebox{90}{\textbf{Health Issue}}} & \multirow{3}{*}{OASIS} & \texttt{DeepSeek-V4.1-Flash} & .021 & .084 & .355 & .792 & 1.524 & .618 & .053 & .215 & .035 & .139 & .177 & .133 & .346 \\
 &  & \texttt{gpt-5.4-mini} & .129 & .178 & .435 & .792 & 1.524 & .646 & .150 & .451 & .174 & .104 & .309 & .026 & .410 \\
 &  & \texttt{gpt-4o-mini} & .114 & .230 & .465 & .792 & 1.524 & .776 & .228 & 1.058 & .658 & .139 & .454 & .030 & .539 \\
\cmidrule(l){2-16}
 & \multirow{4}{*}{SynthPAI} & \texttt{DeepSeek-V4.1-Flash} & .041 & .067 & .013 & .760 & 1.056 & .297 & .084 & .107 & .258 & .078 & .341 & .112 & .268 \\
 &  & \texttt{gpt-5.4-mini} & .066 & .071 & .013 & .845 & 1.253 & .165 & .123 & .521 & .274 & .030 & .318 & .062 & .312 \\
 &  & \texttt{gpt-4o-mini} & .024 & .129 & .190 & .648 & .890 & .698 & .207 & .565 & .197 & .123 & .312 & .168 & .346 \\
\cmidrule(l){2-16}
 & \multirow{3}{*}{\textbf{ThreadPDG}} & \texttt{DeepSeek-V4.1-Flash} & .006 & .008 & .017 & .095 & .166 & .051 & .020 & .062 & .101 & .043 & .128 & .038 & .061 \\
 &  & \texttt{gpt-5.4-mini} & .004 & .003 & .020 & .093 & .163 & .070 & .013 & .108 & .118 & .037 & .055 & .034 & .060 \\
 &  & \texttt{gpt-4o-mini} & .007 & .034 & .040 & .089 & .168 & .177 & .051 & .218 & .072 & .025 & .097 & .048 & .085 \\
\midrule
\multirow{10}{*}{\rotatebox{90}{\textbf{Movies}}} & \multirow{3}{*}{OASIS} & \texttt{DeepSeek-V4.1-Flash} & .013 & .039 & .258 & .968 & 2.191 & .331 & .031 & .266 & .169 & .111 & .031 & .055 & .372 \\
 &  & \texttt{gpt-5.4-mini} & .050 & .143 & .392 & .968 & 2.191 & .533 & .127 & .412 & .156 & .083 & .296 & .096 & .454 \\
 &  & \texttt{gpt-4o-mini} & .061 & .209 & .534 & .968 & 2.191 & .623 & .171 & .539 & .731 & .124 & .523 & .055 & .561 \\
\cmidrule(l){2-16}
 & \multirow{4}{*}{SynthPAI} & \texttt{DeepSeek-V4.1-Flash} & .034 & .060 & .022 & .836 & .771 & .215 & .072 & .063 & .180 & .087 & .315 & .166 & .235 \\
 &  & \texttt{gpt-5.4-mini} & .051 & .064 & .018 & .886 & .858 & .128 & .035 & .296 & .187 & .034 & .259 & .115 & .244 \\
 &  & \texttt{gpt-4o-mini} & .011 & .117 & .252 & .525 & .328 & .605 & .152 & .125 & .125 & .117 & .238 & .342 & .245 \\
\cmidrule(l){2-16}
 & \multirow{3}{*}{\textbf{ThreadPDG}} & \texttt{DeepSeek-V4.1-Flash} & .003 & .002 & .008 & .122 & .223 & .099 & .014 & .040 & .021 & .008 & .056 & .023 & .052 \\
 &  & \texttt{gpt-5.4-mini} & .004 & .004 & .013 & .145 & .266 & .057 & .014 & .041 & .046 & .015 & .037 & .022 & .055 \\
 &  & \texttt{gpt-4o-mini} & .007 & .030 & .082 & .147 & .346 & .061 & .025 & .143 & .100 & .048 & .043 & .080 & .093 \\
\midrule
\multirow{10}{*}{\rotatebox{90}{\textbf{News}}} & \multirow{3}{*}{OASIS} & \texttt{DeepSeek-V4.1-Flash} & .008 & .068 & .332 & 1.060 & 2.421 & .571 & .065 & .309 & .273 & .177 & .134 & .069 & .457 \\
 &  & \texttt{gpt-5.4-mini} & .073 & .156 & .370 & 1.060 & 2.421 & .802 & .097 & .642 & .088 & .073 & .164 & .144 & .508 \\
 &  & \texttt{gpt-4o-mini} & .041 & .212 & .532 & 1.060 & 2.421 & .779 & .111 & .345 & .559 & .129 & .382 & .128 & .558 \\
\cmidrule(l){2-16}
 & \multirow{4}{*}{SynthPAI} & \texttt{DeepSeek-V4.1-Flash} & .034 & .061 & .026 & .679 & .595 & .346 & .047 & .176 & .077 & .144 & .232 & .092 & .209 \\
 &  & \texttt{gpt-5.4-mini} & .058 & .061 & .016 & .843 & .817 & .144 & .028 & .125 & .097 & .111 & .238 & .048 & .215 \\
 &  & \texttt{gpt-4o-mini} & .013 & .125 & .223 & .435 & .299 & .754 & .101 & .056 & .061 & .182 & .264 & .327 & .237 \\
\cmidrule(l){2-16}
 & \multirow{3}{*}{\textbf{ThreadPDG}} & \texttt{DeepSeek-V4.1-Flash} & .005 & .003 & .018 & .170 & .314 & .096 & .013 & .056 & .021 & .037 & .077 & .039 & .071 \\
 &  & \texttt{gpt-5.4-mini} & .003 & .004 & .010 & .130 & .230 & .030 & .006 & .057 & .045 & .048 & .080 & .027 & .056 \\
 &  & \texttt{gpt-4o-mini} & .015 & .033 & .089 & .135 & .269 & .152 & .008 & .336 & .042 & .060 & .022 & .069 & .103 \\
\midrule
\multirow{10}{*}{\rotatebox{90}{\textbf{TV Series}}} & \multirow{3}{*}{OASIS} & \texttt{DeepSeek-V4.1-Flash} & .016 & .051 & .298 & .705 & 1.848 & .531 & .129 & .452 & .163 & .137 & .052 & .016 & .367 \\
 &  & \texttt{gpt-5.4-mini} & .044 & .128 & .409 & .705 & 1.848 & .663 & .193 & .399 & .123 & .125 & .265 & .134 & .420 \\
 &  & \texttt{gpt-4o-mini} & .034 & .176 & .526 & .705 & 1.848 & .670 & .221 & .224 & .527 & .159 & .315 & .233 & .470 \\
\cmidrule(l){2-16}
 & \multirow{4}{*}{SynthPAI} & \texttt{DeepSeek-V4.1-Flash} & .027 & .061 & .035 & .673 & .759 & .242 & .112 & .104 & .140 & .093 & .277 & .184 & .226 \\
 &  & \texttt{gpt-5.4-mini} & .045 & .061 & .051 & 1.035 & 1.138 & .142 & .097 & .321 & .150 & .064 & .231 & .124 & .288 \\
 &  & \texttt{gpt-4o-mini} & .011 & .115 & .215 & .669 & .437 & .640 & .201 & .139 & .102 & .151 & .267 & .379 & .277 \\
\cmidrule(l){2-16}
 & \multirow{3}{*}{\textbf{ThreadPDG}} & \texttt{DeepSeek-V4.1-Flash} & .004 & .005 & .008 & .110 & .164 & .100 & .013 & .079 & .020 & .026 & .041 & .011 & .048 \\
 &  & \texttt{gpt-5.4-mini} & .009 & .003 & .008 & .106 & .153 & .060 & .013 & .098 & .025 & .023 & .015 & .020 & .044 \\
 &  & \texttt{gpt-4o-mini} & .008 & .016 & .051 & .108 & .148 & .063 & .024 & .169 & .078 & .047 & .039 & .081 & .069 \\
\bottomrule
\end{tabular}
}
\end{table*}

\begin{table*}[htbp]
\centering
\footnotesize
\large\setlength{\tabcolsep}{13pt}\renewcommand{\arraystretch}{0.90}
\caption{Absolute Cliff's delta $|\delta_{\mathrm{Cliff}}|$ between simulated and held-out real threads, over the same comparisons as Table~\ref{tab:full_domain_wasserstein}. The measure is scale-free, so columns are comparable; $|\delta|<0.147$, the conventional negligible band, is shown in bold.}
\label{tab:full_domain_cliffs_delta}
\resizebox{\textwidth}{!}{%
\begin{tabular}{@{}lll|ccc|ccc|cc|cccc|c@{}}
\toprule
& & & \multicolumn{3}{c}{\textit{Uniformity}} & \multicolumn{3}{c}{\textit{Interaction}} & \multicolumn{2}{c}{\textit{Expression}} & \multicolumn{4}{c}{\textit{Tone}} & \\
\cmidrule(lr){4-6}\cmidrule(lr){7-9}\cmidrule(lr){10-11}\cmidrule(lr){12-15}
\textbf{Domain} & \textbf{Method} & \textbf{Model} & \rotatebox{90}{Self-BLEU} & \rotatebox{90}{Self-BERT} & \rotatebox{90}{Sem.Cos} & \rotatebox{90}{Depth} & \rotatebox{90}{Virality} & \rotatebox{90}{Len.CV} & \rotatebox{90}{Story} & \rotatebox{90}{Emotion} & \rotatebox{90}{Polite} & \rotatebox{90}{Neutral} & \rotatebox{90}{Impolite} & \rotatebox{90}{Disagr.} & \textbf{Mean} \\
\midrule
\multirow{10}{*}{\rotatebox{90}{\textbf{Camera}}} & \multirow{3}{*}{OASIS} & \texttt{DeepSeek-V4.1-Flash} & .703 & .981 & .977 & .947 & .947 & .980 & .665 & .260 & .657 & .852 & \textbf{.122} & \textbf{.050} & .678 \\
 &  & \texttt{gpt-5.4-mini} & .982 & 1.000 & .994 & .947 & .947 & 1.000 & .972 & .829 & .201 & .692 & .532 & .314 & .784 \\
 &  & \texttt{gpt-4o-mini} & .978 & 1.000 & .999 & .947 & .947 & .998 & .959 & .405 & .957 & .805 & .901 & .224 & .843 \\
\cmidrule(l){2-16}
 & \multirow{4}{*}{SynthPAI} & \texttt{DeepSeek-V4.1-Flash} & .947 & .851 & \textbf{.122} & .179 & \textbf{.060} & .612 & .235 & .485 & .857 & .406 & .933 & .551 & .520 \\
 &  & \texttt{gpt-5.4-mini} & .976 & .921 & \textbf{.040} & .456 & .412 & \textbf{.050} & .179 & .691 & .824 & \textbf{.145} & .848 & .425 & .497 \\
 &  & \texttt{gpt-4o-mini} & .808 & 1.000 & .736 & \textbf{.138} & \textbf{.013} & 1.000 & .794 & .307 & \textbf{.129} & .738 & .338 & .369 & .531 \\
\cmidrule(l){2-16}
 & \multirow{3}{*}{\textbf{ThreadPDG}} & \texttt{DeepSeek-V4.1-Flash} & \textbf{.051} & \textbf{.007} & \textbf{.027} & \textbf{.012} & \textbf{.017} & \textbf{.085} & \textbf{.097} & \textbf{.045} & \textbf{.020} & \textbf{.097} & .173 & \textbf{.108} & .062 \\
 &  & \texttt{gpt-5.4-mini} & \textbf{.103} & \textbf{.074} & \textbf{.024} & \textbf{.002} & \textbf{.026} & \textbf{.026} & \textbf{.091} & \textbf{.030} & .161 & \textbf{.124} & \textbf{.015} & \textbf{.086} & .064 \\
 &  & \texttt{gpt-4o-mini} & .150 & .689 & .332 & \textbf{.056} & \textbf{.083} & .399 & .279 & .270 & .489 & .254 & .491 & .236 & .311 \\
\midrule
\multirow{10}{*}{\rotatebox{90}{\textbf{Celebrity}}} & \multirow{3}{*}{OASIS} & \texttt{DeepSeek-V4.1-Flash} & .537 & .966 & 1.000 & .939 & .939 & .918 & .801 & .614 & .425 & .699 & \textbf{.033} & .438 & .692 \\
 &  & \texttt{gpt-5.4-mini} & .667 & 1.000 & 1.000 & .939 & .939 & .962 & .940 & .760 & .714 & .584 & .822 & \textbf{.082} & .784 \\
 &  & \texttt{gpt-4o-mini} & .630 & 1.000 & 1.000 & .939 & .939 & .953 & .973 & .522 & .794 & .655 & .704 & \textbf{.123} & .769 \\
\cmidrule(l){2-16}
 & \multirow{4}{*}{SynthPAI} & \texttt{DeepSeek-V4.1-Flash} & .756 & .960 & .522 & .430 & .312 & .646 & .535 & .204 & .680 & .180 & .725 & .323 & .523 \\
 &  & \texttt{gpt-5.4-mini} & .878 & .976 & .529 & .726 & .509 & \textbf{.119} & .706 & .602 & .722 & .167 & .692 & .165 & .566 \\
 &  & \texttt{gpt-4o-mini} & .218 & 1.000 & .955 & .603 & .270 & .963 & .963 & .375 & .535 & .766 & .670 & .721 & .670 \\
\cmidrule(l){2-16}
 & \multirow{3}{*}{\textbf{ThreadPDG}} & \texttt{DeepSeek-V4.1-Flash} & \textbf{.021} & \textbf{.022} & \textbf{.009} & \textbf{.040} & \textbf{.076} & \textbf{.142} & \textbf{.043} & \textbf{.031} & \textbf{.027} & \textbf{.030} & \textbf{.030} & \textbf{.021} & .041 \\
 &  & \texttt{gpt-5.4-mini} & \textbf{.112} & \textbf{.076} & \textbf{.008} & \textbf{.041} & \textbf{.081} & \textbf{.098} & \textbf{.085} & \textbf{.050} & \textbf{.142} & \textbf{.040} & \textbf{.072} & \textbf{.059} & .072 \\
 &  & \texttt{gpt-4o-mini} & .386 & .511 & .521 & \textbf{.022} & \textbf{.008} & .160 & .217 & .270 & .232 & .232 & \textbf{.050} & .366 & .248 \\
\midrule
\multirow{10}{*}{\rotatebox{90}{\textbf{Cellphone}}} & \multirow{3}{*}{OASIS} & \texttt{DeepSeek-V4.1-Flash} & .505 & .993 & .961 & 1.000 & 1.000 & .976 & .871 & .665 & .452 & .826 & .305 & \textbf{.014} & .714 \\
 &  & \texttt{gpt-5.4-mini} & .998 & 1.000 & .998 & 1.000 & 1.000 & .989 & .982 & .647 & .300 & .856 & .702 & \textbf{.008} & .790 \\
 &  & \texttt{gpt-4o-mini} & .977 & 1.000 & .999 & 1.000 & 1.000 & .991 & .937 & .175 & .913 & .847 & .705 & .173 & .810 \\
\cmidrule(l){2-16}
 & \multirow{4}{*}{SynthPAI} & \texttt{DeepSeek-V4.1-Flash} & .982 & .979 & \textbf{.113} & .398 & .225 & .839 & .557 & \textbf{.112} & .905 & .698 & .972 & .626 & .617 \\
 &  & \texttt{gpt-5.4-mini} & .998 & .983 & \textbf{.038} & .515 & .385 & .150 & .621 & .491 & .881 & .334 & .799 & .528 & .560 \\
 &  & \texttt{gpt-4o-mini} & .866 & 1.000 & .812 & .241 & \textbf{.041} & .983 & .863 & \textbf{.137} & .563 & .912 & .733 & .547 & .641 \\
\cmidrule(l){2-16}
 & \multirow{3}{*}{\textbf{ThreadPDG}} & \texttt{DeepSeek-V4.1-Flash} & \textbf{.139} & \textbf{.100} & \textbf{.010} & \textbf{.065} & \textbf{.106} & .150 & \textbf{.023} & \textbf{.127} & \textbf{.109} & \textbf{.003} & .160 & .250 & .103 \\
 &  & \texttt{gpt-5.4-mini} & .262 & \textbf{.006} & \textbf{.135} & \textbf{.028} & \textbf{.010} & \textbf{.091} & \textbf{.009} & \textbf{.018} & .267 & \textbf{.100} & \textbf{.123} & \textbf{.121} & .098 \\
 &  & \texttt{gpt-4o-mini} & \textbf{.050} & .645 & .273 & .151 & .206 & .315 & .216 & .320 & \textbf{.062} & \textbf{.106} & .197 & \textbf{.136} & .223 \\
\midrule
\multirow{10}{*}{\rotatebox{90}{\textbf{Game}}} & \multirow{3}{*}{OASIS} & \texttt{DeepSeek-V4.1-Flash} & \textbf{.135} & .979 & .998 & .987 & .987 & .987 & .673 & .556 & .685 & .812 & .223 & \textbf{.131} & .679 \\
 &  & \texttt{gpt-5.4-mini} & .908 & 1.000 & 1.000 & .987 & .987 & .999 & .933 & .750 & .460 & .825 & .545 & \textbf{.107} & .792 \\
 &  & \texttt{gpt-4o-mini} & .922 & 1.000 & 1.000 & .987 & .987 & 1.000 & .948 & .760 & .984 & .897 & .962 & .391 & .903 \\
\cmidrule(l){2-16}
 & \multirow{4}{*}{SynthPAI} & \texttt{DeepSeek-V4.1-Flash} & .877 & .963 & .450 & .812 & .763 & .769 & .373 & .598 & .727 & .436 & .789 & \textbf{.122} & .640 \\
 &  & \texttt{gpt-5.4-mini} & .931 & .957 & \textbf{.121} & .645 & .386 & \textbf{.136} & .450 & .655 & .751 & \textbf{.061} & .667 & .159 & .493 \\
 &  & \texttt{gpt-4o-mini} & .585 & 1.000 & .946 & .428 & \textbf{.050} & .995 & .875 & .281 & .380 & .822 & .392 & .587 & .612 \\
\cmidrule(l){2-16}
 & \multirow{3}{*}{\textbf{ThreadPDG}} & \texttt{DeepSeek-V4.1-Flash} & \textbf{.023} & \textbf{.118} & \textbf{.104} & \textbf{.099} & .161 & \textbf{.027} & \textbf{.027} & \textbf{.076} & \textbf{.102} & \textbf{.038} & \textbf{.011} & .276 & .089 \\
 &  & \texttt{gpt-5.4-mini} & \textbf{.061} & .148 & \textbf{.071} & \textbf{.087} & \textbf{.123} & \textbf{.082} & \textbf{.006} & \textbf{.026} & \textbf{.145} & \textbf{.032} & \textbf{.049} & .247 & .090 \\
 &  & \texttt{gpt-4o-mini} & .447 & .681 & .513 & \textbf{.131} & .174 & .439 & \textbf{.028} & .254 & .417 & .164 & .295 & \textbf{.030} & .298 \\
\midrule
\multirow{10}{*}{\rotatebox{90}{\textbf{Health Issue}}} & \multirow{3}{*}{OASIS} & \texttt{DeepSeek-V4.1-Flash} & .655 & .969 & .979 & .893 & .893 & .972 & .169 & .262 & \textbf{.129} & .680 & .457 & .401 & .622 \\
 &  & \texttt{gpt-5.4-mini} & .959 & .997 & .992 & .893 & .893 & .977 & .565 & .499 & .613 & .485 & .727 & \textbf{.053} & .721 \\
 &  & \texttt{gpt-4o-mini} & .974 & 1.000 & .999 & .893 & .893 & 1.000 & .887 & .849 & .999 & .677 & .919 & \textbf{.022} & .843 \\
\cmidrule(l){2-16}
 & \multirow{4}{*}{SynthPAI} & \texttt{DeepSeek-V4.1-Flash} & .892 & .927 & \textbf{.050} & .680 & .690 & .605 & .290 & \textbf{.048} & .866 & .282 & .817 & .370 & .543 \\
 &  & \texttt{gpt-5.4-mini} & .954 & .927 & \textbf{.055} & .711 & .726 & \textbf{.030} & .458 & .571 & .913 & \textbf{.104} & .765 & .223 & .536 \\
 &  & \texttt{gpt-4o-mini} & .783 & .992 & .736 & .606 & .589 & .995 & .800 & .532 & .684 & .564 & .760 & .489 & .711 \\
\cmidrule(l){2-16}
 & \multirow{3}{*}{\textbf{ThreadPDG}} & \texttt{DeepSeek-V4.1-Flash} & .157 & .151 & \textbf{.056} & \textbf{.047} & \textbf{.058} & \textbf{.022} & \textbf{.059} & \textbf{.062} & .313 & \textbf{.146} & .330 & \textbf{.070} & .123 \\
 &  & \texttt{gpt-5.4-mini} & \textbf{.052} & \textbf{.007} & \textbf{.085} & \textbf{.048} & \textbf{.055} & \textbf{.083} & \textbf{.020} & \textbf{.127} & .383 & .152 & \textbf{.133} & \textbf{.063} & .101 \\
 &  & \texttt{gpt-4o-mini} & .167 & .561 & .253 & \textbf{.054} & \textbf{.055} & .352 & .178 & .262 & .245 & \textbf{.074} & .270 & .168 & .220 \\
\midrule
\multirow{10}{*}{\rotatebox{90}{\textbf{Movies}}} & \multirow{3}{*}{OASIS} & \texttt{DeepSeek-V4.1-Flash} & .538 & .884 & .972 & .974 & .974 & .814 & \textbf{.146} & .394 & .558 & .724 & \textbf{.091} & .216 & .607 \\
 &  & \texttt{gpt-5.4-mini} & .897 & 1.000 & .997 & .974 & .974 & .977 & .698 & .546 & .715 & .553 & .791 & .222 & .779 \\
 &  & \texttt{gpt-4o-mini} & .937 & 1.000 & 1.000 & .974 & .974 & .998 & .986 & .689 & .996 & .821 & .979 & .148 & .875 \\
\cmidrule(l){2-16}
 & \multirow{4}{*}{SynthPAI} & \texttt{DeepSeek-V4.1-Flash} & .888 & .978 & .155 & .624 & .420 & .579 & .348 & \textbf{.044} & .830 & .550 & .902 & .585 & .575 \\
 &  & \texttt{gpt-5.4-mini} & .958 & .966 & \textbf{.133} & .674 & .489 & .178 & .148 & .427 & .870 & .192 & .790 & .412 & .520 \\
 &  & \texttt{gpt-4o-mini} & .483 & 1.000 & .921 & .442 & \textbf{.129} & .991 & .862 & .173 & .572 & .768 & .666 & .747 & .646 \\
\cmidrule(l){2-16}
 & \multirow{3}{*}{\textbf{ThreadPDG}} & \texttt{DeepSeek-V4.1-Flash} & \textbf{.003} & \textbf{.020} & \textbf{.016} & \textbf{.026} & \textbf{.078} & .197 & \textbf{.053} & \textbf{.018} & \textbf{.041} & \textbf{.022} & .169 & \textbf{.067} & .059 \\
 &  & \texttt{gpt-5.4-mini} & \textbf{.107} & \textbf{.073} & \textbf{.063} & \textbf{.036} & \textbf{.095} & \textbf{.129} & \textbf{.031} & \textbf{.024} & .184 & \textbf{.069} & \textbf{.122} & \textbf{.049} & .082 \\
 &  & \texttt{gpt-4o-mini} & .306 & .513 & .453 & \textbf{.059} & .148 & .163 & \textbf{.078} & .198 & .338 & .292 & \textbf{.128} & .275 & .246 \\
\midrule
\multirow{10}{*}{\rotatebox{90}{\textbf{News}}} & \multirow{3}{*}{OASIS} & \texttt{DeepSeek-V4.1-Flash} & .241 & .945 & 1.000 & .960 & .960 & .970 & .566 & .402 & .553 & .915 & \textbf{.063} & \textbf{.054} & .636 \\
 &  & \texttt{gpt-5.4-mini} & .931 & .998 & .997 & .960 & .960 & 1.000 & .864 & .728 & .641 & .366 & .504 & .529 & .790 \\
 &  & \texttt{gpt-4o-mini} & .871 & 1.000 & 1.000 & .960 & .960 & 1.000 & .981 & .469 & .917 & .702 & .722 & .191 & .814 \\
\cmidrule(l){2-16}
 & \multirow{4}{*}{SynthPAI} & \texttt{DeepSeek-V4.1-Flash} & .858 & .937 & .293 & .544 & .302 & .784 & .378 & .297 & .518 & .811 & .803 & .274 & .567 \\
 &  & \texttt{gpt-5.4-mini} & .935 & .937 & .148 & .653 & .429 & .196 & .164 & .174 & .666 & .656 & .798 & \textbf{.130} & .491 \\
 &  & \texttt{gpt-4o-mini} & .501 & .995 & .938 & .373 & \textbf{.055} & .999 & .908 & \textbf{.053} & .367 & .946 & .841 & .798 & .648 \\
\cmidrule(l){2-16}
 & \multirow{3}{*}{\textbf{ThreadPDG}} & \texttt{DeepSeek-V4.1-Flash} & \textbf{.102} & \textbf{.032} & .167 & \textbf{.060} & \textbf{.019} & .179 & \textbf{.100} & \textbf{.041} & \textbf{.115} & .162 & .263 & .168 & .117 \\
 &  & \texttt{gpt-5.4-mini} & \textbf{.054} & \textbf{.087} & \textbf{.064} & \textbf{.038} & \textbf{.001} & \textbf{.051} & \textbf{.006} & \textbf{.084} & .269 & .220 & .272 & \textbf{.106} & .104 \\
 &  & \texttt{gpt-4o-mini} & .562 & .713 & .657 & \textbf{.051} & \textbf{.005} & .343 & \textbf{.027} & .513 & .210 & .325 & \textbf{.041} & .235 & .307 \\
\midrule
\multirow{10}{*}{\rotatebox{90}{\textbf{TV Series}}} & \multirow{3}{*}{OASIS} & \texttt{DeepSeek-V4.1-Flash} & .514 & .909 & .991 & .987 & .987 & .978 & .656 & .512 & .466 & .741 & \textbf{.121} & \textbf{.016} & .656 \\
 &  & \texttt{gpt-5.4-mini} & .772 & 1.000 & .999 & .987 & .987 & 1.000 & .908 & .462 & .599 & .667 & .680 & .305 & .780 \\
 &  & \texttt{gpt-4o-mini} & .744 & 1.000 & 1.000 & .987 & .987 & .999 & .990 & .279 & .899 & .874 & .661 & .445 & .822 \\
\cmidrule(l){2-16}
 & \multirow{4}{*}{SynthPAI} & \texttt{DeepSeek-V4.1-Flash} & .736 & .950 & .305 & .481 & .373 & .622 & .601 & \textbf{.011} & .648 & .459 & .795 & .562 & .545 \\
 &  & \texttt{gpt-5.4-mini} & .853 & .952 & .380 & .804 & .664 & \textbf{.124} & .554 & .401 & .714 & .293 & .697 & .422 & .571 \\
 &  & \texttt{gpt-4o-mini} & .223 & 1.000 & .920 & .668 & .347 & .996 & .953 & \textbf{.019} & .455 & .818 & .680 & .778 & .655 \\
\cmidrule(l){2-16}
 & \multirow{3}{*}{\textbf{ThreadPDG}} & \texttt{DeepSeek-V4.1-Flash} & \textbf{.073} & \textbf{.116} & \textbf{.049} & \textbf{.044} & \textbf{.032} & .187 & \textbf{.037} & \textbf{.003} & \textbf{.034} & \textbf{.086} & \textbf{.116} & \textbf{.030} & .067 \\
 &  & \texttt{gpt-5.4-mini} & .183 & \textbf{.079} & \textbf{.000} & \textbf{.050} & \textbf{.037} & \textbf{.125} & \textbf{.022} & \textbf{.058} & \textbf{.087} & \textbf{.053} & \textbf{.025} & \textbf{.075} & .066 \\
 &  & \texttt{gpt-4o-mini} & .319 & .324 & .378 & \textbf{.036} & \textbf{.016} & \textbf{.137} & \textbf{.125} & .198 & .322 & .197 & \textbf{.095} & .287 & .203 \\
\bottomrule
\end{tabular}
}
\end{table*}

\newpage
\section{AI Assistance}
\label{app:ai_assistance}

Generative AI tools were used to assist with writing and language editing,
including improving grammar, clarity, and phrasing. They were also used to
assist with code development and debugging. All AI-assisted writing and code
were reviewed, tested, and verified by the authors. The authors take full
responsibility for the final content, implementation, and results of the paper.

\end{document}